# Brain Computer Interface (BCI) based on Electroencephalographic (EEG) patterns due to new cognitive tasks

**SAYED SAKKAFF ZAHMEETH**

**M.Phil.**                                        **2009**

S. S. ZAHMEETH

BRAIN COMPUTER INTERFACE (BCI) BASED ON ELECTROENCEPHALOGRAPHIC (EEG) PATTERNS DUE TO NEW COGNITIVE TASKS

M.Phil.

2009

# Brain Computer Interface (BCI) based on Electroencephalographic (EEG) patterns due to new cognitive tasks

A THESIS PRESENTED BY

SAYED SAKKAFF ZAHMEETH

to the Board of Study in Statistics and Computer Science of the

**POSTGRADUATE INSTITUTE OF SCIENCE**

*in partial fulfillment of the requirement*
*for the award of the degree of*

**MASTER OF PHILOSOPHY**

of the

**UNIVERSITY OF PERADENIYA**

**SRI LANKA**

**2009**

*This thesis is dedicated to……,*

*My dearly loved mummy and to the memory of my beloved daddy*

# DECLARATION

I do hereby declare that the work reported in this project thesis was exclusively carried out by me under the supervision of Prof. Asiri Nanayakkara and Dr. S. R. Kodituwakku. It describes the results of my own independent research except where due reference has been made in the text. No part of this project thesis has been submitted earlier or concurrently for the same or any other degree.

Date: ………………                    ………………………………….

                                                              Signature of the Candidate

Certified by:

1.  Supervisor (Name):  **Prof. Asiri Nanayakkara**              Date: ………………

        (Signature): ………………….………..

2.  Supervisor (Name): **Dr. S. R. Kodituwakku**              Date: ………………

        (Signature): ………………….………..

**PGIS Stamp:**

# Brain Computer Interface (BCI) based on Electroencephalographic (EEG) patterns due to new cognitive tasks

**S. S. Zahmeeth**

Artificial Intelligence Unit

Institute of Fundamental Studies (IFS)

Hanthana Road

Kandy

Sri Lanka

Several new mental tasks were investigated to find their suitability in Brain Computer Interface (BCI). Electroencephalography (EEG) signals were collected after amplification while subjects were performing certain mental tasks. Later, collected EEG signals were analyzed to identify changes in EEG due to these mental tasks. Recordings were carried out using an electro-cap containing 20 EEG electrodes which were placed according to the standard 10 -20 system.

Microsoft windows based software with user friendly features was developed for analyzing and classifying recorded EEG data. With this software, unnecessary frequencies were filtered out with Bandpass filtering.  In order to identify the best feature vector construction method for a given mental task, feature vectors were constructed using Bandpower, Principal Component Analysis, and Downsampling separately. These feature vectors were then classified with Linear Discriminant Analysis, Linear Support Vector Machines, Critical Distance Classifier, Nearest Neighbor Classifiers and their Non-Linear counter parts to find the best performing classifier.

For comparison purposes, performances of already well known mental tasks in BCI community were computed along with that of new mental tasks introduced in this thesis. In the preliminary studies it was found that most promising new mental tasks which could be identified by a BCI system is *imagination of hitting a given square by an imaginary arrow from above (or below) and right, (or left) to the screen*. The group of these mental tasks was named as "*Hit Series*" (HS). A detail investigation of HS was carried out and compared it with the performance of Motor Imagery (MI) events which are the most heavily used mental tasks in EEG based BCI systems.

One subject achieved the maximum average performance for HS, 100% in the binary classifications while 99% in overall combined performance. The best average performances of other two subjects for the same mental tasks were 93% and 87% with the over all performance of 89% and 78%. Performances of same three subjects for mental tasks in MI were relatively poor. The average performances were 92%, 78% and 92% while over all performances were 87%, 69% and 88%.

# ACKNOWLEDGMENTS

At this place I would like to thank many people who helped me to complete this thesis. First of all, Prof. Asiri Nanayakkara, who supervised me throughout this research and introduced me to the challenging BCI research area. I am very grateful to him for the immense amount of time that he dedicated for supervision, for his patience, for his continual support and motivation and knowledge leading me into the appropriate methods and techniques needed in this work. Your knowledge and enormous support has been the key to all my work, thanks! Next, I would also like to thank Dr. S. R. Kodituwakku, who supervised me during this study, for his important help and support. I would like to thank all the volunteers, in particular, Dhamika Wijethunga, Gihan Senerath and Chaturanga Thotawattage who participated in this research as subjects. Very special thanks go to the Director of Institute of Fundamental Studies (IFS), Kandy for encouragement and support for BCI research. Finally, I am very grateful to my family members, who were close to me during these years.

Thank you everybody!

# TABLE OF CONTENTS

# LIST OF TABLES

# LIST OF FIGURES

# LIST OF ABBREVIATIONS

| | |
|---|---|
| A | Auricular |
| A/D | Analog to Digital |
| A1 | Left ear lobe |
| A2 | Right ear lobes |
| AAR | Adaptive Autoregressive |
| AC | Air-Conditioning |
| *Ag* | Silver |
| *AgCl* | Silver-Chloride |
| AI | Auditory Imagery |
| AIMTE | AutoIMTE |
| ALS | Amyotrophic Lateral Sclerosis |
| AR | Autoregressive |
| BCI | Brain–Computer Interface |
| BL | Baseline |
| BLR NN | Bayesian Logistic Regression Neural Network |
| BP | BandPowers |
| C | Central |
| Ca | Calcium |
| Cl | Chloride |
| CNS | Central Nerves System |
| Cz | Central Midline |
| DCon | Data conversion program |
| DH | Down Hit |
| ECG / EKG | Electrocardiogram |
| ECoG | Electrocorticography |
| EEG | Electroencephalography |
| EMG | Electromyogram |
| EOG | Electrooculography |
| ERP | Event Related Potentials |
| F | Frontal |
| FFT | Fast Fourier Transform |
| FIR | Finite Impulse Response |
| FIRNN | Finite Impulse Response Neural Network |
| fMRI | functional Magnetic Resonance Imaging |
| Fp | Frontopolar |
| Fz | Frontal Midline |
| G | Ground electrode |
| GBS | Guillain-Barré Syndrome |
| GUI | Graphical User Interface |
| HMM | Hidden Markov Method |
| HS | Hit Series |
| ICA | Independent Component Analysis |
| IIR | Infinite Impulse Response |
| IMTE | Identification of mental tasks through EEG |
| IOHMM | Input–Output Hidden Markov Method |
| K | Potassium |
| kNN | k Nearest Neighbors |
| LDA | Linear Discriminant Analysis |
| LDC | MindMeld24 Live Data Capture |
| LFM | Left Middle Finger Movement |
| LH | Left Hit |
| LPC | Linear Predictive Coding |
| LVQ NN | Learning Vector Quantization (LVQ) Neural Network |
| MEG | Magnetoencephalography |
| MI | Motor Imagery |
| MLP | Multilayer Perceptron |
| MSE | Mean Square Error |
| Na | Sodium |
| NI | Nasion to Inion |
| NIRS | Near Infrared Spectroscopy |
| O | Occipital |
| OLS | Ordinary Least Square |
| P | Parietal |
| PA | Pre-Auricular |
| PCA | Principal Component Analysis |
| PET | Proton Emission Tomography |
| PSD | Power Spectral Density |
| Pz | Parietal Midline |
| QP | Quadratic Programming |
| Rbf | Gaussian radial basis function kernel |
| RBFNN | Gaussian Radial Basis function kernel neural networks |
| RFM | Right Middle Finger Movement |
| RH | Right Hit |
| SCP | Slow Cortical Potentials |
| SFA | Signal fractional analysis |
| SNI | Spatial Navigation Imagery |
| SNR | Signal to Noise Ratio |
| SPECT | Single Photon Emission Tomography |
| SVM | Support Vector Machines |
| T | Temporal |
| UH | Up Hit |
| z | Midline placement |

# CHAPTER 1

# INTRODUCTION

The principal motivation for me to engage in research in Brain Computer Interface is to provide physically impaired individuals, who lack accurate muscle control but have brain capabilities intact, with an alternative way of communicating with the outside world.

Many physiological disorders such as Amyotrophic Lateral Sclerosis (ALS) or injuries like high level spinal cord injury can disrupt the communication path between the brain and the body. Especially, patients diagnosed with neurological diseases such as Guillain-Barré Syndrome (GBS), subcortical stroke, brainstem stroke, or severe cerebral palsy, may lead to severe or complete motor paralysis. On the other hand healthy individuals may also lose their muscle movements partially or completely due to accidents while cognitive and sensory functions remain intact [1-2]. Those who are paralyzed or having restricted motor abilities may have lost all voluntary muscle control and hence, regrettably, cannot interact with their environment like others do [3-11]. As a result, they are enforced to accept a reduced quality of life and become totally depend on caretakers.

In order to lend a hand to individuals who have lost their muscle movements partially, many effective communication aids have been constructed taking advantage of whatever motor abilities the individuals retain in an intelligent way. However, persons who are completely paralyzed cannot benefit from these devices or technologies since they do not retain any motor abilities [6]. The only possible way for these individuals to communicate effectively with outside world is to make use of the retained somatic sensation, cognition and audition which may still be intact. This is where the Brain Computer Interface technology becomes invaluable.

A Brain–Computer Interface (BCI) sometimes called a direct neural interface or a brain – machine interface, is literally a direct technological interface between a brain and a computer not requiring any motor input from the user [5-6]. It is a system that uses electric, magnetic, or hemodynamic brain signals to control external devices as switches, wheelchairs, computers, or neuroprosthesis.

The foundation of current BCI work was formed in the 1970's and 80's. The idea of Electroencephalography (EEG) - based communication system was first introduced by Jacques Vidal during the period 1973-70. In his research it was shown how brain signals could be used to build up a mental prosthesis. Vidal showed that visual evoked potentials could provide a communication channel to control the movement of a computer cursor [12-13] which is the first appearance of the expression of *brain–computer interface* in the scientific literature. Galin and Ornstein [14] studied the asymmetry of the EEG in the two hemispheres and were able to distinguish two cognitive modes - the verbal (left hemisphere) and spatial (right hemisphere). Using a collection of the mental tasks used in these early studies, Keirn and Aunon [15] showed that it is possible to distinguish between pairs of mental tasks using only the EEG to a high degree of accuracy.

One of the first demonstrations of a BCI system was presented by Farwell and Donchin [4] who based their system on the P300 component of event - related brain potentials. The P300 component is a positive change in the measured EEG signal that occurs about 300 milliseconds following the presentation of a type of stimulus that is anticipated but occurs relatively rarely [16]. During last ten years scientists have developed BCI systems which are based on various characteristics of brain potentials. These include use of Event Related Potentials (ERP), Slow Cortical Potentials (SCP) and mental tasks etc. Birbaumer, et al., [17-18] have shown that subjects can learn to control slow cortical potentials that last from one-half second to several minutes. Babiloni, et al., [19] investigated a BCI for recognizing imagined right or left hand, middle finger movements. Studies performed by Wolpaw, et al., [20] provide details on the mu and beta rhythms that are often used in BCI based on a variety of imagined movement tasks. Presently, several research laboratories around the world and various research groups in Europe and USA are working on such BCI systems, which allow for a direct dialog between man and machine [21-23].

Certain actions in a normal human brain can generate various responses such as metabolic activities or electromagnetic signals which can be detected by appropriate sensors and hence used for controlling BCI systems. As an example, brain activity can produce magnetic fields that can be detected using Magnetoencephalography (MEG). Certain brain activities may produce electrical signals that can be detectable on the scalp or cortical surface or within the brain. Some brain activities can result in metabolic consequences in terms of changes in the blood flow and metabolism and they can be imaged by imaging methods such as functional Magnetic Resonance Imaging (fMRI) [24]. At present, among aforementioned detection methods, because of the cost and physical dimensions, methods that measure the electrical activities of the brain are more favored.

BCI can be either noninvasive or invasive based on what type of method is used to capture the brain signals to build the system [25]. Invasive approach uses Electrocorticography (ECoG) signals recorded from the surface of the brain or action potentials from single neurons in the cerebral cortex which is the thin, folded outer surface of the user's brain, using implanted microelectrodes. Moreover, invasive BCI derives the user's intent from neural action potential or local field potential recorded from within the cerebral cortex or from its surface. This requires the direct implantation of hundreds of electrodes, fixed into tiny array, which is placed in or on the surface of the cortex. These electrodes record the electrical signals from the cortical neurons which are then translated by computer algorithms to drive a specific action. Since recording electrodes are directly implanted in or on the cortex [25-28], the signals produced by them are very high quality. However, invasive BCI face substantial technical difficulties and involve clinical risks. Surgery is required to implant the electrodes which causes unnecessary burden to the user. Further, electrodes are prone to scar-tissue build-up, causing the signal to become weaker or even loose as the body reacts to a foreign object in the brain. As these implants are expected to function well for a long period, they create a risk of infection and may cause other damages to the brain. The drive to develop invasive BCI methods is based in part on the widespread conviction that only invasive BCI will be able to provide users with real – time multidimensional sequential control of a robotic arm or neuroprosthesis [26-29].

In non-invasive BCI, the system records brain signals from scalp of the user by means of various signal capturing techniques such as, Electroencephalography (EEG), Magnetoencephalography (MEG), functional Magnetic Resonance Imaging (fMRI), or Near Infrared Spectroscopy (NIRS) [24]. These signal capturing techniques along with suitable signal processing methods can provide basic communication and control to people with severe disabilities. Unlike invasive systems, which entail the risk associated with any brain surgery, noninvasive systems are basically harmless. Perhaps due to this reason, noninvasive BCI systems have become the most popular and shown the highest assurance in practical neurological rehabilitation.

Non–invasive BCI systems mostly use the EEG signal as the source of information. Since MEG, fMRI, and NIRS are expensive or bulky, and fMRI and NIRS present longtime constants in that they do not measure neural activity directly, but relying instead upon the hemodynamic coupling between neural activity and regional changes in blood flow, they cannot be deployed as ambulatory or portable BCI systems. As a result the majority of BCI systems to date exploit EEG signals [5, 8, 30-36].

Studies using both invasive and noninvasive BCI recording of brain activity have successfully demonstrated the feasibility of controlling mechanical devices. Both methodologies will need improved speed, accuracy, and reliability to provide truly useful system to the general public. A recent study showed that only 12% of published BCI studies use implanted electrodes, 5% use microelectrodes arrays, and more than 80% use EEG signals [21]. The main reason is that the EEG recording equipment is commercially produced and their cost is lower than other brain signal recording technologies. It has a relatively fine temporal resolution (on the millisecond scale), enabling rapid estimates of the user's mental state. Also since no surgery is required for placing electrodes, more individuals are willing to participate in such BCI experiments. In addition, the most EEG data acquisition systems are portable, economically affordable, and importantly non-invasive.

EEG signals are recorded by means of electrodes placed over the scalp and have good temporal resolution. However, their spatial resolution is not as good as that of other recording techniques.

Currently there are two main types of EEG-based BCI systems, namely the systems which use brain activities generated in response to specific visual or auditory stimuli and systems which use activities generated spontaneously by the user. Steady – state - visual evoked potentials [37] and P300 based BCI systems [38] are two examples of the first type. Donchin, et al. [38] have used a common stimulus-driven BCI system based on P300 activities for controlling a virtual keyboard in this system where the user looks at the letter on the keyboard he/she wishes to communicate. The system randomly highlights parts of the keyboard and when, by chance, that part of the keyboard corresponding to the user's choice is highlighted, a so-called P300 mental response is evoked. This response is relatively robust and easy to recognize in the EEG recordings. A disadvantage with this kind of stimulus-driven BCI systems is that the user cannot operate the system in a free manner.

Examples of the latter type (BCI based on spontaneous brain activities) are Event Related Desyncronization (motor imagery) [10, 39-41], Slow Cortical Potential (SCP) [42] and non-motor imagery [43-44]. Spontaneous approach consists of two methods. In the first method, the user is asked to imagine one of the mental tasks of motor imagery, Auditory Imagery (AI), or Spatial Navigation Imagery (SNI). Based on the EEG recordings, the recognized thoughts can be used to control a cursor or provide an alternative interface to a virtual keyboard. In the second method, called the operant conditioning approach, the subject may think of anything or nothing to control the cursor on a computer screen. A review of use of EEG components to control BCI systems can be found in Curran et al [45].

The advantage of the spontaneous activity approach is that the interface is potentially more immediate and flexible to operate since the system may, in principle, be used to directly recognize the mental state of the user. However, compared to stimulus-driven EEG systems, spontaneous EEG systems present some additional difficulties, such as inconsistencies in the user's mental state, due to change of strategies, fatigue, motivation and other physiological and psychological factors. These issues make the correspondence between electrode activity and mental state more difficult to achieve than with stimulus-driven systems. Despite above mentioned difficulties, this thesis primarily concentrates on BCI systems which are based on set of mental tasks.

Mental tasks such as multiplication, imaginary letter composing, cubic rotation, mental counting and various imaginary motor movements are known to alter EEG with various degrees. Most widely used and successful in identifying mental tasks are the imaginary motor movements [40, 44, 46]. However, it has been observed in some severely paralyzed patients that when a person looses his/her ability to move arms or legs, over the time, imaginary motor movements becomes less effective in altering the EEG to a identifiable level [47-48].

In this study, our aim is twofold. The first aim of this investigation is to find new (non-motor) mental tasks which can alter EEG signals and hence can be used in BCI systems effectively. Therefore, mental tasks such as identifying symmetries of an object, imagining inner tone music, imagining various smells, and imagining an arrow hitting a square from up, down, left or right (HS) were tested in this investigation for their suitability to be used in BCI. In addition, for comparison purposes, effectiveness of mental tasks involved in imaginary motor movements (MI) were also investigated.

The performances of above mentioned mental tasks, HS and MI, were tested with three healthy individuals and all of them showed good performances for HS.

The second aim is to investigate several computational techniques for analyzing and classifying EEG signals according to the signatures of mental tasks hidden in them.

This includes finding suitable preprocessing methods, effective and efficient feature vector construction techniques and most accurate classification methods for recognizing mental tasks. In this study several preprocessing methods and feature vector construction methods were investigated for various classification schemes. Parameters in these methods and schemes were optimized to obtain the best overall performance.

Out line of the thesis is as follows. In Chapter 2, human brain structure and its activities are described. Basics of neuroanatomy and physiology including functional cortex divisions and the identification of mental states in EEG are discussed in detail. An introduction to EEG and details of placement of electrodes are given in Chapter 3. EEG signal analyzing methods including preprocessing and feature vector construction methods alone with classifications schemes used in BCI are discussed in Chapter 4. In Chapter 5, hardware used for recording EEG and software developed for analyzing and classifying mental tasks are presented. Experimental methodology and results are discussed in Chapter 6 while discussion and conclusion are made in Chapter 7.

# CHAPTER 2

# BRAIN STRUCTURE AND ACTIVITIES

The research work presented in this thesis describes our effort to construct a BCI system using EEG devices. Although our ability to identify activities of the living human brain is very limited, it is known that some thoughts can be recognized by analyzing EEG signals recorded from the scalp. EEG can only provide information about an extremely small fraction of processes which are due to our thoughts, actions, or consciousness.

## 2.1 Basics of neuroanatomy and physiology

In order to give an idea about how the brain activities and EEG signals are interrelated, first part of this chapter reviews, briefly, basics of neuroanatomy and physiology which are relevant to EEG and BCI. A special emphasis is put on how the anatomical and physiological processes of the brain are causing to generate the EEG signals. In the second part of this chapter, we describe the functional divisions of the cortex and how they contribute to EEG signals according to mental activities.

EEG reflects electrical activity of a multitude of neural population in the Brain. Since EEG is generated as a superposition of different simultaneously active dynamical systems, this signal is extremely complex [1]. We will start from the simplest dynamical system called neuron.

The specialized cells that constitute the functional units of the nervous system are called neurons or nerve cells. It is the conducting cell of the nervous system and present in all complex metazoans. The nervous system of man is made up of innumerable neurons. The total number of neurons in the adult human brain is estimated approximately more than 100 billion neurons. The neurons are linked together in a highly intricate manner to be regarded not merely as simple networks, but as cells that are dedicated for reception, integration, interpretation and transmission of information [3, 5].

Neurons are specialized cells designed mainly for processing and conducting information in the form of electrical signals. Each neuron consists of three distinct parts; the cell body or soma, dendrites and the axon (Figure 2.1).

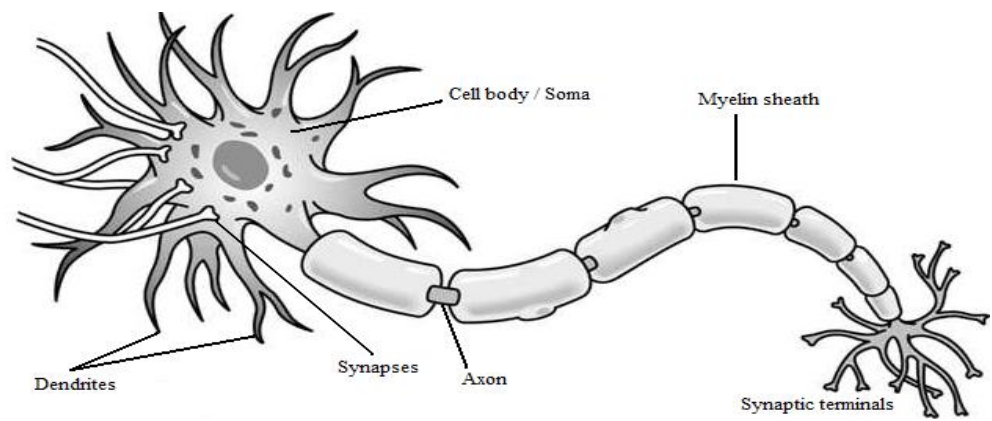

Figure 2.1. Structure of a neuron

The cell body or soma is the enlarged region of the neuron which is surrounded by a cell membrane and contains the nucleus embedded in the granular cytoplasm. Within neuron cytoplasm, typical cell organelles such as mitochondria and Golgi apparatus are present. However, unlike in other cell cytoplasm, centrosomes are not present in neurons. There are also other structures such as the nissl bodies and neurofibrils present in neurocytoplasm which are unique to neurons [2-4].

Dendrites are the extensions of the cell body of a neuron that act to conduct the electrochemical stimulation received from other neural cells to the cell body. Usually a neuron has several dendrites which are highly branched, thread like short processes. Dendrites are connected to either the axons or dendrites of other cells and receive impulses from other neurons or relay the signals to other neurons. In the human brain each nerve cell is connected to approximately 10,000 other nerve cells, mostly through dendritic connections.

The axon of a neuron is also a cytoplasmic process which originates as small conical elevation of the cell body called the axon hillock. The axon is thin, long cylinder and usually long, but length may vary. In certain axons, especially in the larger ones, there is a fatty covering termed the myelin sheath. The axons with myelin sheath are referred to as myelinated axons. These axons are enveloped along their length by flattened glial cells called schwann cells. In myelinated axons the Schwann cell membrane encircles and winds around the axon many times, forming several layers of the myelin sheath. The myelin sheath does not enclose the axon completely, but is interrupted at approximately one millimeter intervals by gaps which are unmyelinated. These gaps are called the nodes of Ranvier.

There is no cytoplasmic continuity between two neurons. The close junction between two neurons is termed synapses. It is across these synapses that action potential pass from one neuron to another. The structure of the synapses contains three distinct regions (Figure 2.2).

1.  The presynaptic region (membrane / knob)

2.  The postsynaptic region (membrane / cell)

3.  The synaptic cleft

In general, the axon terminal of a synapse is known as the presynaptic region. The presynaptic ending of an axon is usually divided into numerous fine branches which end in synaptic knobs. Inside these knobs there are many mitochondria and membrane bound spherical vesicles called synaptic vesicles. They are more numerous in the part of the presynaptic knob which is closest to the synaptic cleft. These vesicles contain chemical substances that can act as transmitters and play an important role in conduction an action potential across a synapse [4].

The cell body (soma), dendrites or some other part of a neuron constitutes the postsynaptic region of a synapse. In general, there are no vesicles in this region, instead specific receptors are found in the post synaptic membrane. These receptors can bind the transmitter substances released from the presynaptic knob.

The gap that separates the pre and postsynaptic membranes is the synaptic cleft. It is generally 20 – 25nm width. This gap is filled with a mucopolysaccharide that "glues" the pre and postsynaptic membrane together. In the human brain there are estimated to be one hundred trillion synapses.

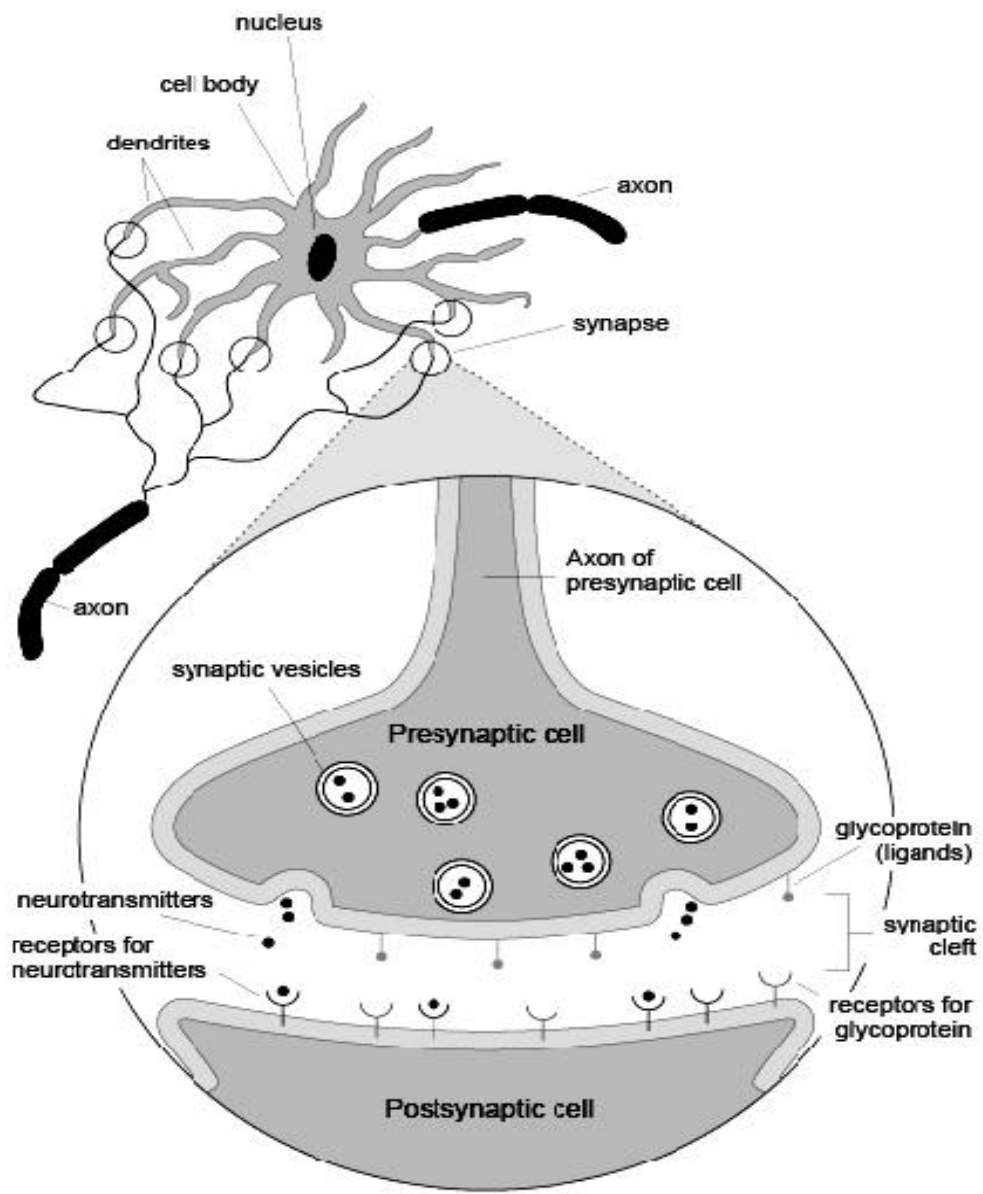

Figure 2.2. Model of the Synapse

Neurons have physiological specializations. Most prominent among these are a wide variety of membrane spanning ion channels that that allow ions, mainly Sodium ($Na^+$) Potassium ($K^+$), Calcium ($Ca^{2+}$), and Chloride ($Cl^-$) to move into and out of the cell. By opening and closing in response to voltage changes and both internal and external signals, ion channels control the flow of ions across the cell membrane. The electrical potential difference between the interior of a neuron and the surrounding extracellular medium is responsible for electrical signals in the central nervous system. Under the resting conditions, the potential inside the cell membrane of a neuron is about -70mV with respect to that of the surrounding medium which is conventionally taken as 0 mV (reference) and the cell is considered to be polarized. Ion pumps located in the cell membrane sustain concentration gradients that support this potential difference in the membrane. $K^+$ ion is more concentrated inside a neuron than in the extracellular medium while $Na^+$ ion is significantly higher outside the neuron than the inside. Therefore, ions flow into or out of a cell is as a result of both voltage and concentration a gradient. Flow of positively charged ions (current) from inside the cell to outside or negatively charged ions from outside to inside through open channels makes the membrane potential more negative and this process is called hyperpolarization. Similarly positively charged ions flowing into the cell makes the membrane potential less negative or some times even positive. This is called depolarization. Generation of field potentials is a direct result of currents that flow through the extracellular space. These field potentials usually have frequencies less than 100 Hz. When there are no changes in the signal average, these potentials are called EEG [5-6].

As it is mentioned above, under the membrane of the cell body, approximately a potential of 70 mV with negative polarity may be recorded. This potential changes with variations in synaptic activities. If an action potential is received by an *excitatory synapse*, then an excitatory post synaptic potential is generated in the following neuron. On the other hand, if the synapse is inhibitory then the hyperpolarization will occur in the following neuron indicating an inhibitory synaptic potential.

The activities in the Central Nerves System (CNS) are mainly related to the synaptic currents transferred between the junctions of axons (synapse) and dendrites, or dendrites and dendrites of cells. Nerve cells respond to stimuli and transmit information over long distances. Neurons are highly irritable (responsive to stimuli). When a neuron is adequately stimulated, an electrical impulse (in the form of an electrochemical change) is conducted along the length of its axon. This response is always the same, regardless of the source or type of stimulus. This electrical phenomenon is called the action potential or nerve impulse [5] and its propagation associates with electrical changes in the axon (Figure 2.3).

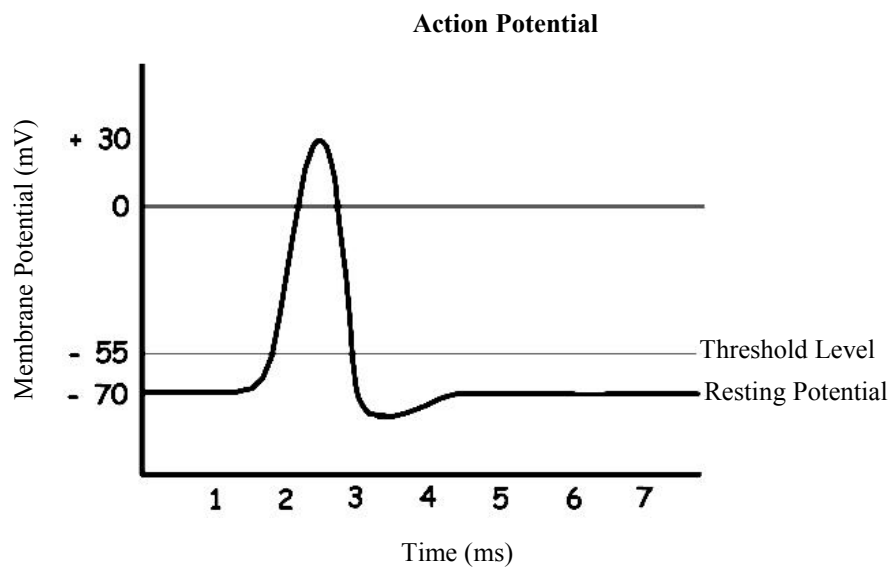

Figure 2.3. Action potential of the signal propagation in axon.

The conduction velocity of action potentials is usually between 1 ms$^{-1}$ and 100 ms$^{-1}$. Action potentials are initiated by various types of stimuli. As an example sensory neurons respond to stimuli such as light, electricity, pressure, touch and stretching. On the other hand, the nerve cells with in the Central Nervous System are mostly simulated by chemical activity at synapses.

A stimulus has to be above a threshold value to initiate an action potential. Very weak stimuli cause only a small electrical disturbance but they are not large enough to produce an action potential. However, when the strength of the stimulus becomes larger than the threshold, an action potential is generated and it travels down the nerve. The neuron needs approximately 2 ms before another is presented. During this time no action potentials can be generated. This is called the refractory period. The membrane potential changes with variations in synaptic activities. If an action potential is received by an *excitatory synapse*, then an excitatory post synaptic potential is generated in the following neuron. On the other hand, if the synapse is inhibitory then the hyperpolarization will occur in the following neuron indicating an inhibitory synaptic potential.

## 2.2 Functional cortex divisions and the identification of mental states in EEG

Cerebral cortex is the outer sheet of gray matter or mantle of cells (1 – 4mm thick) covering the two hemispheres of the brain which constitutes the outermost portion of the cerebrum [7-10]. The gray matter is formed by neurons and their unmyelinated fibers whereas white matter below the grey matter of the cortex is formed mainly by myelinated axons interconnecting different regions of the central nervous system. The cerebral cortex plays a key role in memory, attention, perceptual awareness, thought, language, and consciousness. Therefore the cortex is believed to be the seed of the mind. The elaboration of the cortex into different areas is one of the most distinguishing characteristics of the human brain (Figure 2.4).

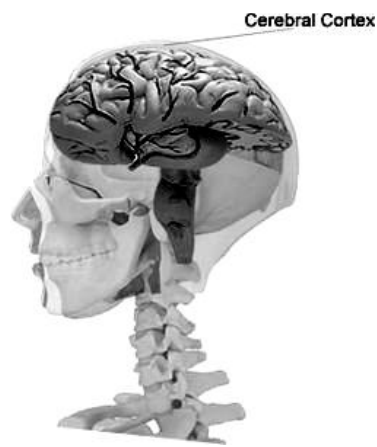

Figure 2.4. Location of the human cerebral cortex

Cortex can be divided into four areas by lobes (Figure 2.5); Frontal lobe, Parietal lobe, Occipital lobe, and Temporal lobe. The names given for these lobes are derived from bones that overlie the area of the cortex [7, 10].

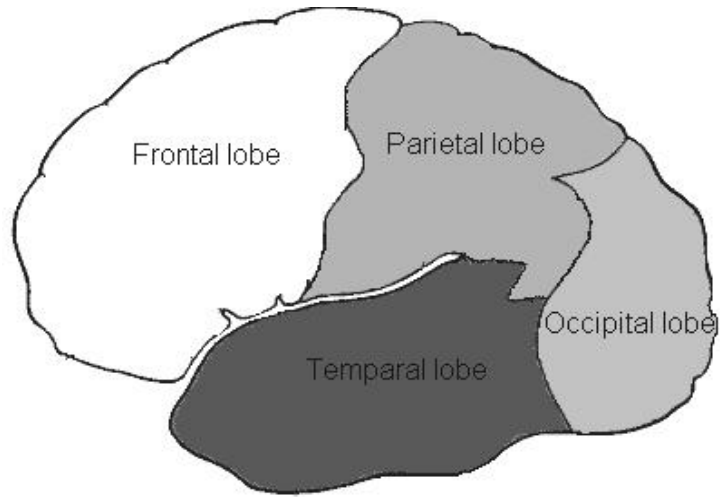

Figure 2.5 Lobes of the Cerebral Cortex

German neurologist Korbinian Brodmann (1868 - 1918) divided the cerebral cortex into 52 distinct regions by analyzing cellular structure in each area (cytoarchitectonic) starting from the central sulcus [the boundary between the frontal and parietal lobes] (Figure 2.6) and labeled them with integers from 1 to 52.

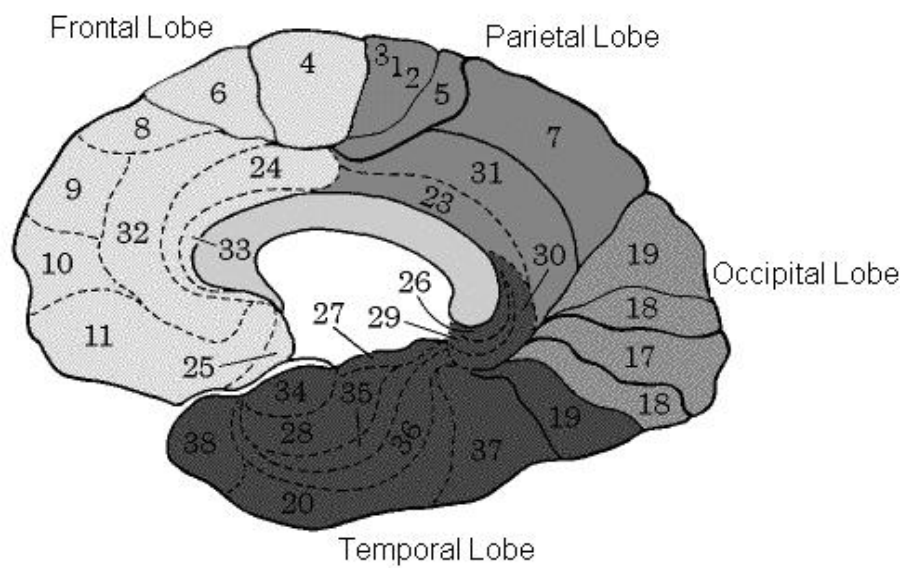

Figure 2.6. Brodmann's classification system

It was found later that each area in the Brodmann system represents functionally distinct region of the brain. Although human cortex is extremely complex, Brodmann numbers are still very useful when referring to an area of the cortex.  Now we discuss how some brain functions are related to Brodmann's regions.

*(1) Vision*

The area 17 is the primary visual cortex, located in the Occipital lobe which is the first place in the cortex receives visual signals from the thalamus. Secondary visual cortex is located at Brodmann areas 18, 19, 20, 21, and 37.

*(2) Audition*

Primary auditory cortex is located in the area 41 in the Temporal lobe. The secondary auditory cortex is located in the areas 22 and 42.

*(3) Body Sensation*

Primary Somatosensory cortex (sensory information coming from the body - e.g. pain, temperature) is located in the regions 1, 2, and 3 in the Parietal Lobe. The secondary somatosensory cortex is located in the regions 5, 7. Tertiary Somatosensory cortex is located in the regions 22, 37, 39, 40.

*(4) Motor*

Primary motor cortex is located in the area 4 in Frontal lobe where the motor movements initiated. The area 6 is located in the premotor cortex where planning of motor movements take place. Area 8 is responsible for eye movements while area 44 is responsible for speech. Higher order tertiary motor cortex is located in the areas 9, 10, 11, 45, 46, 47.

It is also possible to group functions of the brain according to the four lobes described earlier [9].

### Frontal lobe

In addition to motor activities, Frontal lobe is involved in executive functions. The executive functions of the frontal lobes involve the ability to recognize future consequences resulting from current actions, to choose between good and bad actions (or better and best), override and suppress unacceptable social responses, and determine similarities and differences between things or events. Therefore, it is involved in higher mental functions. The frontal lobes also play an important part in retaining longer term memories which are not task-based. These are often memories associated with emotions derived from input from the brain's limbic system. The frontal lobe modifies those emotions to generally fit socially acceptable norms.

### Parietal Lobe

The Parietal lobe plays important roles in integrating sensory information from various parts of the body, knowledge of numbers and their relations, and in the manipulation of objects. Parietal lobe in the left hemisphere (Dominating Hemisphere) is responsible for written languages i.e., read and write.

### Temporal lobe

Temporal lobe contains the primary auditory cortex where auditory signals from the cochlea first reach the cerebral cortex. It also contains high-level auditory processing including speech. The underside (ventral) part of the temporal cortices is involved in high-level visual processing of complex stimuli such as faces and scenes. Anterior parts of this ventral stream for visual processing are involved in object perception and recognition. Further, Neocortical area (Entorhinal Cortex) and the old cortical area (Hippocampus) of the temporal lobe are involved in learning and memory.

### Occipital lobe

The occipital lobe mainly contains primary and secondary visual cortices (Brodmann areas 17, 18, and 19).

In general, information tends to flow from the back of the brain to the front. The brain expends most of its energy packaging sensory input from all available modalities into a coherent view of the environment. Vision is combined with somatosensory information to give a sense of where one's body is in space. Memory functions in the temporal lobe allow for recognition of the visual perceptions. The processed sensory input finally makes its way to the frontal lobe where decisions are made regarding the various stimuli [9].

Generalizations about the cerebral cortex:

- The cerebral cortex contains three kinds of functional areas; sensory are that provide for conscious awareness of sensation; motor areas that control voluntary motor functions; and association areas that act mainly to integrate diverse information for purposeful action.

- Each hemisphere is chiefly concerned with sensory and motor functions of the opposite side of the body.

- Although largely symmetrical in structure, the two hemispheres are not equal in function; i.e. there is a specialization in cortical function in each hemisphere.

- No functional area of the cortex acts alone, and conscious behavior involves the entire cortex in one way or the other

Different sensory inputs are processed at different parts of the cortex. Furthermore particular cortex areas are involved in higher mental functions such as memory, control of attention, complex planning and reasoning. When sensory inputs are processed or when other mental processing takes place, the corresponding regions of the cortex are particularly active. These activities may alter EEG signal. In the next chapter we discuss how the EEG and various mental tasks are correlated.

# CHAPTER 3

# INTRODUCTION TO ELECTROENCEPHALOGRAM (EEG) AND PLACEMENT OF ELECTRODES

In the preceding Chapter the neural mechanisms connected to brain activities have been explained in detail. In order to monitor and analyze brain activities, various techniques such as Electroencephalography (EEG), Functional Magnetic Resonance Imaging (fMRI), Magnetoencephalogtaphy (MEG), Single Photon Emission Tomography (SPECT) and Proton Emission Tomography (PET) are commonly applied in the medical and research domains [2, 5]. These techniques can be classified using characteristics based on spatial resolution, temporal resolution, intrusiveness, resources required for operation of the monitoring device, physiological parameters which are monitored and applicability as portable devices [3]. Since the experimental part of our research is based on EEG, it is reviewed in detail in this Chapter.

In particular, general description of EEG and rhythmic patterns at characteristic frequencies associated with the brain, correlation between EEG and brain activities, conventional electrode positioning of the standard 10-20 system, amplification of EEG Signal, noise, and artifacts are described in the current Chapter.

First EEG Recordings were made as early as 1875. But the first human scalp recordings were made in 1929 by Hans Berger, who discovered that characteristic patterns of EEG activity are associated with different levels of consciousness [2]. Berger showed that electrical signals can be recorded externally from the scalp of human subjects. From that time on, EEG has been mainly used for evaluating neurological disorders and analyzing brain functions.

EEG measures the electrical activity on the scalp produced by firing of large numbers of neurons within the brain. EEG is noninvasive, trouble-free, and does not interfere much with a human subject's ability to move or perceive stimuli, and is relatively low-cost. EEG has amplitudes between 0 μV and 80 μV and a frequency range between 0Hz and 80Hz [6].

## 3.1 Electrical activities of the brain and EEG generation

Action potentials, which are distinct electrical signals, produced by neurons, that travel down axons and cause the release of chemical neurotransmitters at the synapse. These neurotransmitters then fit into a receptor in the dendrites of the cell body of the neuron that is on the other side of the synapse, the post-synaptic neuron [4]. The neurotransmitter, when combined with the receptor, usually causes an electrical current within dendrite or cell body of the post-synaptic neuron. Thousands of post-synaptic currents from dendrites of a single neuron and its cell body then sum up to cause the neuron to generate an action potential or not. This neuron then synapses on other neurons, and so on. EEG reflects correlated synaptic activities caused by post-synaptic potentials of cortical neurons. The ionic currents involved in the generation of fast action potentials may not contribute greatly to the averaged field potentials representing the EEG. Particularly, electrical potentials at the scalp that produce EEG are generally thought to be due to the extracellular ionic currents caused by dendritic electrical activity. These electrical activities are propagated through volume conduction [3]. The EEG recordings are the summation of electrical fields produced by millions of interconnected neurons. Dendrites, Axons and Cell bodies of the neuronal components produce the currents due to following reasons,

1. All cortical neurons have the same orientation; dendrites near the surface and axons projecting inward.

2. $Na^+$ entering dendrites during neuronal firing leaves the outside of the dendrites negatively charged.

3. If enough neurons beneath an electrode are activated at the same time, the resulting electric field they produce can be detected through the tissue of the scalp as shown in (Figure 3.1).

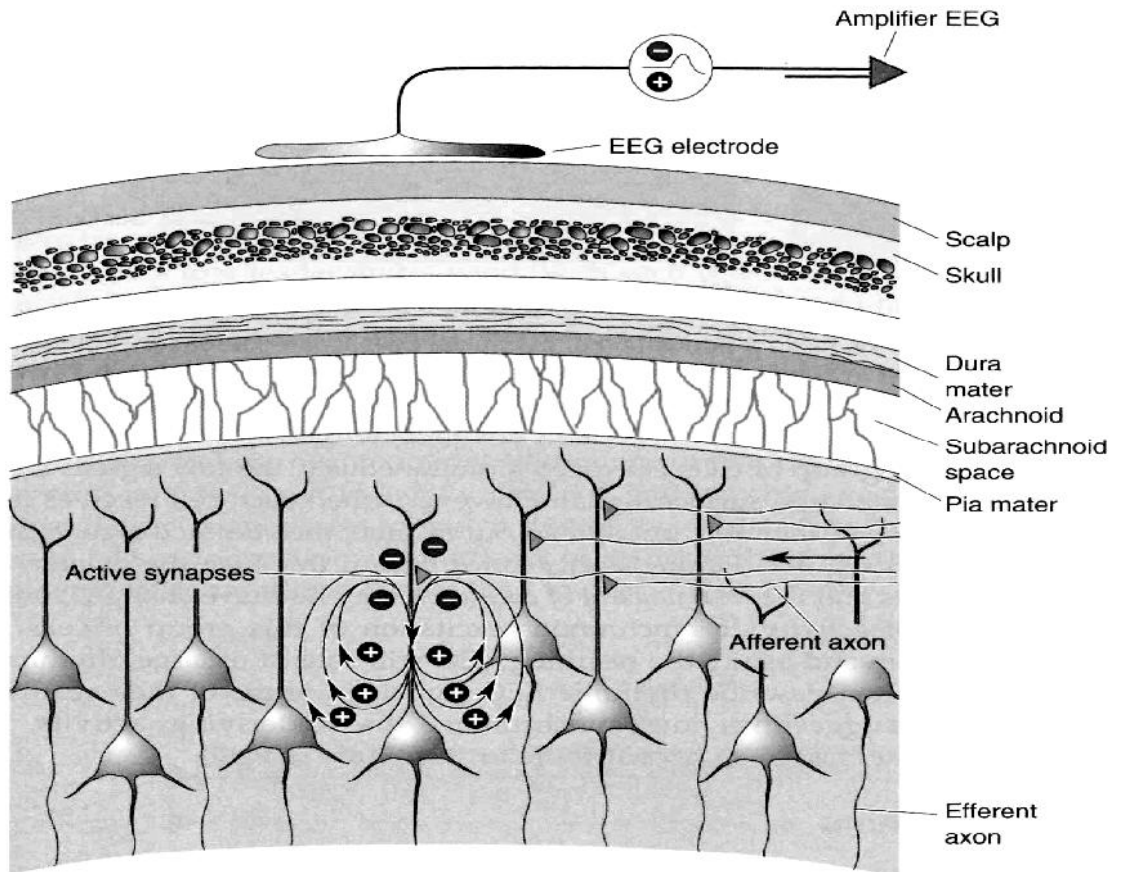

Figure 3.1. Synaptic excitation of cortical pyramidal cells generates the electrical activity along the scalp produced by summation of thousands of synapses in the tissue.

The architecture of the brain is not uniform but varies with different locations. Thus the EEG can vary depending on the location of the recording electrodes. The potential differences which can be measured between two points of the scalp are very different. The reasons for this disparity can be summarized as follows [5].

1. Since brain tissue and the fluid are conductive, a superposition of potentials generated in different areas of the cortex contributes to the voltages measured at scalp electrodes

2.  The amplitude of the originally generated potential differences is attenuated because of the resistive properties of tissues (e.g. fluid, skin, bone of the skull) between the potential generators and the electrode.

3.  Capacitance caused by cell membranes and other inhomogeneities (e.g. fluid-skull, skull-skin) between potential generators and electrodes influence the amplitude of the EEG signals as a function of their frequency.

The electric potentials induced by single neurons are far too small to be picked by EEG. Many neurons need to sum their activity in order to be detected by EEG electrodes. The timing of their activity is crucial. Synchronized neural activity produces larger signals (Figure 3.2). EEG activity therefore always reflects the summation of the synchronous activity of thousands or millions of neurons that have similar spatial orientation, radial to the scalp [7-8]. Currents that are tangential to the scalp are not picked up by the EEG. The EEG therefore benefits from the parallel, radial arrangement of apical dendrites in the cortex.

EEG is usually described in terms of rhythmic activities. The rhythmic activities are divided into bands by frequencies. As mentioned above collective synchronous activities of thousands of neurons are needed to create an EEG wave. More synchronous activity leads waves with larger amplitudes and slower frequencies while less synchronous activities indicates strong brain activities [9-11].

EEG rhythms correlate with patterns of behavior (level of attentiveness, sleeping, waking, seizures, and coma).

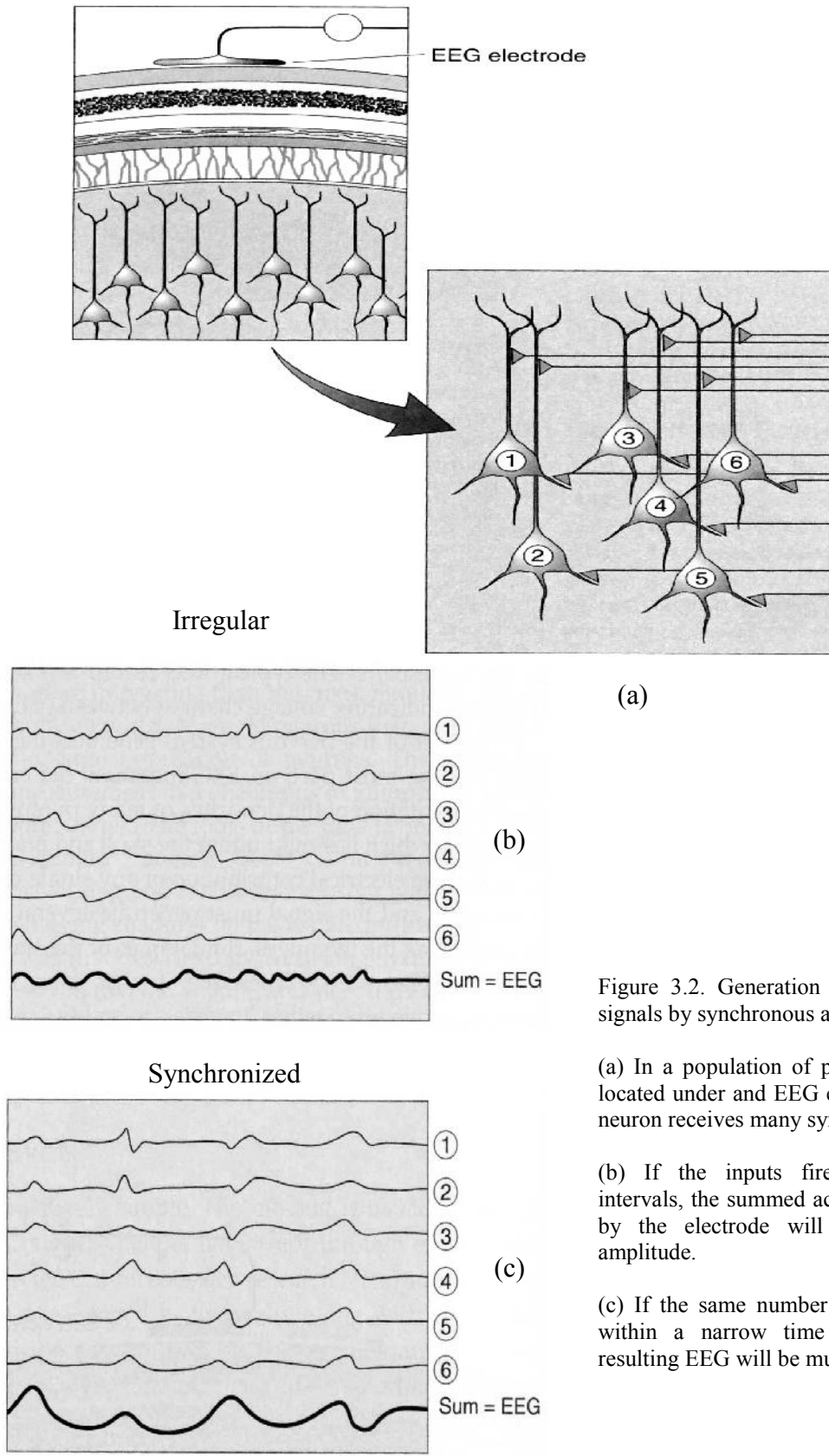

Irregular

Synchronized

(a)

(b)

(c)

Figure 3.2. Generation of large EEG signals by synchronous activity.

(a) In a population of pyramidal cells located under and EEG electrode, each neuron receives many synaptic inputs.

(b) If the inputs fire at irregular intervals, the summed activity detected by the electrode will be of small amplitude.

(c) If the same number of inputs fire within a narrow time window, the resulting EEG will be much larger.

In healthy adults, the amplitudes and frequencies of such signals change from one state of a human to another, such as sleep and wakefulness. The characteristic of the waves also varies depending on the state of mind and change with age [2, 12]. EEG potentials are good indicators of global brain state.  They often display rhythmic patterns at characteristic frequencies as shown in the figure below (Figure 3.3).

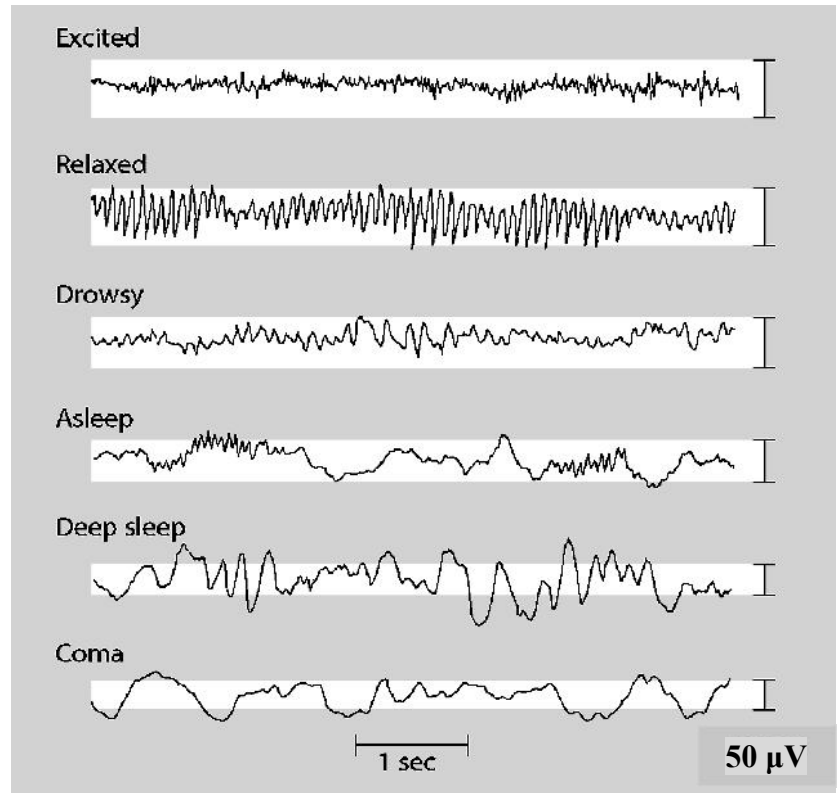

Figure 3.3. EEG waveform of a normal subject at different states.

Waves are categorized into five general types (Figure 3.4) distinguished by their rhythms occur in distinct frequency ranges and are given in the bellow Table 3.1. Higher frequencies refer to active processing, relatively de-synchronized activity (alert wakefulness, dream sleep). Lower frequencies refer to strongly synchronized activity (non - dreaming sleep, coma) [12].

Table 3.1. Different EEG rhythms with distinct frequency ranges

| Type | Frequency (Hz) | Object condition | Main scalp area |
|---|---|---|---|
| Delta (δ) | 0.5 - 4 or less than 4 | Sleep stages, especially deep sleep or present in the waking state. adults slow wave sleep, in babies | Frontally in adults, posteriorly in children; high amplitude waves |
| Theta (θ) | 4 – 7.5 | young children, drowsiness or arousal in older children and adults, idling,  sleep stages | |
| Alpha (α) | 8 - 13 | quiet waking, relaxed/reflecting, closing the eyes | Posterior half of the head, both sides, and occipital region, higher in amplitude on dominant side (left side). Central sites (c3-c4) at rest. |
| Beta (β) | 14 - 26 | Waking rhythm , active, busy thinking and attention, solving problem, alert/working | Both sides, symmetrical distribution, most evident frontally; low amplitude waves |
| Gamma(γ) | 30 – 45 or above 30 | Cognitive  frequency band, certain cognitive or motor functions | |

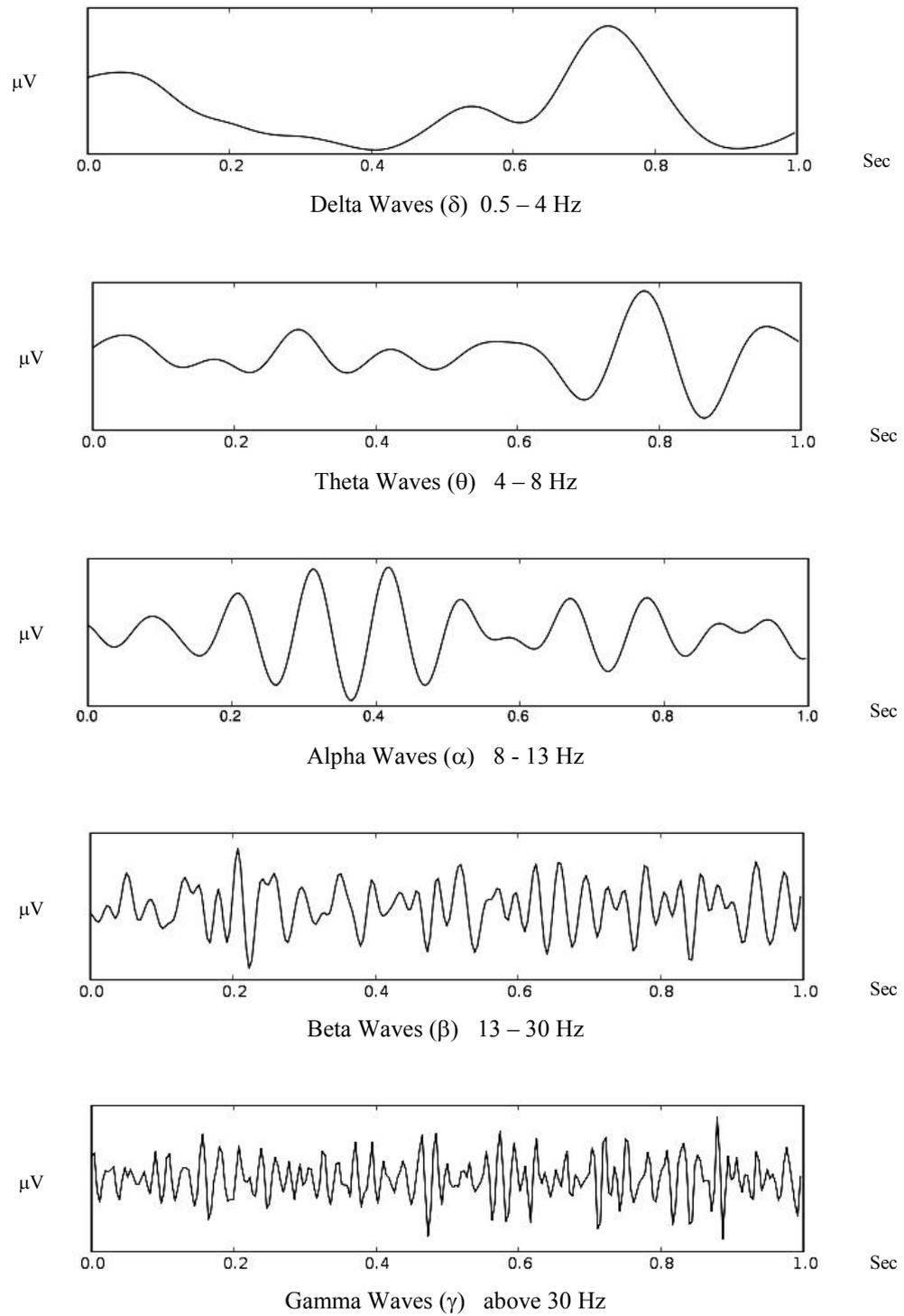

Delta Waves (δ)  0.5 − 4 Hz

Theta Waves (θ)   4 − 8 Hz

Alpha Waves (α)   8 - 13 Hz

Beta Waves (β)   13 − 30 Hz

Gamma Waves (γ)   above 30 Hz

Figure 3.4. Five typical dominant brain normal rhythms, from high to low frequencies.

The typical setup [13] involved in measuring the EEG electrical activities is as follows:

a.  Several electrodes are taped to scalp at standard positions.

b.  Signals read by electrodes are then amplified.

c.  Amplified signal at each electrode with respect to the reference electrode is recorded and displayed on a graph as amplitude versus time or Signal get digitized by Analog to Digital (A/D) converter  such that it can be recorded by a computer for analyzing.

### 3.2 Electrode positions

The positions for EEG electrodes should be chosen in a way that all cortical regions, which might show interesting EEG patterns, are covered. For most applications this is usually the entire cortex. An internationally accepted standard for electrode placements is the 10-20 system introduced in 1957 by the International EEG Federation [14]. During all experiments conducted for this research work, electrodes were placed according to the 10-20 system.

Three anatomical reference points must be determined before the 10-20 system electrode positions can be introduced (Figure 3.5):

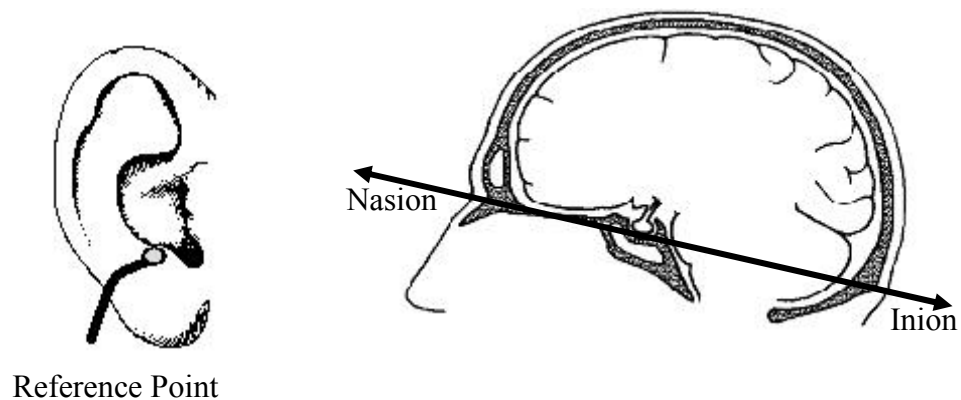

Reference Point

Figure 3.5. Placement of reference electrode and anatomical reference points which represent the starting points for finding the electrode positions defined by the 10-20 system.

*Nasion:* The start of the nose on the skull, nasion is the intersection of the frontal and two nasal bones of the human skull.

*Inion:* The bony lump which marks the transition between skull and neck, prominent projection of the occipital bone at the lower rear part of the skull.

*Preauricular reference point:* indentation to the forepart of the ear between the convoluted cartilage and the protective cartilage.

Electrode placement has been standardized by an international $10 - 20$ system that uses anatomical landmarks on the skull. These sites are then subdivided by specific interval of 10% to 20% and to designate the site where an electrode will be placed [14]. Figure 3.6 shows the 20 electrode positions (apart from the earlobe electrode) which follow the electrode placement system on the scalp [15].

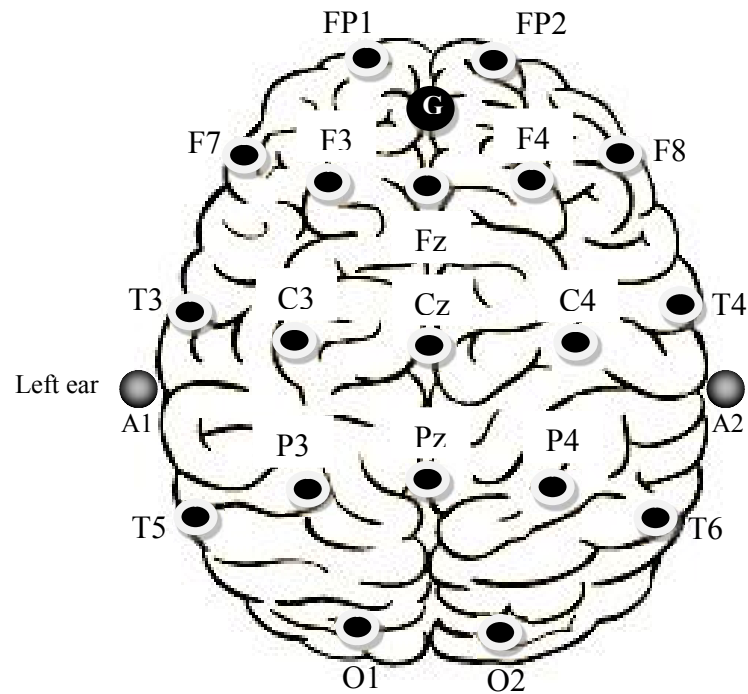

Figure 3.6. Electrode positions of the 10-20 system

The $10 - 20$ system avoids both eyeball placements. A1 and A2 stands for earlobe electrode connected respectively to the left and right earlobes which are used as the reference electrodes in the recording. The designation; Fp stands for frontopolar, F (frontal), T (temporal), P (parietal), C (central), O (occipital), A (auricular), and G terms ground electrode are utilized in the 10 - 20 system. Subsequently, numbers combined following the letters for location reflect either the left (odd numbers) or right (even numbers) hemisphere of electrode placement. The "z" designation reflects midline placement i.e., Fz = frontal midline, Cz = central midline and Pz = parietal midline respectively. The following procedure can be used to find the electrode positions using the above defined anatomical references as initial points:

a) The link from nasion to inion (NI) passing through the top of the head is measured. Start from the nasion, the point Fpz is placed after 10% of NI, the electrodes Fz, Cz, and the point Pz are placed each after proceeding 20% of NI from the previous position. The point Oz is placed before 10% of inion (Note that no electrodes are located at the points Fpz and Oz but they are used to find the positions of other electrodes.)

b) The electrodes A1 and A2 are sited at the left and right ear lobes. The connection from one pre-auricular (PA) reference point to the other one via the electrode position Cz is measured. After 10% of PA above both reference points the electrodes T3 and T4 are located. 20% of PA above T3 and T4 the electrodes C3 and C4 are placed.

c) The connection $FO_1$ between the point Fpz and Oz is measured via the position T3. Starting at Fpz the electrode Fp1 is placed after 10% of $FO_1$, the electrodes F7, T5 and O1 are placed each after proceeding 20% of $FO_1$ from the previous electrode. In an analogous way the connection $FO_2$ between Fpz and Oz via T4 is measured and the electrodes Fp2, F8, T6 and O2 are placed accordingly.

d) The electrode F3 is placed at half the distance between F7 and Fz, the electrode P3 is placed at half distance between T5 and PZ. In the same fashion the electrodes F4 and P4 are placed on the other side of the head (Figure 3.7).

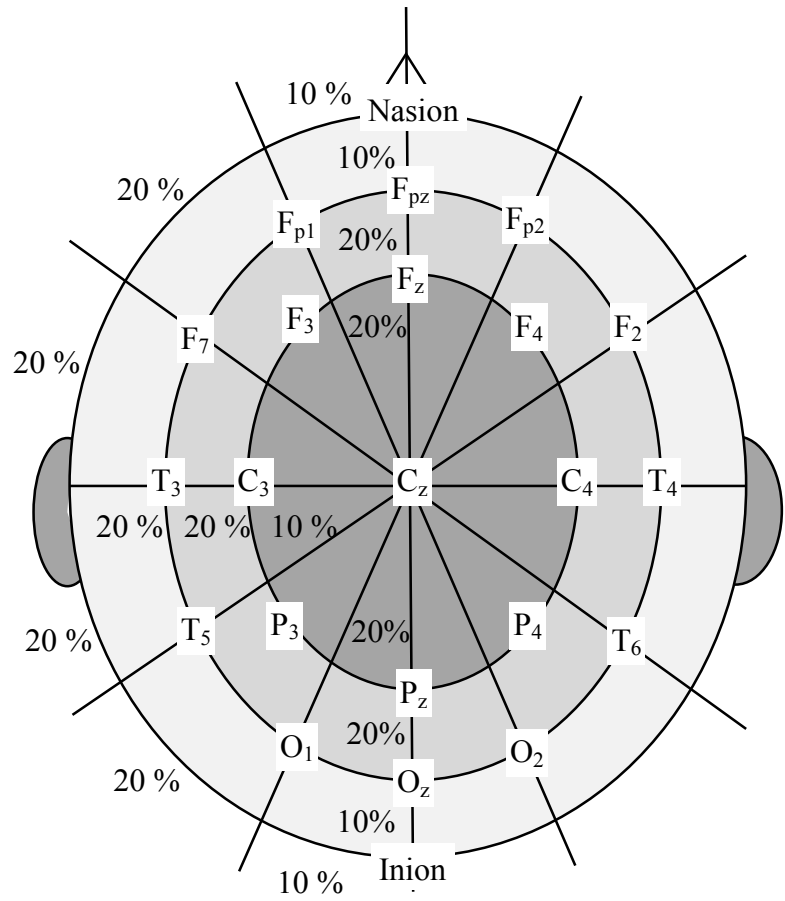

Figure 3.7. Conventional 10 -20 EEG electrode positions for the placement of 21 electrodes

Apart from the above conventional system, for special application purpose much more additional electrodes (up to 256) can be added to the standard set-up more or less evenly spaced around the scalp that can be used with improved amplifier technology, which usually placed according to various non-standardized conventions in each EEG laboratory [2].

Other addition electrodes are sometime used for measurement of Electrooculography (EOG) of the eyelid and eye surrounding muscles. Electrodes may include Electrocardiogram (EKG) recommended with every EEG which provides information about relationship between the heart and the brain, and to monitor the eye movement, Electromyogram (EMG) is used [4].

Finally in brief, ground electrodes [8] are essential to reduce noise from the AC power lines which are present in the whole body and hence contaminate the measured signals. The ground electrode gives the noise signal the possibility of flowing out of the body against a comparatively small resistance. Therefore the noise takes preferably the path via the ground electrode and not via other electrodes so that the measured signal is contaminated less. In the 10-20 system the position for the ground electrode is close in front of the position Fz (see figure 3.5). However, other positions for ground electrodes are possible only if they are captured as little muscular activity as possible, since this would introduce artifacts in all electrode channels.

### 3.3 EEG Recording Electrode types

In clinical EEG recordings the entire procedure as described above is usually carried out prior to each session which is extremely time consuming. However when recording EEG with commonly available EEG electrode caps or other standard recording devices, the procedure for determining the electrode positions needs not to be repeated prior to each data acquisition session. It is important that electrodes can be attached very quickly for EEG recording in the context of user state identification and task demand assessment. At present, flexible, multichannel electrode-caps and other recording devices are available in the markets which are more comfortable to wear, available with various sizes and even easier to attach to the individuals [2].

The EEG recoding electrodes and their proper functions are the most critical components which determine the EEG signal quality. It is their task to mediate between the ion based transport of electrical charge in the tissue and the electron based charge transport in copper wires which lead to the amplifier. In the tissue an electrolyte is responsible for the conduction of electricity. This is typically *NaCl* which dissociates to $Na^+$ and $Cl^-$ ions in a watery solution [5]. The electrode however gets only in contact with the skin which is somewhat dry and usually not permeable for the electrolyte in the body. Therefore it is essential to use an additional electrolyte in form of gel, which permeates the skin and thus establishes a connection between body electrolyte and the metal phase of the electrode. As a consequence the resistance and the capacitance of the skin decrease by some orders of magnitude when using electrode gel. Only this makes EEG recording possible.

Two types of electrodes are available; Polarized (or reversible) and non - polarized (or irreversible) electrodes. Polarized electrodes are usually made of precious metals, mostly gold or of stainless steel. Non-polarized electrodes which are made of silver (*Ag*) with a thin silver-chloride (*AgCl*) layer on top (*Ag / AgCl* electrodes) are state of the art since their resistive component is much lower.

Different types of electrodes are used in the EEG recording such as headband and electrode caps, disposable (gel less and pre-gelled types), reusable disc electrodes (tin, steel, or silver) etc. Commonly used electrodes made of *Ag – AgCl* disks, less than 3mm in diameter, with long flexible leads that can be directly plugged in to an amplifier. To obtain a satisfactory recording, the electrode impedances must be adjusted such a way that it should read less than $3k\Omega$ and be balanced to within $1k\Omega$ of each of them. The electrodes with high impedance as well as high impedance between the cortex and the electrode can lead to ruin or distortion, which can even mask the original signals.

### 3.4 Amplification of EEG Signal

EEG machines use a differential amplifier to construct each channel or trace of activity. Each amplifier has two inputs. An electrode is connected to each of the inputs. Differential amplifiers measure the voltage difference between the two signals at each of its inputs. The resulting signal is amplified and then displayed as a channel of EEG activity. Currently, commercial EEG recording systems are often equipped with impedance monitors. The signals need to be amplified to make them compatible with A/D converters [5]. Amplifiers which are adequate to measure these signals have to satisfy very specific requirements. They have to provide amplification selective to the physiological signal, guarantee protection from damages through voltage and current surges for both subject and electronic equipment, and reject superimposed noise and interference signals. The basic requirements that a bio potential amplifier has to satisfy are:

- The measured signal should not be distorted.

- The physiological process to be monitored should not be influenced in any way by the amplifier.

- The amplifier should provide the best possible separation of signal and interferences.

- The amplifier itself has to be protected against damages that might result from high input voltages as they occur during the application of defibrillators or electrosurgical instrumentation.

- The amplifier has to offer protection of the subject from any hazard of electric shock.

Electrode montage represents the various ways in which the EEG signals from different electrode channels can be combined before amplification. Typically, each montage will use one of three standard recording montages: bipolar electrode, common-reference electrode or average reference. In first two cases the EEG measures the difference between the potential of two electrodes or between one electrode and the averaged potential of a set of electrodes [4].

For a common- reference montage the EEG signals represent the difference of the potentials of each electrode and one or more reference points which all electrodes (or at least a subset of electrodes) have in common. Each amplifier records the difference between a scalp electrode and a reference electrode. The same reference electrode is used for all channels. Electrodes regularly used as the reference electrodes are A1, A2, the ear electrodes, or A1 and A2 linked together.

The idea of bipolar electrode montages is that always the difference between neighboring (two adjacent) electrodes are taken. These sequentially link electrodes together usually in straight lines from the front to the back of the head or transversely across the head. For example the first amplifier may have electrodes FP1 and F3 connected to it and the second amplifier F3 and C3 connected to it.

Finally in average reference montage the activity from all the electrodes are measured, summed together and averaged before being passed through a high value resistor. The resulting signal is then used as a reference electrode and connected to input 2 of each amplifier and is essentially inactive. All EEG systems will allow the user to choose which electrodes are to be included in this calculation. For multichannel montages, electrode caps are preferred, with number of electrodes installed on the scalp surface (Figure 3.8).

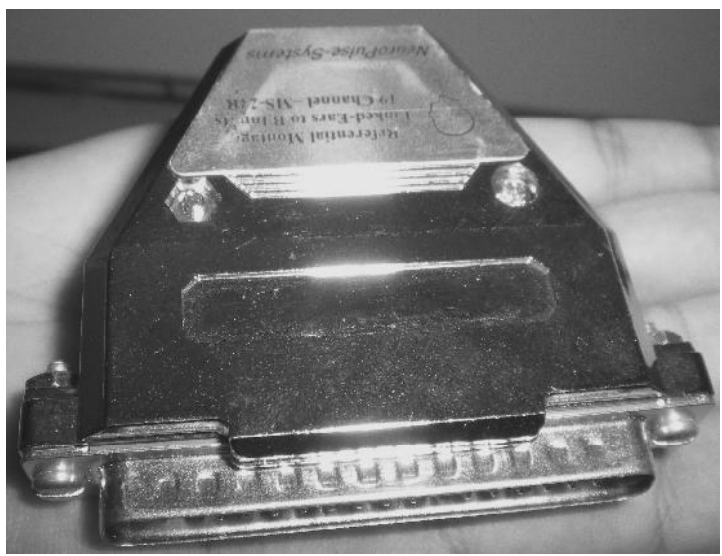

Figure. 3.8 Example of a montage: NPS 19 channel Referential Montage for the amplifier MS - 24R, manufactured by NeroPulse Systems, USA.

### 3.5 Noise and Artifacts

Different kinds of interfering waveforms added to the EEG signal during the recording sessions. These interfering waveforms are generally termed as an Artifact or Noise. One can observe artifacts, as they are waves or group of waves that are produced by technical or other disturbances, which are not due to brain activity [16-17]. The most important reasons for occurrence of the artifacts are the movements of the subject during recording session and the normal electrical activity of the heart, muscles and eyes.

Generally two types of artifacts can be observed during the recording process.

1. Biological artifacts, which are caused by the recorded subject.

2. Technical artifacts caused by the EEG recording device.

Most of the biological artifacts generated by sources external to the brain must be removed after the recording process. This is the only way of handling this type of artifacts. The sources of many biological artifacts are dipoles originating, for example from muscular activity which are much stronger than the EEG related dipoles. A superposition of both types of dipoles causes artifacts in the signal which are often characterized by large peaks. However they can hardly be distinguished from the actual EEG.

List of following activities can cause biological artifacts.

- Potential differences caused by facial muscles introduce peaks in the EEG.

- The dipole caused by the electrical activity of the heart is one order of magnitude larger than the dipoles caused by neural brain activity.

- The neurons in the retina generate dipoles. During vertical eye movements and also during eye blinks; the eye moves up- and downwards which changes the direction of these dipoles. This introduces saw tooth shaped artifacts in the EEG signal. Artifacts introduced by horizontal eye movements are more rectangular shaped.

- Sensory neurons of the tongue also generate dipoles. Therefore tongue movements can be seen as slow and irregular fluctuations in the EEG.

- The state of the subject's hair (greasy, treated with hairspray etc.) can influence resistance between skin an electrode and thus cause a change of amplitudes across sessions.

- Eye blinks, sweating, jaw clenching and swallowing.

- Sweat or too much conductive paste can cause shortcuts between electrodes which lead to a strong decrease of signal amplitudes.

Improving technology can decrease the externally generated technical artifacts (non biological) such as line noise.

List of following external activities can also cause contamination in EEG signals.

- Dirty or corroded contacts of cables, cable movements, or even broken cables can cause a decrease of the signal amplitude or even a zero signal.

- Too much electrode paste or dried pieces

- Impedance fluctuation

- Broken or dirty electrodes can cause sudden large potential changes.

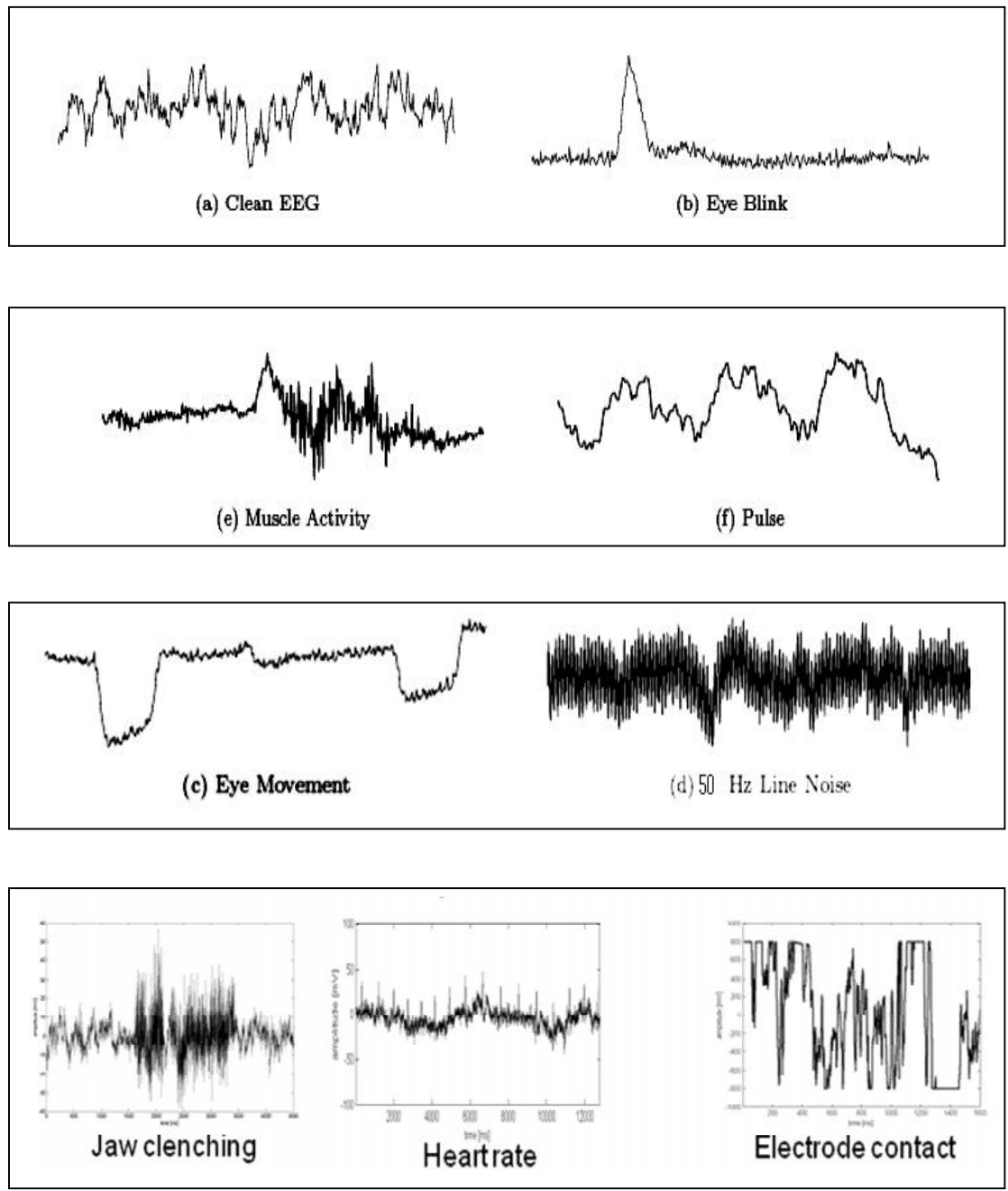

Figure 3.9. Different Artifact Waveforms.

- Strong signals from A/C power supply call line noise (50/60 Hz)

- Badly attached ground electrodes or a too strong asymmetry between the impedances of EEG and reference electrodes

- Low battery

Technical artifacts and also some biological artifacts can be generally avoided when the recording equipment is operated correctly and the subject to be recorded is prepared carefully. Other artifacts are unavoidable. Figure 3.9 show waveform of some of the most common EEG artifacts.

## 3.6 Changes in EEG due to mental tasks

It is known that certain mental activities can alter EEG signals up to a level where it can be detected by signal processing methods. Some of mental tasks (cognitive tasks) which are known to alter EEG signals consist of, Multiplication, Letter composing, Cubic rotation, Mental counting and Imaginary hand and leg moments. These mental tasks are discussed in detail in Chapter 5.

# CHAPTER 4

# EEG SIGNAL PROCESSING TECHNIQUES

Brain Computer Interface involves five major steps given below (See Figure 4.1).

(1) Cognitive tasks of the subject initiate activities in the cerebral cortex.

(2) Activities in the cortex alter the EEG signals recorded from the scalp.

(3) EEG signals are then amplified and digitized.

(4) These digitized signals are read by a computer as time series data.

(5) Data will then be analyzed with help of Digital Signal Processing methods.

Steps from one to four have been discussed in previous Chapters. In this Chapter we mainly focus our attention on the step five and describe in detail Digital Signal Processing techniques which are used in identifying mental activities through digitized EEG signals in BCI.

Signal processing step involves preprocessing of data, feature vector construction, classification, and post-processing.

In the Pre - processing stage, the acquired EEG signals will go through various filters such that other processing stages become free of unnecessary contaminations. In the next stage, feature extraction is commonly used to reduce the dimensionality of EEG signal while keeping the important features of the original signal intact.

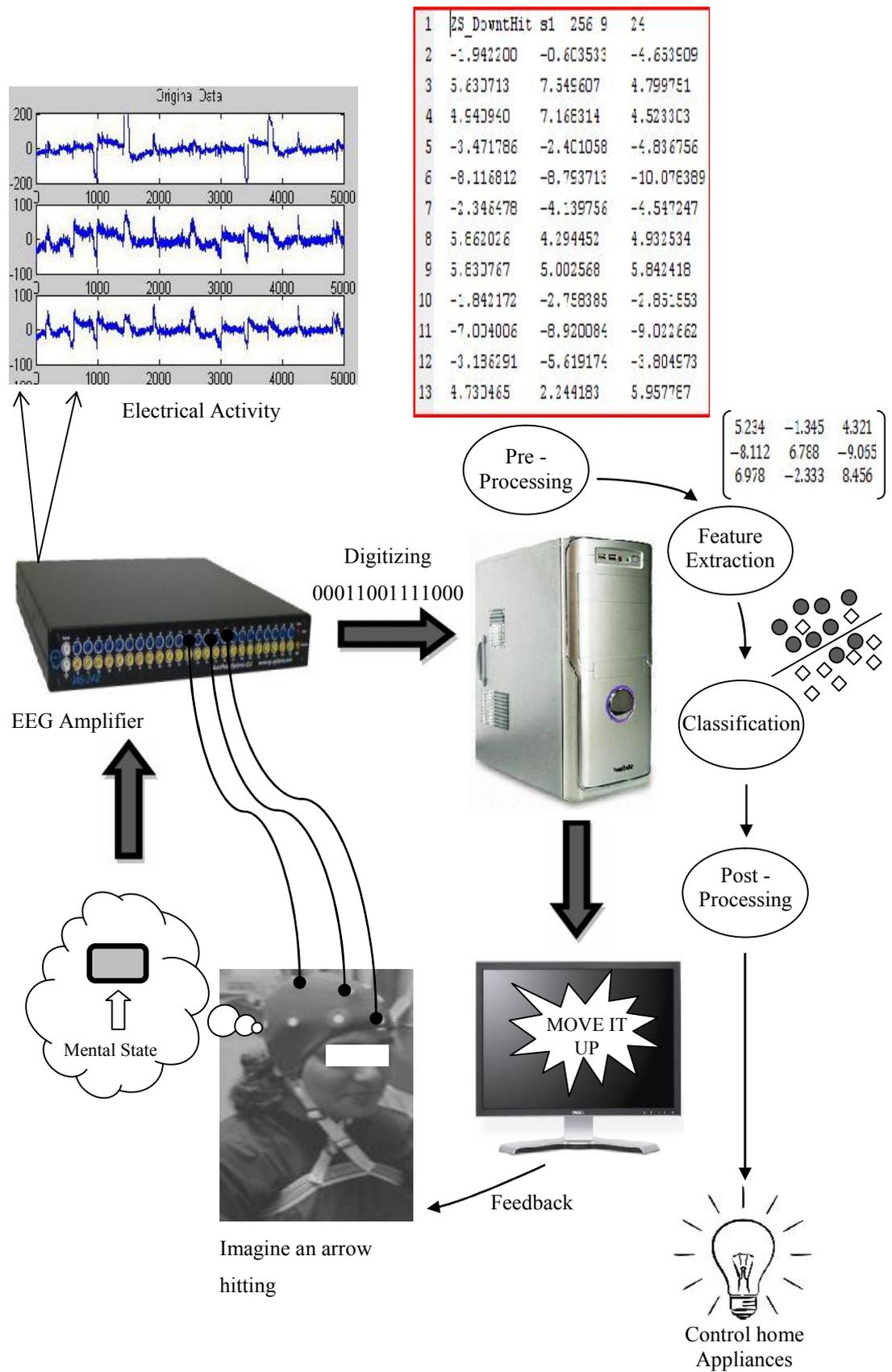

Electrical Activity

Pre - Processing

Feature Extraction

Digitizing
00011001111000

Classification

EEG Amplifier

Post - Processing

Mental State

MOVE IT UP

Imagine an arrow hitting

Feedback

Control home Appliances

Figure 4.1. A typical scheme of simple BCI system.

Once the signals are cleaned and feature vectors are constructed, they will be classified according the mental tasks the subject is performing. Once the signals are classified, at post-processing stage, the computer takes actions according to the mental tasks in a preprogrammed manner.

## 4.1 Preprocessing

The preprocessing stage is responsible for,

(1) Elimination of contamination presented in the raw EEG data due to technical and biological artifacts.

(2) Removal of unnecessary frequencies using appropriate filters.

Usually, external electrical interference can be removed at the hardware level by inbuilt filters of the EEG amplifier. Technical artifacts, such as AC (50/60 Hz) power line noise, can be lowered by decreasing electrode impedance and by shorter electrode wires. In addition, the 50/60 Hz signal noise may be eliminated by built in notch filters in the EEG amplifier. Most of the common technical artifacts can be avoided by proper grounding of the subject. Fluorescent lights emit radiation noise and can be avoided by replacing them with special incandescent or white LED bulbs. Whatever the remaining artifacts that cannot be removed at hardware level have to be removed by software at the preprocessing stage.

In the last segment of the previous Chapter, various types of biological artifacts have been described. The recognition of the eye blinks and eye moment artifacts are generally made by detecting a voltage increase in the EOG channel above a threshold, generally 100 µV. For better discrimination of different physiological artifacts, placing additional electrodes for monitoring eye blinks, eye moments, and electrical contamination due to heart beats and muscle activities are important.

In BCI research, usually two main techniques are used to take care of artifacts. The first method is to discard the affected segments of EEG due to Electrocardiogram (ECG / EKG - pulse, pace-maker), Electrooculographic (EOG) and other artifacts. Usually, short time segments of the signal (between 0.25 seconds and 2 seconds length) containing eye activity are simply discarded. Eye activity is identified by visual inspection [1] or automatically by rejecting all segments where the signal exceeds a certain threshold [2]. Another method for automatic eye activity detection and rejection uses a separate channel which records the eye activity only (i.e. the EOG) [3-5]. Since very little EEG is contained in the EOG data, artifacts are identified more reliably using EOG when it exceed of a threshold is used as criterion [6]. This method can greatly decrease the amount of data available for analysis.

In the second method, various software preprocessing tools such as spatial filters [7], blind source separation [8], and linear regression models [9] in both time domain and frequency domain are used. The blind source separation method is the most popular method used in removing artifacts. This method is based on multivariate statistical analysis techniques such as Principal Component Analysis (PCA), Signal fractional analysis (SFA) and Independent Component Analysis (ICA). However, for a small number of EEG channels, it is very likely that regression methods perform better than principal component analysis and independent component analysis. On the other hand, some recent work suggests that the regression methods are appropriate for EOG reduction [10].

### 4.1.1 Digital Filters

The digital EEG signal is stored electronically and can be filtered for display. Typical settings in EEG signal for the high-pass filter and a low-pass filter are 0.5-1 Hz and 35–70 Hz, respectively. The high-pass filter typically filters out slow artifacts, such as electrogalvanic signals and movement artifacts, whereas the low-pass filter filters out high-frequency artifacts, such as electromyography signals. When filtering the signal, frequencies between 48-62Hz should be removed, because technical equipment tends to send out waves in that specific frequency interval. To separately remove those frequencies a notch filter can be used. The notch filter does not filter frequencies below and above the specified frequency threshold. In additional notch filter is commonly used for removing artifacts caused by electrical power lines (50/60 Hz)[14].

Filters are useful for removing the unnecessary frequencies present in digital EEG signals which are not directly relevant to BCI applications. EEG signals can be filtered using *high – pass* or *low – pass* filters to eliminate noise at low or high frequencies respectively. Sometimes the interesting part of an EEG signal may exist only in a certain frequency band. Such instances, one can use *bandpass filters* to filter out uninteresting frequencies.

In the design of frequency – selective filters, the desired filter characteristics are specified in the frequency or Fourier domain in terms of magnitude and phase response. Filter in the Fourier domain take the whole data record, convert it to frequency domain using Fast Fourier Transform (FFT), multiply the FFT output by a filter function, and then do an inverse FFT to get back a filtered data set in time domain. Following additional information should be considered, when implementing Fourier methods for filtering [11].

(1) Define the filter function $F$ ( ) for both positive and negative frequencies, and that the magnitude of the frequency extremes is always the Nyquist frequency $1/(2$ ), where    is the sampling interval. The magnitude of the smallest nonzero frequencies in the FFT is $\pm 1/(N$ ), where $N$ is the number of data points in the FFT. The positive and negative frequencies to which this filters are applied are arranged in wrap-around order.

(2) If the measured data are real, the need the filtered output also to be real, then arbitrary filter function should obey $F(-$ $) = F($ $)^{*}$. Arrange this most easily by picking an $F$ that is real and even in    .

(3) If chosen $F$ ( ) has sharp vertical edges in it, then the *impulse response* of filter (the output arising from a short impulse as input) will have damped "ringing" at frequencies corresponding to these edges. Such situations pick a smoother $F$ ( ). To get a first – hand look at the impulse response of the filter, take the inverse FFT of the $F$ ( ). Once smoothed the all edges of the filter function over some number $k$ of points, then the impulse response function of the filter will have a span on the order of a fraction $1/k$ of the whole data record.

(4) If the data set is too long for FFT all at once, then break it up into segments of any convenient size, as long as they are much longer than the impulse response function of the filter. Use zero – padding, if necessary.

(5) Remove any trend from the data, by subtracting from it a straight line through the first and last points (i.e., make the first and last points equal to zero). When segmenting data one can pick overlapping segments and use only the middle section of each, comfortably distant from edge effect.

(6) A digital filter is said to be casual or physically realizable if its output for a particular time-step depends only on inputs at that particular time-step or earlier. It is said to be *acausal* if its output can depend on both earlier and later inputs. In general Filtering in the Fourier domain is, acuasal, since the data are processed "in a batch," without regard to time ordering. Acuasal filters generally give superior performance (e.g., less dispersion of phases, sharp edges, less asymmetric impulse response functions). Causal filters are better, but some situations don't allow access to out-of-time order data. Time domain filters can be either causal or acuasal, heavily depends on the application where physical realization is a constraint.

In real-time applications, Time-domain filters are favored because of its flexibility in implementing and its ability to provide output filtered values at the same rate as continuous data stream it receives. Time-domain filters are also useful when quantity of data to be processed is so large that only very small number of floating operations can be carried out on each data point and even a modest-sized FFT (with a number of floating operations per data point several times the logarithm of the number of points in the data set or segment) cannot be afforded [11].

Most general time domain filter is the linear filter which takes $x_k$ of input points and produces a sequence $y_n$ of output points by the formula,

$$y_n = \sum_{k=0}^{M} c_k \, x_{n-k} + \sum_{j=1}^{N} d_j y_{n-j}$$  (4.1)

Here the $M + 1$ coefficients $c_k$ and the $N$ coefficients $d_j$, are fixed and define the filter response. The filter (4.1) produces each new output values from the current and $M$ previous input values, and from its own $N$ previous output values. If $N = 0$, there is no second sum in (4.1), then the filter is called *non-recursive* or *finite impulse response (FIR)*. If $N \quad 0$, then it is called *recursive* or *infinite impulse response (IIR)*. (The term "IIR" connotes only that such filters are *capable* of having infinite long impulse responses, not that their impulse response is necessarily long in a particular application. Usually the response of an IIR filter will drop off exponentially at late times, rapidly becoming negligible.)

The relation between the $c_k$'s and $d_j$'s and the filter response function $F$ ( ) is

$$F(\omega) = \frac{\sum_{k=0}^{M} c_k \, exp^{-2\pi i k(f\Delta)}}{1 - \sum_{j=1}^{N} d_j exp^{-2\pi i j(f\Delta)}}$$  (4.2)

Where    is, the sampling interval. The Nyquist interval corresponding to $f$ between -1/2 and 1/2. For FIR filters the denominator of (4.2) is just unity.

Equation (4.2) tells how to determine $F(\omega)$ from the $c$'s and $d$'s. To design a filter, need a way of doing the inverse, getting a suitable set of $c$'s and $d$'s as small set as possible, to minimize the computational burden from a desired $F(\omega)$.

One clearly has to make compromise, since $F(\omega)$ is a full continues function while the short list of $c$'s and $d$'s represents only a few adjustable parameters. Digital filter design concerns itself with the various ways of making these compromises.

However, in this study, for preprocessing we have implemented Butterworth bandpass filter alone with upper and lower cutoff frequencies which will passes frequencies within a certain range and rejects (attenuates) frequencies outside that range. A 3$^{rd}$ order forward-backward Butterworth bandpass filter was used to filter the data. Cutoff frequencies were set to approximately 1.0 Hz and 50.0 Hz. The MATLAB$^®$ function *butter* was used to compute the filter coefficients and the function *filtfilt* was used for filtering.

When using offline EEG data for analysis, it is important to have the ability to group data at different time segments. Functions are needed not only for cutting out time segments, but also for modifying the sample frequency. If an EEG data set is sampled at 256Hz, the data can easily be modified into 64Hz sample rate, just by keeping every fourth sample and throwing away the rest. This technique is named as down sampling which will be discussed under the feature vector construction. When down sampling the signal from 256Hz into 64Hz important information can be lost and disturbance in the data could occur. To prevent disturbances from high frequencies that remain in the down sampled data, a Butterworth Filter is selected to filter the high frequencies away before the down sampling is made.

We performed zero-phase digital filtering on EEG data in both the forward and the reverse directions. As mentioned above, we have used *filtfil* function available in the MATLAB$^®$ Software for this purpose. After filtering in the forward direction, it reverses the filtered sequence and runs it back through the filter. The resulting signal has:

- Precisely zero-phase distortion
- Magnitude that is the square of the filter's magnitude response
- Filter order that is double the order of the filter specified by b and a

- Minimizes start-up and ending transients by matching initial conditions, and works for both real and complex inputs.

During this research, our intention has been only to record EEG signals without any eye blink or eye movement. Even, when the eye activities are involved during the recording, that particular trial was discarded. As a result eye activity related artifacts are unnecessary to remove when it is analyzed. Out of many artifact removal methods PCA is one of the popular techniques implemented in BCI and will be discussed in the following section alone with its application in feature vector constructions.

## 4.2 Feature Vector Construction

After filtering out both artifacts and unwanted frequencies, the time consuming part of the procedure in BCI is the construction of feature vectors. Even after filtering, digitized signal contains large number of data points. Usually, multichannel EEG are used in BCI for better performance. Therefore, when digitized signals corresponding to all the channels are combined, very large number of data points is generated and hence all the data cannot directly be used for classification. Therefore dimensional reduction is needed while keeping the important features of the original signal as intact as possible. Also information required to distinguish or classify the signal may not be apparent in the time domain. Therefore, various digital signal processing methods are used to extract features and construct the feature vectors.

In order to select the most appropriate classifier for a given BCI system, it is essential to clearly understand what features are used, what their properties are and how they are used. The following section will discuss what are the two main factors involving in feature extraction for BCI and describe the common BCI features [12].

*(a) Feature properties:* A great variety of features vector construction methods have been attempted in BCI system design. Among the large number of methods, amplitude values of EEG signals, Downsampling, Band Powers (BP), Power

Spectral Density (PSD) values, Autoregressive (AR) and Adaptive Autoregressive (AAR) parameters, time-frequency features and inverse model-based features, Principal components analysis and Independent components analysis are the most popular methods. In the design of a BCI system, some critical properties of features in EEG signals must be considered and they are as follows,

(i) <u>High dimensionality</u>: In BCI systems, often feature vectors are made of high dimensionality, in fact, various features are commonly extracted from several channels and from several time segments before being concatenated into a single feature vector.

(ii) <u>Noise and outliers</u>: BCI features are noisy or contain outliers since EEG signals have a poor signal-to-noise ratio.

(iii) <u>Time information</u>: BCI features should contain time information as brain activity patterns are generally related to specific time variations of EEG.

(iv) <u>Non-stationarity</u>: BCI features are non-stationary whereas EEG signals may rapidly vary over time and more especially over sessions.

(v) <u>Small training sets:</u> the training sets are fairly small, since the training process is time consuming and demanding for the subjects.

Above properties are verified for most features currently used in BCI research laboratories. However, it should be noted that it may no longer be true for BCI used in clinical practice. For instance, the training sets obtained for a given patient would no longer be small as a huge quantity of data would have been acquired during sessions performed over days and months.

*(b) Time variations of EEG*: Most brain activity patterns used in BCI are related to particular time variations of EEG, possibly in specific frequency bands. Hence, the time variation of EEG signals should be considered during feature extraction stage. To use this temporal information, three main approaches have been proposed:

(i)     *Use of different time segments for concatenation of features:* this consists of extracting features from several time segments and concatenating them into a single feature vector.

(ii)    *Combination of classifications at different time segments:* it consists of performing the feature extraction and classification steps on several time segments and then combining the results of the different classifiers.

(iii)   *Dynamic classification:* it consists of extracting features from several time segments to build a temporal sequence of feature vectors. This sequence can be classified using a dynamic classifier [12].

First approach is the most widely used and results in feature vectors that are often of high dimensionality. In this investigation, we have implemented PCA, Bandpowers, and Downsampling with and without scaling for feature vector construction.

## 4.2.1 Principal Components Analysis (PCA)

PCA is a way of identifying patterns in data, and expressing the data in such a way as to highlight their similarities and differences. Since patterns in data can be hard to find in data of high dimension, PCA is a one of the powerful tool for reducing the dimensionality. The other main advantage of PCA is that once the patterns are found in the data, the data can be compressed, i.e., by reducing the number of dimensions having less information. PCA is used commonly in all forms of analysis, from neuroscience to computer graphics, because it is a simple, non-parametric method of extracting relevant information from confusing data sets. With minimal additional effort PCA provides a proper guidance for how to reduce a complex data set to a lower dimension to reveal the sometimes hidden, simplified structure that often underlie it. In other words, the goal of principal component analysis is to compute the most meaningful basis to re-express a noisy data set. The hope is that this new basis will filter out the noise and reveal hidden structure.

Principle component analysis finds a projection axis that can be used to reduce the high dimensional data into low dimensional data while retaining the major characteristics of the data. The idea behind PCA can be explained with a two dimensional example. The left panel of Figure 4.2 shows how the data is distributed in the two dimensional plane. Suppose that we want to reduce the two-dimensional data to one-dimension. The right panel of Figure 4.2 shows the same data with respect to another set of orthogonal axes $w_1$ and $w_2$ which were found by principle component analysis.

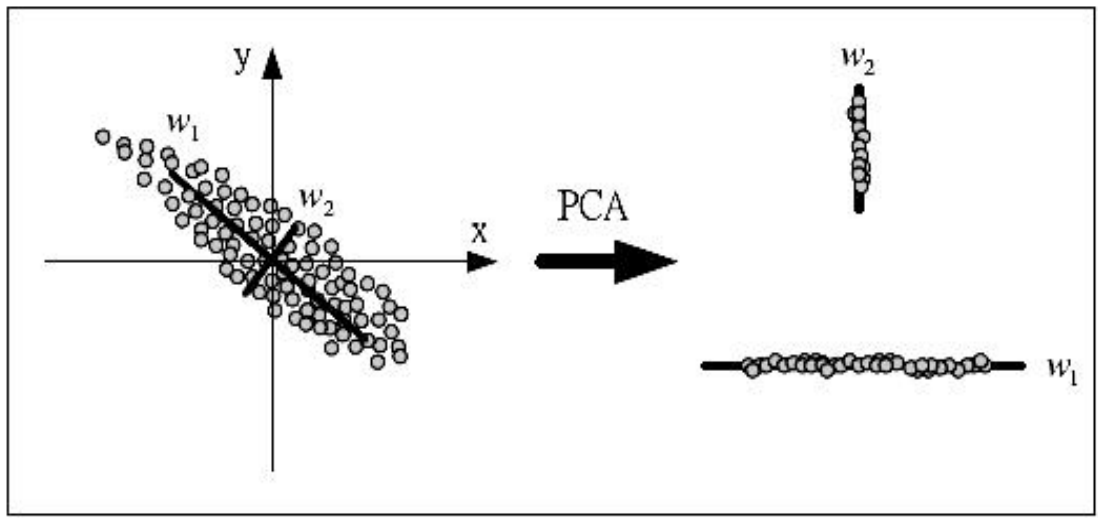

Figure 4.2. The left panel shows the distribution of two-dimension data. Right panel shows the data has been projected to two orthogonal axes via principle component analysis. Because the projected data on axis $w_2$ has smaller variance than that on axis $w_1$, the projection axis $w_1$ was a better choice than $w_2$.

In this particular example, these two components explain most of the cloud's trends. In a more complex data set, more components might add information about interesting trends in the data.

Principle component analysis can be applied to EEG data for finding the optimal projection axis that reduces data dimensionality by performing a covariance analysis between factors.

Before the PCA procedure is discussed, let us consider what types of assumptions and limitations are involving in PCA techniques [13].

(I)     Linearity: Necessary change of basis can be carried out through linear transformations. Linearity vastly simplifies the problem by, restricting the set of potential bases and formalizing the implicit assumption of continuity in a data set.

(II)    Mean and variance are sufficient statistics: The formalism of sufficient statistics captures the idea that the mean and the variance entirely describe a probability distribution. The only classes of probability distributions that are fully described by the first two moments are exponential distributions (e.g. Gaussian, Exponential, etc). This assumption guarantees that the Signal to Noise Ratio (SNR) and the covariance matrix fully characterize the noise and redundancies.

(III)   Large variances have important dynamics: This assumption is based on the fact that the data has a high SNR. Hence, principal components with larger associated variances represent interesting dynamics, while those with lower variances represent noise.

(IV)    The principal components are orthogonal. This assumption provides an intuitive simplification that makes PCA soluble with linear algebra decomposition techniques.

With first assumption PCA is now limited to re-expressing the data as a linear combination of its basis vectors.

Assume that we have a set of single channel EEG data containing $m$ number of trials and each trial consist of $n$ number of data points as a column vector $\mathbf{X}$ (i.e., $\overrightarrow{X}$). $\mathbf{X}$ is an $m \times n$ matrix (4.3) as,

$$\mathbf{X} = \begin{bmatrix} X_1 \\ X_2 \\ X_3 \\ \vdots \\ X_m \end{bmatrix} \tag{4.3}$$

Let $\mathbf{Y}$ be another $m \times n$ matrix related by a linear transformation $\mathbf{P}$. $\mathbf{X}$ is the original recorded data set and $\mathbf{Y}$ is a re-representation of that data set (4.4) as,

$$\mathbf{P}\,\mathbf{X} = \mathbf{Y} \tag{4.4}$$

Now we define the following quantities,

- $\mathbf{p_i}$ are the *rows* of $\mathbf{P}$.
- $\mathbf{x_i}$ are the *columns* of $\mathbf{X}$.
- $\mathbf{y_i}$ are the *columns* of $\mathbf{Y}$.

Equation 4.4 represents a *change of basis* and can have following interpretations.

1.  $\mathbf{P}$ is a matrix that transforms $\mathbf{X}$ into $\mathbf{Y}$.
2.  Geometrically, $\mathbf{P}$ is a rotation and a stretch which again transforms $\mathbf{X}$ into $\mathbf{Y}$.
3.  The rows of $\mathbf{P}$, $\{\mathbf{p_1}, \ldots, \mathbf{p_m}\}$, are a set of new basis vectors for expressing the *columns* of $\mathbf{X}$.

The latter interpretation is not obvious but can be seen by writing out the explicit dot products of $\mathbf{PX}$ (4.5).

$$\mathbf{PX} = \begin{bmatrix} \mathbf{p_1} \\ \vdots \\ \mathbf{p_m} \end{bmatrix} \begin{bmatrix} \mathbf{x_1} & \cdots & \mathbf{x_n} \end{bmatrix}$$

$$\mathbf{Y} = \begin{bmatrix} \mathbf{p_1 \cdot x_1} & \cdots & \mathbf{p_1 \cdot x_n} \\ \vdots & \ddots & \vdots \\ \mathbf{p_m \cdot x_1} & \cdots & \mathbf{p_m \cdot x_n} \end{bmatrix} \tag{4.5}$$

The form of each column of **Y** (4.6) is

$$\mathbf{y}_i = \begin{bmatrix} \mathbf{p_1 \cdot x_i} \\ \vdots \\ \mathbf{p_m \cdot x_i} \end{bmatrix} \tag{4.6}$$

It is important to recognize that each coefficient of $\mathbf{y}_i$ is a dot-product of $\mathbf{x_i}$ with the corresponding row in **P**. Therefore, the *rows* of **P** are a new set of basis vectors for representing of *columns* of **X**.

By assuming linearity, the problem reduces to finding the appropriate change of basis. The row vectors $\{\mathbf{p_1}, \ldots, \mathbf{p_m}\}$ in this transformation will become the principal components of **X**. In order to obtain **P** satisfying all the assumptions described above, first we construct the covariance matrix $C_X$: $C_X \equiv \dfrac{1}{n-1} XX^T$. The matrix $C_X$ has the following properties.

(a) $C_X$ is a $m \times m$ square symmetric matrix

(b) The diagonal terms of $C_X$ are the variance of particular measurement types.

(c) The off-diagonal terms of $C_X$ are the covariance between measurement types.

Now aim is to find some orthonormal matrix $P$ where $Y = PX$ such that $C_Y \equiv \dfrac{1}{n-1} YY^T$ and all off-diagonal terms in $C_Y$ are zero (i.e. covariance between principal components must be a minimum). Thus, $C_Y$ is a diagonal matrix having diagonal elements as maximum variances. In PCA the transformation matrix P is found by finding eigen vectors of the covariance matrix $C_X$. The principal components of X are the eigenvectors of $XX^T$; or the rows of P.

In other words, the procedure for finding principal components is first to construct the covariance matrix of X and then finding the eigen vectors of the covariance matrix. The rows of the transformation matrix P are the eigen vectors.

### 4.2.2 BandPower

Construction of feature vectors using band powers is also popular in the BCI community. In order to obtain band powers, first, the signal corresponding to each EEG channel has to be Fourier transformed into the frequency domain separately. Then the power spectrum is calculated and whole frequency range is divided into bands having the same band width (the band width is given as the number of frequencies in each band). The band width can be determined such that classification accuracy is a maximal. The total power for each band is calculated and the feature vector is constructed with these band power values as its elements.

In order to describe the BandPower method in detail, let us assume that $X$ is a $n \times m$ matrix constructed with column vectors $X_1, X_2, \ldots, X_m$ containing data from $m$ EEG channels and each vector $X_i$ contains $n$ EEG data points.

$$X = \left( X_1, X_2, X_3, \ldots X_m \right)$$

Now define the Fourier transform $Y_i(\omega)$ of each column vector $X_i$ as

$$Y_i(\omega) = \Im(X_i)$$

The power spectrum matrix $P_i$ of $X_i$ is calculated by

$$P_i(\omega) = \begin{bmatrix} P_i(\omega_1) \\ P_i(\omega_2) \\ . \\ . \\ P_i(\omega_n) \end{bmatrix} = Y_i^* \bullet Y_i = \begin{bmatrix} Y_i^*(\omega_1)Y_i(\omega_1) \\ Y_i^*(\omega_2)Y_i(\omega_2) \\ . \\ . \\ Y_i^*(\omega_n)Y_i(\omega_n) \end{bmatrix}$$

where $Y_i^* = Conjgate \quad (Y_i)$ and $Y_i^* \bullet Y_i$ is the element by element multiplication of matrices $Y_i^*$ and $Y_i$ $i = 1, 2, \ldots m$. $\omega_j$ is the $j^{th}$ frequency.

Now assume that the number of frequencies in each band is $b$. Then the power $\Delta P_i(k)$ of $k^{th}$ band is

$$\Delta P_i(k) = \sum_{j=b(k-1)+1}^{j=bk} P_i(\omega_j)$$

where $k = 1, 2, \ldots \lambda$, ($\lambda$ is the number of bands) and $n = b\lambda$. Here we assume that n is exactly divisible by the band width $b$. If it is not, first $\lambda - 1$ bands will have the same band width $b$ while the last band has rest (less than $b$) of the frequencies.

Then the feature vector is constructed as a column vector having $\Delta P_i(k)$ as its elements where $k = 1, 2, \ldots \lambda$ and $i = 1, 2, \ldots m$. Hence the feature vector $F$ has the dimension $\lambda m$ and the form

$$F = \begin{bmatrix} F_1 \\ F_2 \\ F_3 \\ . \\ F_m \end{bmatrix} \text{ where } F_i = \begin{bmatrix} \Delta P_i(1) \\ \Delta P_i(2) \\ . \\ . \\ \Delta P_i(\lambda) \end{bmatrix}$$

### 4.2.3 DownSampling

Down sampling is a simple way of constructing feature vectors with reduced dimensions [14]. The data is down sampled by collecting data points at regular intervals. Let us assume that the column vector $X_i$ contains $n$ EEG data points of the channel $i$. That is

$$X_i = \begin{bmatrix} x_{1,i} \\ x_{2,i} \\ . \\ . \\ x_{n,i} \end{bmatrix}$$ . Then the down sampling is simply achieved by constructing a vector by

collecting data points from $X_i$ at regular intervals. If the down sampling is done with the rate $\eta$, then the resulting down sampled feature vector $F_i$ for $i^{th}$ channel is

$$F_i = \begin{bmatrix} x_{1,i} \\ x_{\eta,i} \\ x_{2\eta,i} \\ . \\ x_{j\eta,i} \end{bmatrix} \text{ and over all feature vector is } F = \begin{bmatrix} F_1 \\ F_2 \\ F_3 \\ . \\ F_m \end{bmatrix}$$

where $j = n \bmod(\eta)$; remainder on division of $n$ by $\eta$.

## 4.3 Classification Techniques

Several schemes have been employed in classifying cognitive tasks in brain computer interface systems. Most widely used methods are Discriminant Analysis (Linear and nonlinear discriminant analysis) [16-17], support vector machines [15, 18-19], nearest neighbor classifiers [12, 20-21], Neural networks (NN) [1, 22], Multilayer perceptron [23], Nonlinear Bayesian classifiers such as Bayes quadratic [12] and Hidden Markov Model (HMM) [24] and Combinations of classifiers. In this study we have employed Discriminant Analysis (Linear and nonlinear), support vector machines (Linear, nonlinear), and $k^{th}$ nearest neighbor classifiers. In the proceeding sections some of these classification techniques will be discussed in detail.

## 4.3.1 Linear Discriminant Analysis (LDA)

Linear discriminant analysis which has a close relationship with Fisher's linear discriminant is a method used in statistics and machine learning to find the linear combination of features which optimally divide two or more classes of objects or events. This method maximizes the ratio of between-class variance to the within-class variance in any particular data set thereby guaranteeing maximal separability.

First we describe the linear discriminant analysis for two class problems using Fisher criteria. Assume that we have a set of D-dimensional samples $\{x^1, x^2, ..., x^N\}$, $N_1$ of which belong to class $\omega_1$, and $N_2$ to class $\omega_2$. We seek to obtain a scalar y by projecting the samples $x$ onto a line

$$y = w^T x$$

The mean vector of each class in $x$ and $y$ feature space is

$$\mu_i = \frac{1}{N_i} \sum_{x \in \omega_i} x$$

and

$$\widetilde{\mu}_i = \frac{1}{N_i} \sum_{y \in \omega_i} y = \frac{1}{N_i} \sum_{x \in \omega_i} w^T x = w^T \mu_i$$

For each class we define the scatter which is equivalent to the variance as

$$\widetilde{s}_i^2 = \sum_{y \in \omega_i} (y - \widetilde{\mu}_i)^2$$

where the quantity $\left( \widetilde{s}_1^2 + \widetilde{s}_2^2 \right)$ is called the within class scatter.

Now the linear discriminant is defined as the linear function $w^T x$ which maximizes the Fisher criterion function

$$J(w) = \frac{\left| \widetilde{\mu}_1 - \widetilde{\mu}_2 \right|^2}{\left( \widetilde{s}_1^2 - \widetilde{s}_2^2 \right)}$$

We define a measure of the scatter in multivariate feature space $x$

$$S_i = \sum_{x \in \omega_i} (x - \mu_i)(x - \mu_i)^T$$

and $S_B = S_1 + S_2$

The scatter of the projection $y$ can be written in terms of scatter matrix in feature space $x$ as

$$\tilde{s}_i^2 = \sum_{y \in \omega_i} (y - \tilde{\mu}_i)^2 = \sum_{x \in \omega_i} \left(w^T x - w^T \mu_i\right)^2 = \sum_{x \in \omega_i} w^T (x - \mu_i)(x - \mu_i)^T w = w^T S_i w$$

$$\tilde{s}_1^2 + \tilde{s}_2^2 = w^T S_w w$$

and

$$\left(\tilde{\mu}_1 - \tilde{\mu}_2\right)^2 = \left(w^T \mu_1 - w^T \mu_2\right)^2 = w^T (\mu_1 - \mu_2)(\mu_1 - \mu_2)^T w = w^T S_B w$$

Where $S_B = (\mu_1 - \mu_2)(\mu_1 - \mu_2)^T$. Now we write $J(w)$ in terms of $w$ as

$$J(w) = \frac{w^T S_B w}{w^T S_W w}$$

To obtain the maximas of $J(w)$, we differentiate $J(w)$ and equate to zero.

$$\frac{dJ}{dw} = \left(w^T S_W w\right)\frac{d}{dw}\left[w^T S_B w\right] - \left(w^T S_B w\right)\frac{d}{dw}\left[w^T S_W w\right] = 0$$

$$\left[w^T S_W w\right] 2 S_B w - \left[w^T S_B w\right] 2 S_W w = 0$$

dividing by $w^T S_W w$, we have

$$S_B w - J S_W w = 0 \quad \text{and}$$

$$S_W^{-1} S_B w - J w = 0$$

This is equivalent to the eigenvalue problem

$$S_W^{-1} S_B w = J w$$

and by solving the general eigenvalue problem we obtain the eigenvector $w^* = S_W^{-1}(\mu_1 - \mu_2)$ which provides the optimum projection. In other words, it will project samples from the same class very close to each other while, means of the two classes will be kept as farther apart as possible.

LDA method for two class problem described above can be generalized to multi-class problems. Suppose that there are a $c$ total number of classes in the sample. It can be shown that optimal projection matrix $w^*$ is given by

$$\left(S_B - \lambda_i S_w\right)w_i^* = 0$$

where columns of $w^*$ are the eigenvectors corresponding to the largest eigenvalue of the eigenvalue equation given above and

$$S_w = \sum_{i=1}^{C} S_i$$

and $\quad S_i = \sum_{x \in \omega_i}(x - \mu_i)(x - \mu_i)^T$

$$\mu_i = \frac{1}{N_i} \sum_{x \in \omega_i} x$$

$$S_B = \sum_{i=1}^{C} N_i (\mu_i - \mu)(\mu_i - \mu)^T$$

$$\mu = \frac{1}{N} \sum_{x \in \omega_i} N_i \mu_i$$

In multi-class problems matrix $w^*$ provides the optimum projection.

**4.3.2 Support Vector Machines**

*(i)  Linearly Separable Binary classification*

Assume that we have a set of D-dimensional samples *{x¹, x², ..., x^N}*, *N₁* of which belong to class *ω₁*, and *N₂* to class *ω₂*. We represent classes ω₁ and ω₂ by the variable $y_i$ (*i =1,2*) which can have two values  *-1* or *+1* respectively. Then our training sample has of the form

$\{\vec{x}_i, y_i\}$ where *i = 1 ... N,* $y_i \in \{-1,1\}$, $\vec{x} \in R^D$

Here we assume that the data is linearly separable such that we can draw a line on a graph of *x₁* vs *x₂* separating the two classes when *D = 2* and a hyperplane on graphs of *x₁, x₂ . . . x_D* for when *D > 2*. This hyper place is described by the equation $\vec{w}.\vec{x} + b = 0$  where  $\vec{w}$ is normal to the hyperplane and $\frac{b}{\|w\|}$ is the perpendicular distance from the hyperplane to the origin. Support Vectors are the sample points closest to the separating hyperplane and the aim of Support Vector Machines (SVM) is to orient this hyperplane in such a way as to be as far as possible from the closest members of both classes.

As it can be seen in Figure 4.3, implementation of a SVM comes down to selecting the variables *w* and *b* so that our training data can be described by:

$x_i.w + b \geq +1$    *for* $y_i = +1$                                          (4.7)

$x_i.w + b \leq -1$    *for* $y_i = -1$                                          (4.8)

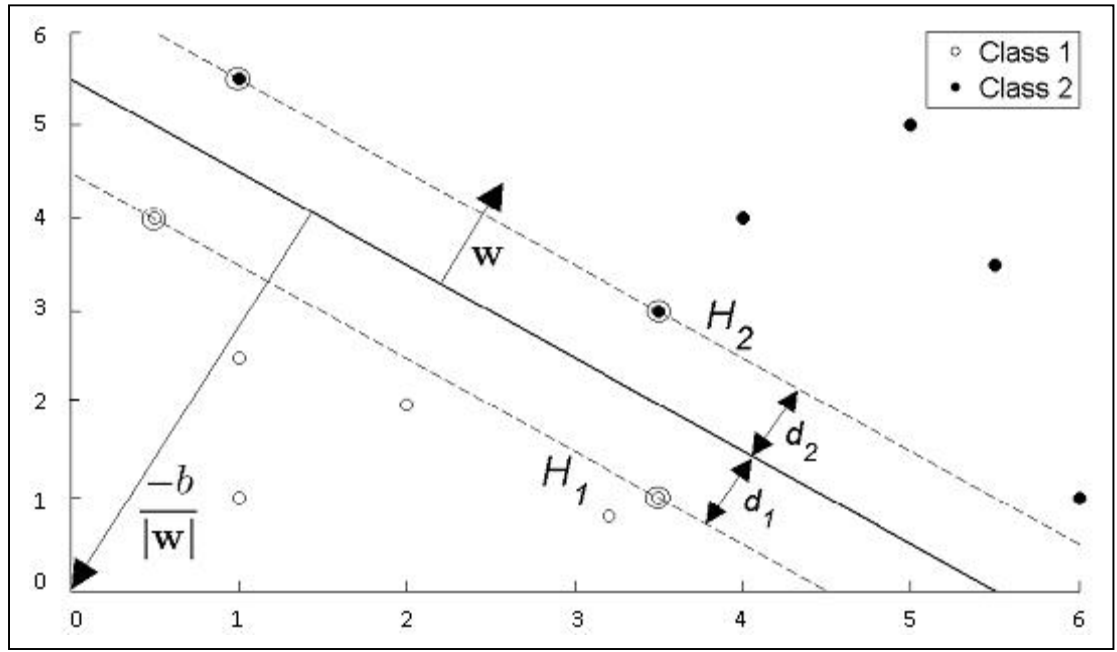

Figure 4.3. Hyperplane through two linearly separable classes

These equations can be combined into

$$y_i(x_i.w+b)-1 \geq 0 \quad \forall i \qquad (4.9)$$

If we now just consider the points that lie closest to the separating hyperplane, i.e. the Support Vectors (shown in circles in the diagram), then the two planes $H_1$ and $H_2$ that these points lie on can be described by:

$$x_i.w+b = 1 \text{ for } H_1$$

$$x_i.w+b = -1 \text{ for } H_2$$

As shown in figure 1, we define $d_1$ as being the distance from $H_1$ to the hyperplane and $d_2$ from $H_2$ to it. The hyperplane is equidistance from $H_1$ and $H_2$ means that $d_1 = d_2$ - a quantity known as the SVM's margin. In order to orient the hyperplane to be as far from the Support Vectors as possible, we need to maximize this margin.

Simple vector geometry shows that the margin is equal to $\dfrac{1}{\|w\|}$ and maximizing it subject to the constraint in (4.9) is equivalent to finding:

$$\min \|w\| \text{ such that } y_i\left(x_i.w+b\right)-1 \geq 0 \quad \forall i$$

Minimizing $\|w\|$ is equivalent to minimizing $\dfrac{1}{2}\|w\|^2$ and the use of this term makes it possible to perform Quadratic Programming (QP) optimization later on. We therefore need to find:

$$\min \frac{1}{2}\|w\|^2 \text{ such that } y_i\left(x_i.w+b\right)-1 \geq 0 \quad \forall i \qquad (4.9.1)$$

In order to cater for the constraints in this minimization, we need to allocate them Lagrange multipliers $\alpha$ where $\alpha_i \geq 0 \; \forall i$

$$L_p \equiv \frac{1}{2}\|w\|^2 - \alpha\left[y_i\left(x_i.w+b\right)-1 \; \forall i\right]$$

$$\equiv \frac{1}{2}\|w\|^2 - \sum_{i=1}^{N}\alpha_i\left[y_i\left(x_i.w+b\right)-1\;\right]$$

$$\equiv \frac{1}{2}\|w\|^2 - \sum_{i=1}^{N}\alpha_i y_i\left(x_i.w+b\right) \;\; - \sum_{i=1}^{N}\alpha_i \qquad (4.10)$$

We wish to find the $w$ and $b$ which minimizes, and the $\alpha$ which maximizes (4.10) (while keeping $\alpha_i \geq 0 \; \forall i$). We can do this by differentiating $L_P$ with respect to $w$ and $b$ and setting the derivatives to zero:

$$w = \sum_{i=1}^{N}\alpha_i y_i x_i \qquad (4.11)$$

$$\sum_{i=1}^{N}\alpha_i y_i = 0 \qquad (4.12)$$

Substituting (4.11) and (4.12) into (4.10) gives a new formulation which, being dependent on $\alpha$, we need to maximize:

$$L_D \equiv \sum_{i=1}^{N} \alpha_i - \frac{1}{2} \sum_{i,j} \alpha_i \alpha_j y_i y_j x_i.x_j \text{ such that } \alpha_i \geq 0 \ \forall i \ , \ \sum_{i=1}^{N} \alpha_i y_i = 0$$

$$\equiv \sum_{i=1}^{N} \alpha_i - \frac{1}{2} \sum_{i,j} \alpha_i H_{ij} \alpha_j \text{ where } H_{ij} \equiv y_i y_j x_i.x_j$$

$$\equiv \sum_{i=1}^{N} \alpha_i - \frac{1}{2} \alpha^T H \alpha \text{ such that } \alpha_i \geq 0 \ \forall i \ , \ \sum_{i=1}^{N} \alpha_i y_i = 0$$

This new formulation $L_D$ is referred to as the Dual form of the Primary $L_P$. It is important noting that the Dual form requires only the dot product of each input vector $x_i$ to be calculated, this is important for the Kernel Trick described in section *(iii)* below.

Having moved from minimizing $L_P$ to maximizing $L_D$, we need to find:

$$\max_{\alpha} \left[ \sum_{i=1}^{N} \alpha_i - \frac{1}{2} \alpha^T H \alpha \right] \text{ such that } \alpha_i \geq 0 \ \forall i \ , \ \sum_{i=1}^{N} \alpha_i y_i = 0 \qquad (4.13)$$

This is a convex quadratic optimization problem, and we run a QP solver which will return $\alpha$ and from (4.11) will give us $w$. What remains is to calculate $b$. Any data point satisfying (4.12) which is a Support Vector $x_s$ will have the form:

$$y_s \left( x_s.w + b \right) = 1$$

Substituting in (4.10):

$$y_s \left( \sum_{m \in S} \alpha_m y_m x_m.x_s + b \right) = 1$$

Where $S$ denotes the set of indices of the Support Vectors. S is determined by finding the indices $i$ where $\alpha_i > 0$. Multiplying through by $y_s$ and then using $y_s^2 = 1$ from (4.7) and (4.8):

$$y_s^2 \left( \sum_{m \in S} \alpha_m y_m x_m . x_s + b \right) = y_s$$

$$b = y_s - \sum_{m \in S} \alpha_m y_m x_m . x_s$$

Instead of using an arbitrary Support Vector $x_s$, it is better to take an average over all of the Support Vectors in $S$:

$$b = \frac{1}{N} \sum_{s \in S} \left( y_s - \sum_{m \in S} \alpha_m y_m x_m . x_s \right)$$

We now have the variables $w$ and $b$ that define our separating hyperplane's optimal orientation and hence our Support Vector Machine.

*(ii) Binary Classification for Data that is not Fully Linearly Separable*

In order to extend the SVM methodology to handle data that is not fully linearly separable, we relax the constraints for (4.7) and (4.8) slightly to allow for misclassified points. This is done by introducing a positive slack variable $\xi_i$ *where i = 1, . . . . N*

$$x_i . w + b \geq +1 - \xi_i \quad \textit{for } y_i = +1$$

$$x_i . w + b \leq -1 - \xi_i \quad \textit{for } y_i = -1$$

$$\xi_i \geq 0 \,\, \forall i$$

These equations can be combined into

$$y_i \left( x_i . w + b \right) - 1 + \xi_i \geq 0 \,\, \text{ where } \,\, \xi_i \geq 0 \,\, \forall i$$

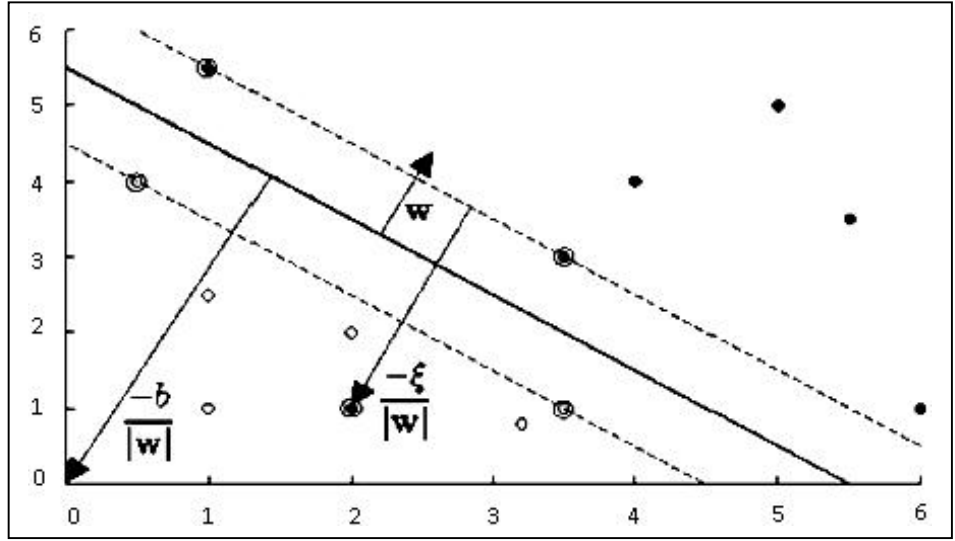

Figure 4.4. Hyperplane through two non-linearly separable classes

Hyperplane through two non-linearly separable classes in this soft margin SVM, data points on the incorrect side of the margin boundary have a penalty that increases with the distance from it (Figure 4.4). As we are trying to reduce the number of misclassifications, a sensible way to adapt our objective function (4.13) from previously, is to find:

$$\min \frac{1}{2}\|w\|^2 + C\sum_{i=1}^{N}\xi_i \text{ such that } y_i\left(x_i.w + b\right) - 1 + \xi_i \geq 0 \quad \forall i$$

Where the parameter C controls the trade-off between the slack variable penalty and the size of the margin. Reformulating as a Lagrangian, which as before we need to minimize with respect to w, b and $\xi_i$ and maximize with respect to $\alpha$ (where $\alpha_i \geq 0$, $\mu_i \geq 0 \ \forall i$)

In order to use an SVM to solve a classification or regression problem on data that is not linearly separable, we need to first choose a kernel and relevant parameters which map the non-linearly separable data into a feature space where it is linearly separable. This can be achieved empirically, as an example, by trial and error. Sensible kernels to start with are the Radial Basis, Polynomial and Sigmoidal kernels. The first step, therefore, consists of choosing our kernel and hence the mapping

$$L_p \equiv \frac{1}{2}\|w\|^2 + C\sum_{i=1}^{N}\xi_i - \sum_{i=1}^{N}\alpha_i\left[y_i(x_i.w + b) - 1 + \xi_i\right] \quad -\sum_{i=1}^{N}\mu_i\xi_i$$

Differentiating with respect to w, b and $\xi_i$ and setting the derivatives to zero:

$$w = \sum_{i=1}^{N}\alpha_i y_i x_i$$

$$\sum_{i=1}^{N}\alpha_i y_i = 0$$

$$c = \alpha_i + \mu_i \tag{4.14}$$

Substituting these in, $L_D$ has the same form as (4.13) before. However (4.14) together with $\mu_i \geq 0 \; \forall i$, implies that $\alpha \geq C$. We therefore need to find:

$$\max_{\alpha}\left[\sum_{i=1}^{N}\alpha_i - \frac{1}{2}\alpha^T H\alpha\right] \text{ such that } C \geq \alpha_i \geq 0 \; \forall i, \; \sum_{i=1}^{N}\alpha_i y_i = 0$$

$b$ is then calculated in the same way as in (4.9.1) before, though in this instance the set of Support Vectors used to calculate $b$ is determined by finding the indices $i$ where $C \geq \alpha_i \geq 0$

*(iii) <u>Nonlinear Support Vector Machines</u>*

When applying our SVM to linearly separable data we have started by creating a matrix H from the dot product of our input variables:

$$H_{ij} = y_i y j k(x_i, x_j) = x_i.x_j = x_i^T.x_j$$

$k(x_i, x_j)$ is an example of a family of functions called *Kernel Functions* ($k(x_i, x_j) == x_i^T.x$ being known as the Linear Kernal) The set of kernel functions is composed of variants of (4.15) in that they are all based on calculating inner products of two vectors. This means that if the functions can be recast into a higher dimensionality space by some potentially non-linear feature mapping function $x \rightarrow \phi(x)$ only inner products of the mapped inputs in the feature space need be determined without us needing to explicitly calculate $\phi$.

The reason that this Kernel Trick is useful is that there are many classification / regression problems that are not linearly separable / regressable in the space of the inputs x, which might be in a higher dimensionality feature space given a suitable mapping $x \rightarrow \phi(x)$.

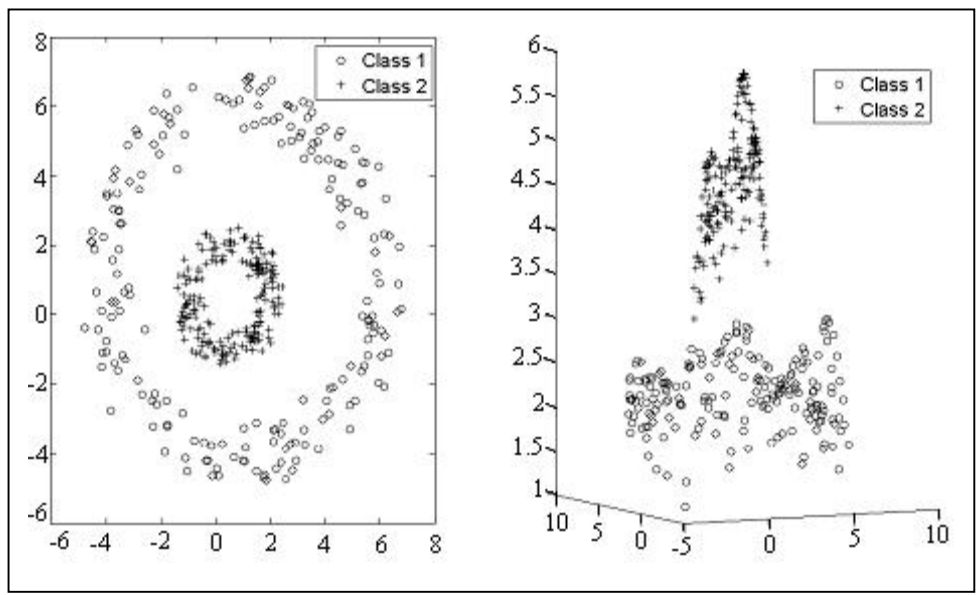

Figure 4.5. Dichotomous data re-mapped using Radial Basis Kernel

Referring to Figure 4.5, if we define our kernel to be:

$$k(x_i, x_j) = exp^{-\left[\frac{\|x_i - x_j\|^2}{2\sigma^2}\right]}$$
(4.15)

then a data set that is not linearly separable in the two dimensional data space x (as in the left hand side of Figure 4.5) is separable in the nonlinear feature space (right hand side of Figure 4.5) determined implicitly by this non-linear kernel function - known as the Radial Basis function Kernel or Gaussian Kernel. Other popular kernels for classification and regression are the Polynomial Kernel

$$k(x_i, x_j) = \left(x_i . x_j + a\right)^b$$

and the Sigmoidal Kernel

$$k(x_i, x_j) = \tanh\left(ax_i . x_j - b\right)$$

### 4.3.3 Nearest Neighbor classification method

Third category of classification methods discussed in this Chapter is nearest neighbor classifications.

The classifiers presented in this section consist in assigning a feature vector to a class according to its nearest neighbor(s). This neighbor can be a feature vector from the training set as in the case of $k$ Nearest Neighbors ($k$NN), or a class prototype as in Mahalanobis distance. They are discriminative nonlinear classifiers.

### k[th] Nearest neighbor ( kNN) classification

The aim of this technique is to assign to an unseen point the dominant class among its k nearest neighbors within the training set [20]. For BCI, these nearest neighbors are usually obtained using a metric distance, e.g., [21]. With a sufficiently high value of k and enough training samples, kNN can approximate any function which enables it to produce nonlinear decision boundaries. kNN algorithms are not very popular in the BCI community, probably because they are known to be very sensitive to the curse-of-dimensionality [25], which made them fail in several BCI experiments [21, 26]. However, when used in BCI systems with low dimensional feature vectors, kNN may prove to be efficient [27].

*Mahalanobis distance*

Mahalanobis distance based classifiers assume a Gaussian distribution $N$ $(\mu_c, M_c)$ for each prototype of the class $c$. Then, a feature vector $x$ is assigned to the class that corresponds to the nearest prototype, according to the so-called Mahalanobis distance $d_c(x)$ [28]:

$$d_c(x) = \sqrt{(x - \mu_c)M_c^{-1}(x - \mu_c)^T}$$

This leads to a simple yet robust classifier, which even proved to be suitable for multiclass [29] or asynchronous BCI systems [28]. Despite its good performances, it is still scarcely used in the BCI literature.

### 4.3.4 Combinations of classifiers

In most papers related to BCI, the classification is achieved using a single classifier. A recent trend, however, is to use several classifiers, aggregated in different ways. The classifier combination strategies used in BCI applications are the following:

*(a) Boosting:*

Boosting consists of using several classifiers in sequence, each classifier focusing on the errors committed by the previous ones. It can build up a powerful classifier out of several weak ones, and it is unlikely to overtrain. To date, in the field of BCI, boosting has been experimented with MLP and Ordinary Least Square (OLS) [12].

*(b) Voting:*

While using voting, several classifiers are being used, each of them assigning the input feature vector to a class [12]. The final class will be that of the majority. Voting is the most popular way of combining classifiers in BCI research, probably because it is simple and efficient. For instance, voting with LVQ NN, MLP or SVM has been attempted.

*(c) Stacking:*

Stacking consists in using several classifiers, each of them classifying the input feature vector. These classifiers are called level-0 classifiers. The output of each of these classifiers is then given as input to a so-called meta-classifier (or level-1 classifier) which makes the final decision. Stacking has been used in BCI research using HMM as level-0 classifiers, and an SVM as meta classifier. The main advantage of such techniques is that a combination of similar classifiers is very likely to outperform one of the classifiers on its own. Actually, combining classifiers is known to reduce the variance and thus the classification error [25].

### 4.3.4 Main classification problems in BCI

While performing a pattern recognition task, classifiers may be facing several problems related to the feature properties such as outliers, overtraining, etc. In the field of BCI, two main problems need mentioning: the curse-of-dimensionality and the bias–variance tradeoff.

*The curse-of-dimensionality*

The amount of data needed to properly describe the different classes increases exponentially with the dimensionality of the feature vectors [12]. In fact, if the number of training data is small compared to the size of the feature vectors, the classifier will most probably give poor results. It is advisable to use, at least, five to ten times as many training samples per class as the dimensionality. Unfortunately this cannot be applied in all BCI systems as generally the dimensionality is high and the training set small. Therefore this 'curse' is a major concern in BCI design.

*The bias–variance tradeoff*

Formally, classification consists in finding the true label y* of a feature vector x using a mapping φ. This mapping is learnt from a training set T. The best mapping *f\** that has generated the labels is, of course, unknown. If we consider the Mean Square rror (MSE), classification errors can be decomposed into three terms [25]:

$$MSE = E\left[(y^* - f(x))^2\right]$$
$$= E\left[(y^* - f^*(x) + f^*(x) - E[f(x)] + E[f(x)] - f(x))^2\right]$$
$$= E\left[(y^* - f^*(x))^2\right] + E\left[(f^*(x) - E[f(x)])^2\right] + E\left[(E[f(x)] - f(x))^2\right]$$
$$= Noise^2 + Bias(f(x))^2 + Var(f(x))$$

These three terms describe three possible sources of classification error:

- *noise:* represents the noise within the system and cannot be reduced.
- *bias:* represents the divergence between the estimated mapping and the best mapping. Therefore, it depends on the method that has been chosen to obtain *f* (linear, quadratic,...)
- *variance:* reflects the sensitivity to the training set *T* used.

To attain the lowest classification error, both the bias and the variance must be low. Unfortunately, there is a 'natural' bias– variance tradeoff. Actually, stable classifiers tend to have a high bias and a low variance, whereas unstable classifiers have a low bias and a high variance. This can explain why simple classifiers sometimes outperform more complex ones. Several techniques, known as stabilization techniques, can be used to reduce the variance. Among them, combination of classifiers and regularization can be mentioned. EEG signals are known to be non-stationary. Training sets coming from different sessions are likely to be relatively different. Thus, a low variance can be a solution to cope with the variability problem in BCI systems.

# CHAPTER 5

# HARDWARE AND SOFTWARE FOR BRAIN COMPUTER INTERFACE

As we have already mentioned in Chapter 1, this project mainly focused on two objectives,

(1) Finding new mental tasks which alter EEG signals such that they can be used directly in BCI applications.

(2) Finding the best computational methods for preprocessing, feature vector construction and classification to recognize above mental tasks correctly and efficiently to be used in real BCI systems.

## 5.1 Mental Tasks

To accomplish the first objective mentioned above, we started our investigation by analyzing the existing mental tasks used by the BCI research community. Here we mainly concentrated on how successfully these mental tasks have been implemented by others in the existing BCI systems. Usually the success rate of a group of mental tasks or a signal processing technique is measured with classification accuracies.

Some of the commonly used known mental tasks are listed bellow with details [28, 11].

*(1)* Baseline $M_1$:

The Subject is asked not to perform a specific mental task. However they can relax and think of nothing in particular while their eyes are open. The baseline signal represents the mental stage of the subject when he/she is not thinking of any specific mental task used in controlling BCI. In some BCI experiments Subject is asked to relax but close and open their eyes in few seconds intervals.

*(2) Mental Arithmetic* **M₂**:

In this case, subjects have to solve non-trivial multiplications without vocalizing or moving. The tasks are non-repeating and designed so that an immediate answer is not obvious. The subjects usually verify at the end of the task whether or not he/she arrived at the solution.

*(3) Mental Letter Composing* **M₃**:

Subject is instructed to mentally compose a letter to a friend or relative, without moving or vocalizing.

*(4) Visual Counting* **M₄** :

Subjects have to imagine a black board and mentally to visualize numbers being sequentially written on the board.

*(5) Geometric Figure Rotation* **M₅**:

Subject is shown a drawing of a complex geometric figure. Then the figure is moved out of sight and Subject is instructed to imagine the rotation of this figure.

*(6) Mental word generation* **M₆**:

Subject is asked to imagine generation of words beginning with the same random letter.

*(7) Visual stimulus driven letter imagination* **M₇** :

Subject is asked to imagine letter according to the visual stimulus.

*(8) Auditory stimulus driven letter imagination* **M₈** :

Subject is asked to imagine letter according to the auditory stimulus.

*(9) Left motor imagery* $M_9$ *:*

Subject is asked to imagine of left hand movement, left finger movement or left foot movement.

*(10) Right motor imagery* $M_{10}$ *:*

Subject is asked to imagine of right hand movement, right finger movement or right foot movement.

*(11) Left imagined hand movements* $M_{11}$ *:*

Subject is asked to imagine of left hand movement.

*(12) Right imagined hand movements* $M_{12}$ *:*

Subject is asked to imagine of right hand movement.

*(13) Imagined foot movements* $M_{13}$ *:*

Subject is asked to imagine of foot movement.

*(14) Imagined tongue movements* $M_{14}$ *:*

Subject is asked to imagine the movement of tongue.

*(15) Music composition* $M_{15}$ *:*

Subject is listening to a tone shortly, and then started imagining the same tone.

Following Tables (5.1, 5.2, and 5.3) present a summary of commonly employed algorithms and show the best performances in EEG-based BCI research done so far [27].

Table.5.1 Accuracy of classifiers in *mental task imagination* based BCI.

| Mental Tasks (as numbered above) | Preprocessing methods | Feature Vector Construction | Classification methods | Accuracy (%) | References |
|---|---|---|---|---|---|
| $M_7$ & $M_8$ | PCA + ICA | LPC spectra | RBF NN | 77.8 | [1] |
| ($M_9$ or $M_{10}$) & $M_2$ | | Lagged AR | BLRNN | $80 \pm 12$ | [2, 3] |
| $M_1$ & $M_2$ | | Multivariate AR | MLP | 91.4 | [4] |
| $M_1$ & best of {$M_2$,$M_3$,$M_4$, $M_5$} | | BP | MLP | 95 | [5] |
| All couples between {$M_1$,$M_2$,$M_3$,$M_4$,$M_5$} on Keirn & Aunon EEG data | | AR | BGN | 90.63 | [6] |
| | | | Bayes quadratic | $90.51 \pm 3.2$ | [7] |
| Best triplet between {$M_1$,$M_2$,$M_3$,$M_4$,$M_5$} | | WK-Parzen PSD | Fuzzy ARTMAP NN | 94.43 | [8] |
| $M_9$ & $M_10$ & $M_1$ | | Cross ambiguity functions | Mahalanobis distance & MLP | 71 | [9] |
| $M_1$ & $M_2$ & $M_3$ & $M_4$ & $M_5$ on Keirn and Aunon EEG data | | PSD with Welch's periodogram | MLP | 90.6–98.6 | [10] |
| | | AR | MLP | up to 71 | [11] |
| | 0.1–100 Hz band pass | AR | Gaussian SVM | 72 | [12] |
| $M_9$ & $M_{10}$ & $M_1$ asynchronous mode | SL & 4–45Hz band pass | PSD with Welch's periodogram | Gaussian classifier | 84 | [13] |
| $M_9$ & $M_{10}$ & $M_6$ in asynchronous mode | SL | PSD | IOHMM | 81.6 | [14] |

Table 5.2. Accuracy of classifiers in *pure motor imagery* based BCI: *two-class* and *synchronous*. The two classes are left and right imagined hand movements ($M_{10}$ and $M_{11}$).

| Mental Tasks (as numbered above) | Preprocessing methods | Feature Vector Construction | Classification methods | Accuracy (%) | References |
|---|---|---|---|---|---|
| On different EEG data | | Correlative time -frequency representation | Gaussian SVM | 86 | [15] |
| | | BP | LDA | 83.6 | [16] |
| | | Fractal dimension | LDA | 80.6 | [16] |
| | | Hjorth parameters | HMM | 81.4 ± 12.8 | [17] |
| | | AAR parameters | FIR NN | 87.4 | [18] |
| | | PCA | HMM+SVM | 78.15 | [19] |
| On BCI competition 2003 data set III | | Morlet wavelet | Bayes quadratic integrated over time | 89.3 | [20] |
| | | AAR | MLP | 84.29 | [7] |
| | | Raw EEG | HMM | up to 77.5 | [21] |
| | SL + 4–45 Hz band pass | PSD | LDA | 65.6 | [22] |

Table 5.3. Accuracy of classifiers in *pure motor imagery* based BCI: *multiclass* and /or *asynchronous* case.

| Mental Tasks (as numbered above) | Pre processing methods | Feature Vector Construction | Classification methods | Accuracy (%) | References |
|---|---|---|---|---|---|
| $M_{10}$ & $M_{11}$ & $M_{12}$ on different data | | BP | HMM | 77.5 | [23] |
| $M_{10}$ & $M_{11}$ & $M_{12}$ & $M_{13}$ on different data | | AAR | Linear SVM | 63 | [24] |
| | | BP | HMM | 63 | [23] |
| $M_{10}$ & $M_{11}$ & $M_{12}$ & $M_{13}$ & $M_1$ | | BP | HMM | 52.6 | [23] |
| $M_{10}$ & $M_{11}$ in asynchronous mode | | Welch power spectrum | Mahalanobis distance | 90 | [25] |
| $M_{10}$ & $M_{11}$ & $M_{12}$ in asynchronous mode | | BP | LDA | 95 | [26] |

In order to find new mental tasks which alter EEG signals significantly enough to be used in BCI systems, we have investigated following mental activities.

## (1) Imagining writing of an English character

Subject is asked to imagine writing English characters. We tried to distinguish thoughts related to writing different characters.

## (2) Imagination of Mirror image of an English character

Given an English character, Subject is asked to imagine the mirror image of it.

## (3) Imagination of hitting a given square by an imaginary arrow from above, below, right, or left to the screen

Subject is asked to imagine one of the following arrows hitting a given square (Figure 5.1).

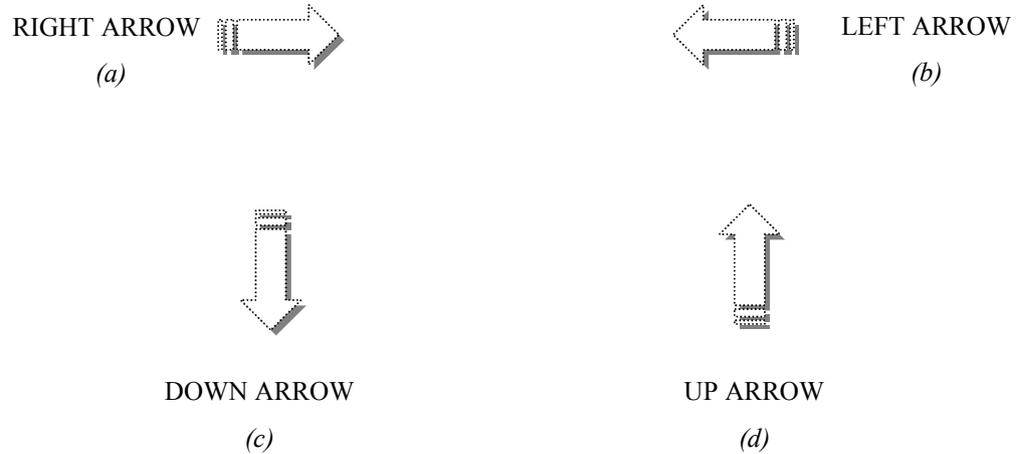

RIGHT ARROW
*(a)*

LEFT ARROW
*(b)*

DOWN ARROW
*(c)*

UP ARROW
*(d)*

Figure 5.1. Imaginary arrow directions

**(4) *Finding possible Symmetries of a given object***

Subject is shown an object and asked to find all possible symmetries of it.

**(5) *Imagining bunch of vertical lines and horizontal lines***

Subject is asked to imagine one of two of the following (Figure 5.2).

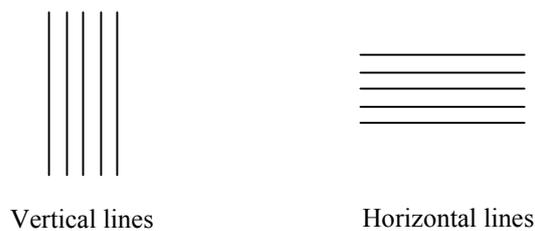

Vertical lines

Horizontal lines

Figure 5.2. Vertical and Horizontal lines

In the preliminary studies it was found that most promising set of mental tasks which could be identified by a BCI system is (3), i.e. *imagination of hitting a given square by an imaginary arrow from above, below, right, or left to the screen*. We call these mental tasks "*Hit Series*" (HS). We have carried out detail investigations of HS and compared it with performance of Left Middle Finger Movement (LFM) and Right Middle Finger Movement (RFM) methods mentioned above.

The major reason for comparing performance of HS with that of LFM and RFM is that currently Motor Imagery (MI) is the most heavily used mental tasks in EEG based BCI systems.

In the next Chapter we will describe how we have tested the performances of HS, and MI.

## 5.2 Hardware

For recording EEG signals from subjects, we have used the Mindset-24R amplifier system manufactured by NeuroPulse-Systems LLC, USA (Figure 5.3) [29]. This amplifier system consists of 24 differential input channels with 90 dB amplifier gain, and 60 dB signals to noise ratio. Signals receiving from each channel are amplified by a separate amplifier. Further it has an inbuilt 16 bit analog to digital converter with software selectable 64, 128, 256, or 512 samples/second/channel sample rate. Common Mode Rejection of the amplifier is 120 dB maximum. Mindset-24R amplifier accepts input voltages between 0 to 120 µV peak-to-peak. Amplifier contains an expansion slot which can be used to insert a 19 channel International 10/20 montage adopter. Input lines of the amplifier are isolated via an Opto-Isolator which has maximum protection up to 2500 volts RMS. Calibration of the amplifier can be accomplished by means of software. Amplifier communicates with the computer through a SCIS connection.

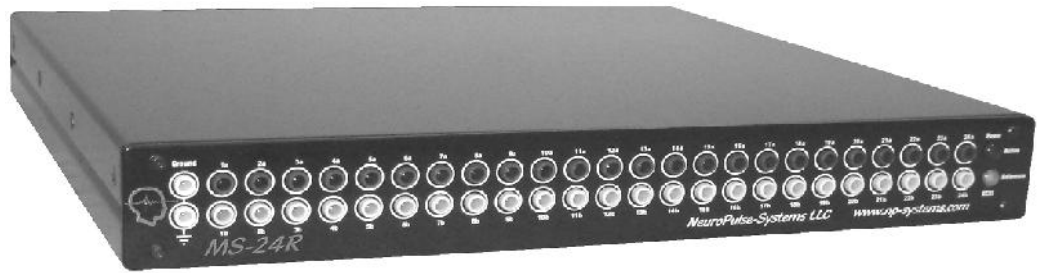

Figure 5.3. Mindset-24R amplifier system.

To capture EEG signals from the scalp we have used ECI Electro-Cap electrode system II ™ manufactured by Electro-Cap International Inc (Figure 5.4) [30].

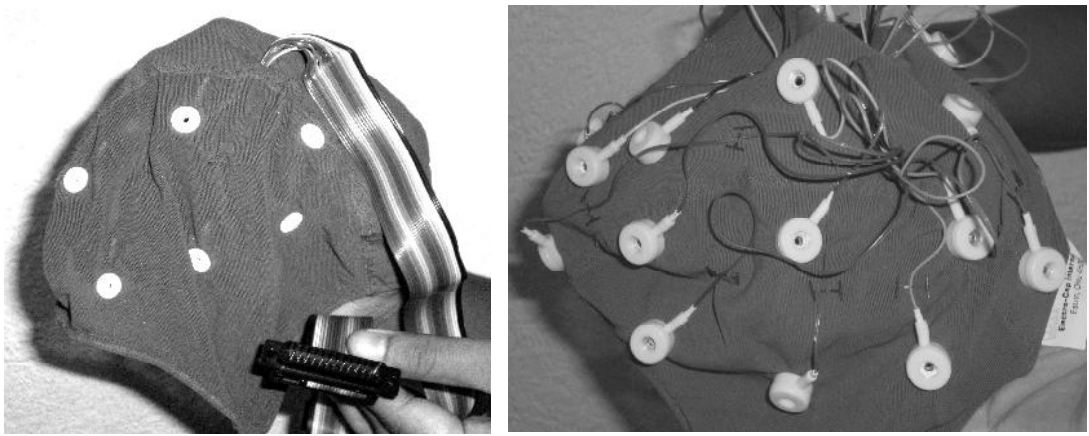

Figure 5.4. ECI Electro-Cap electrode system II ™.

The electrodes on this cap are positioned to the international 10-20 method of electrode placement which we have described in Chapter 03. Contacts between electrodes and the scalp are made by injecting conducting gel "Electro-Gel™" into the electrodes. Electro-cap consists of 20 electrodes linked to 25-pin male connector.

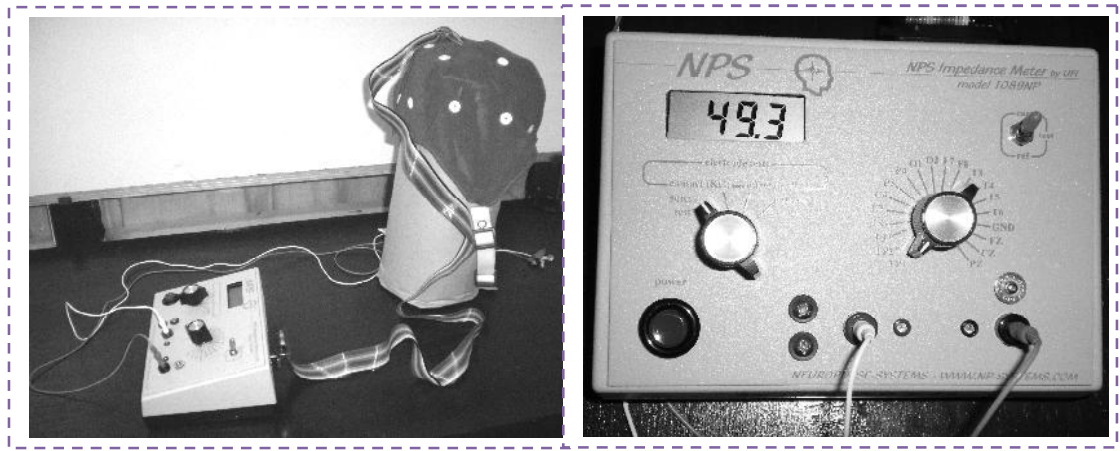

Figure 5.5. Setup of the Electro-Cap and the NPS impedance meter.

In addition to 20 electrodes in the electro-cap, two ear electrodes are also connected to the amplifier as reference electrodes. For EEG recordings impedance between EEG electrodes and these reference electrodes must be below 3 KΩ.

For measuring impedance, we have used NPS impedance meter (model 1089NP) manufactured by NeroPulse Systems (Figure 5.5). All 20 electrodes of the electro-cap can be connected to the above impedance meter via connector pins of the electro-cap and ear electrodes can be directly connected to the impedance meter through two sockets located on top of the meter.

Although the EEG recording systems contains precision components, component tolerance and ageing issues of these components can cause the amplifiers to vary by about 10% from the norm. Therefore, before the EEG recordings are made for the first time, the system should be calibrated. For this purpose NPS calibrator 00230 manufactured by NeuroPulse System LLC has been used. This calibrator injects a very precise 16 Hz, 50 µV signal into all 24 amplifiers in the system. The recording software reads the outputs of the amplifiers and calculates a correction factor (a scalar) to bring the amplifiers to the norm (Figure 5.6).

Power Supply

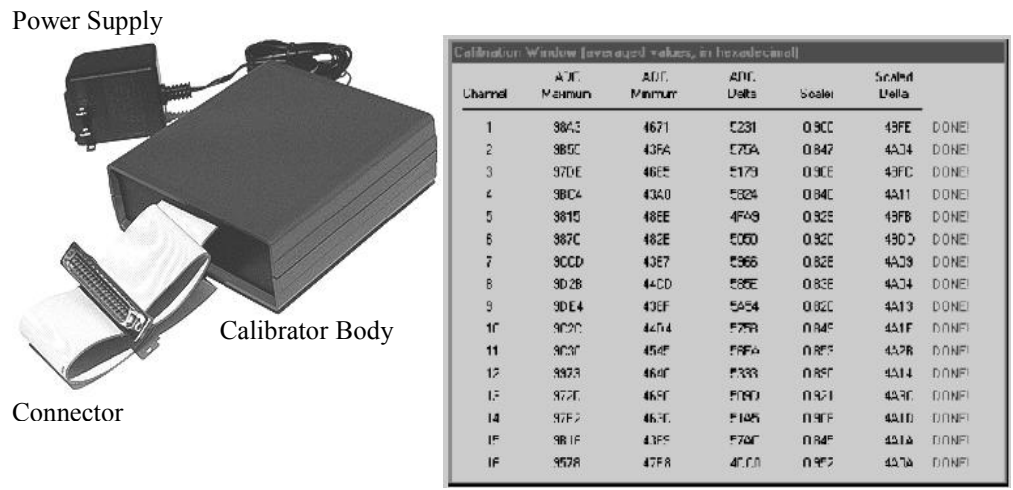

Calibrator Body

Connector

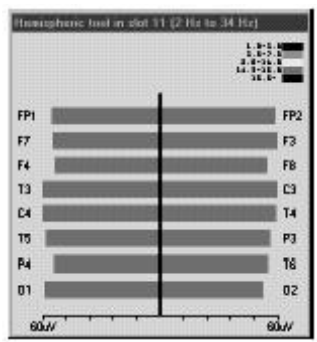

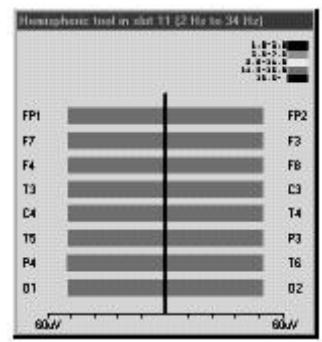

Calibration signal injected into a typical production Mindset. Note small amplification differences between the channels due to component tolerances:

The same signals following calibration. Note how the channels have been been normalized:

Figure 5.6. NPS Calibrator and the Calibration software.

## 5.3 Recording Software and accessories

Three following software programs have been used for recording and manipulating data received from EEG amplifier system through SCSI connection.

1. Alarm program

2. Mindmeld 24 Data acquisition software

3. Data conversion program (DCon)

We have developed a program named *"Alarm Program"* for informing the subject about the forthcoming mental task to be recorded and assisting the subject with when the recording should start and when it should end (Figure 5.7). The program is written in Visual Basic version 6.0 and it informs the subject about the imminent mental task with computer voice while starting and ending of each recording are notified with a beep.

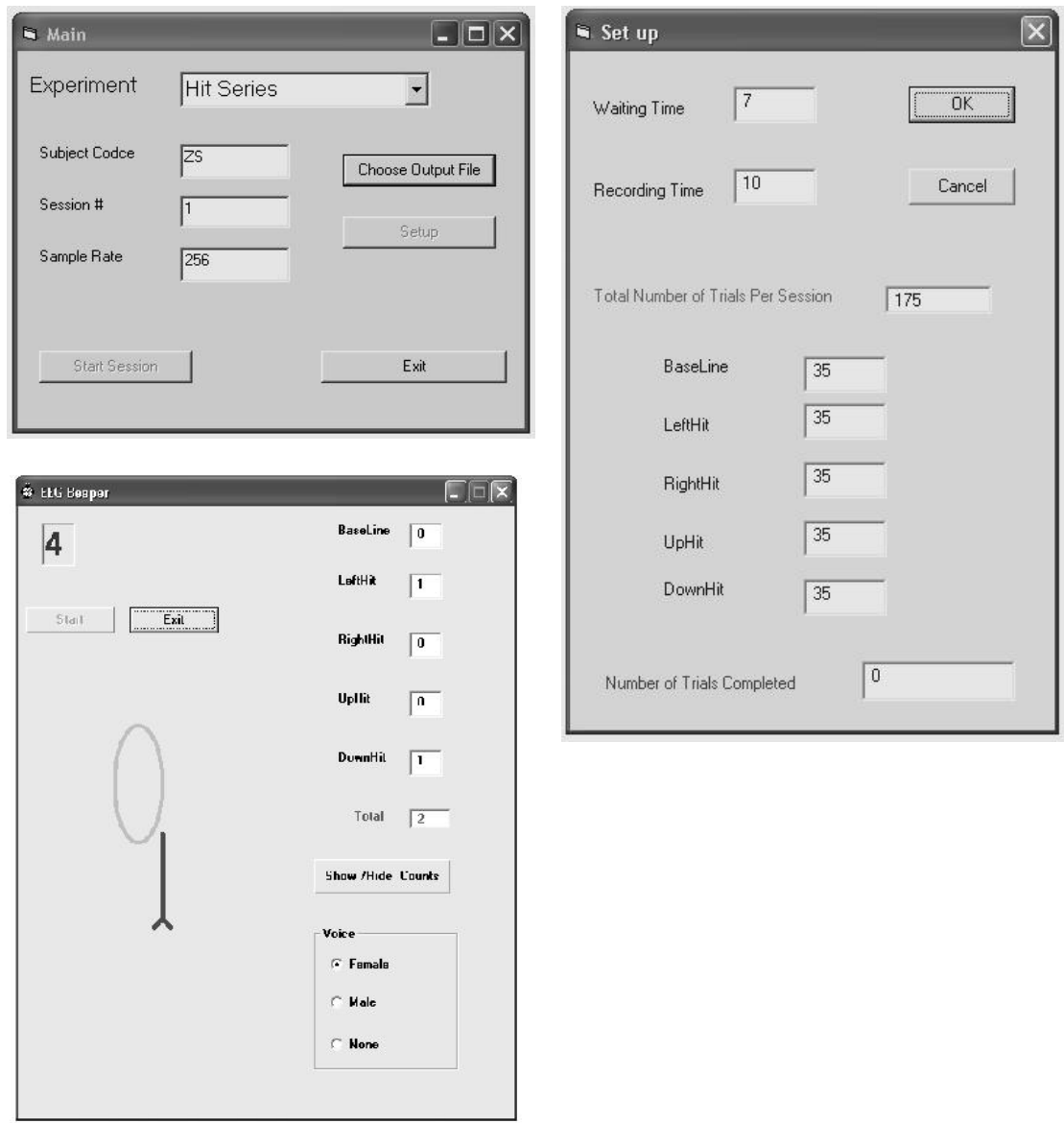

Figure 5.7. Alarm program.

Data acquisition is carried out with Mindmeld 24 Data acquisition software developed by NeuroPulse-Systems LLC, USA (Figure 5.8). MindMeld24 Live Data Capture (LDC) is used for data acquisition. This software provides facilities to set the sample rate at 64, 128, 256, or 512 samples per second and data block size at 96, 192, 384, or 768 bytes. A sample rate of 256 samples per second is sufficient for most EEG experiments. The default value for block size is 768 bytes per block. The block size setting is provided to help eliminate SCSI bus noise which may intrude onto the EEG data.

The EEG signals are displayed as a strip chart in LDC. This software has a facility to modify the number of channels to be displayed on the strip chart and trace color of each channel. Further it has an option to save data for off-line analysis.

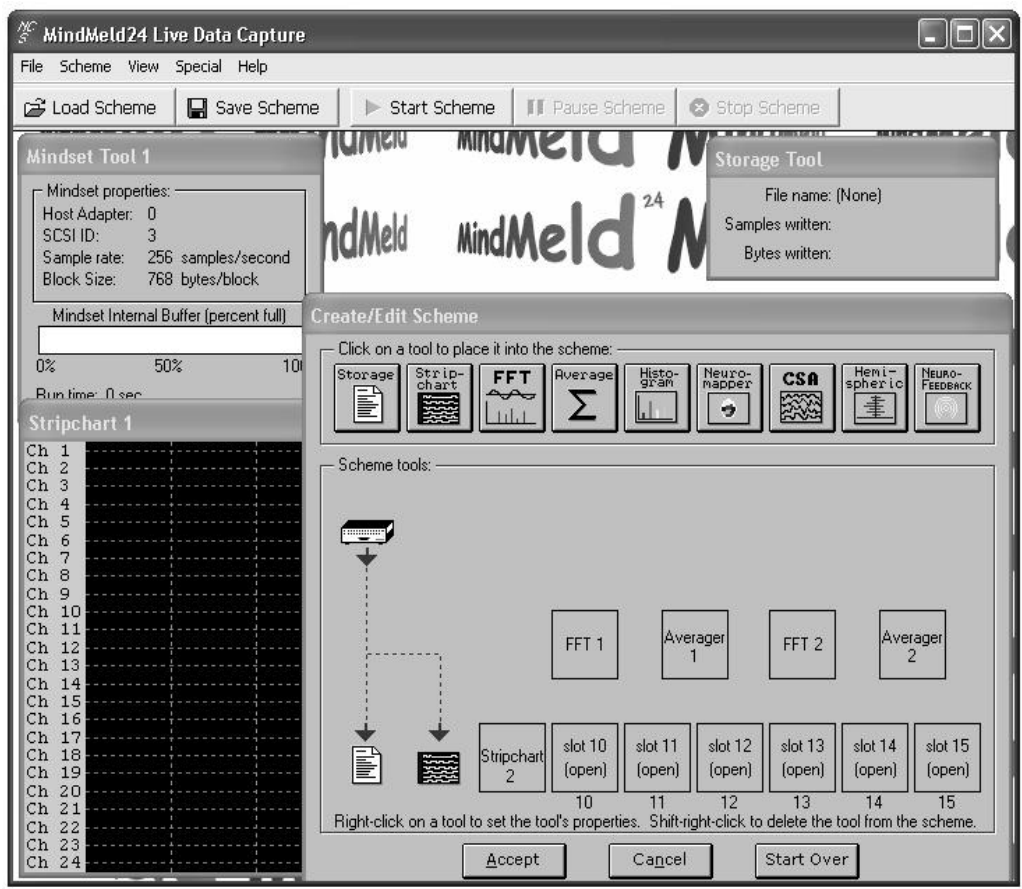

Figure 5.8. MindMeld24 Live Data Capture (LDC)

In addition, LDC software has a feature to calibrate the EEG amplifier system. For this purpose developer of this software has provided a scheme which contains required parameters for calibration. When the calibration is completed, the software automatically saves the calibration scalars in the windows registry.

The Data conversion program (DCon) has been developed to convert binary data structures saved by the LDC to data formats required by the data analysis software which will be described in the next section. DCon stores EEG data of each trial in a separate file in which each channel is stored as an individual column (Figure 5.9).

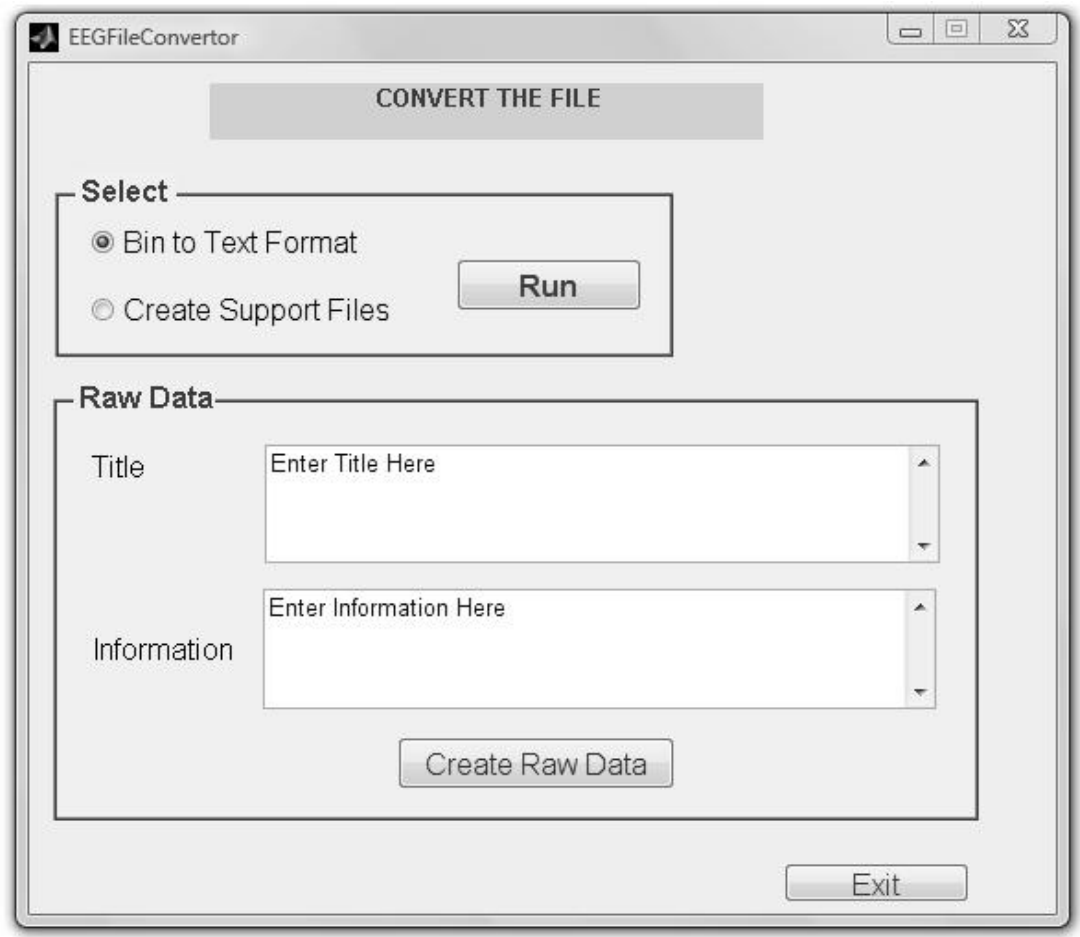

Figure 5.9. Data conversion program (DCon)

## 5.4 Signal processing and classification software

For signal processing and classification of EEG data, software called "IMTE" (Identification of mental tasks through EEG) has been developed with MATLAB® version 7.8.0.347 (R2009a). IMTE has a Graphical User Interface (GUI) with user friendly features and supports various signal processing techniques (Figure 5.10).

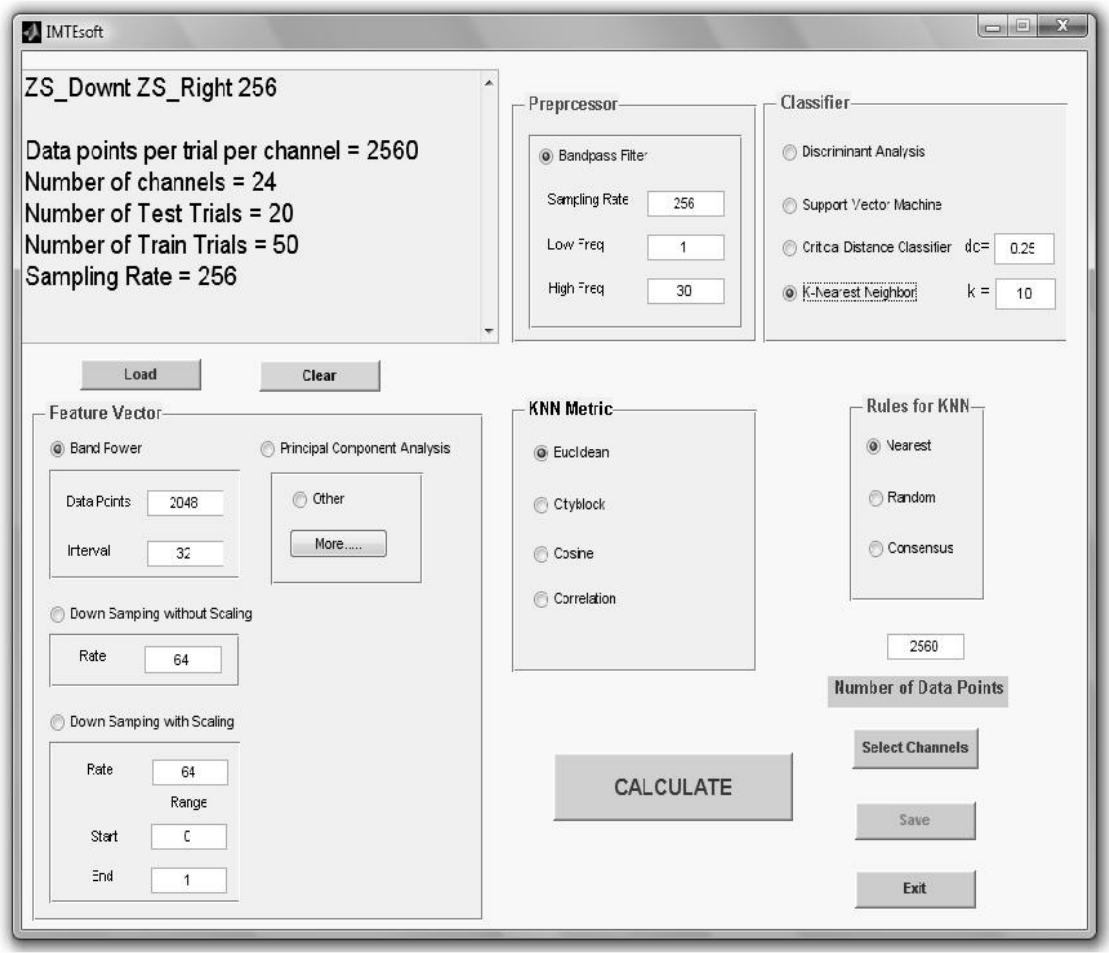

Figure 5.10. Identification of mental tasks through EEG (IMTE) main screen

We developed this software for testing various combinations of preprocessing, feature vector construction, and classification methods for optimizing the performance of identifying mental tasks. The parameters in these methods can be tuned through the facilities provided by GUI of IMTE to find the optimal parameters to be used in BCI systems. Moreover interested channels can be selected via channel selection panel and the number of data points to be used in the calculation can be entered at runtime (Figure 5.11).

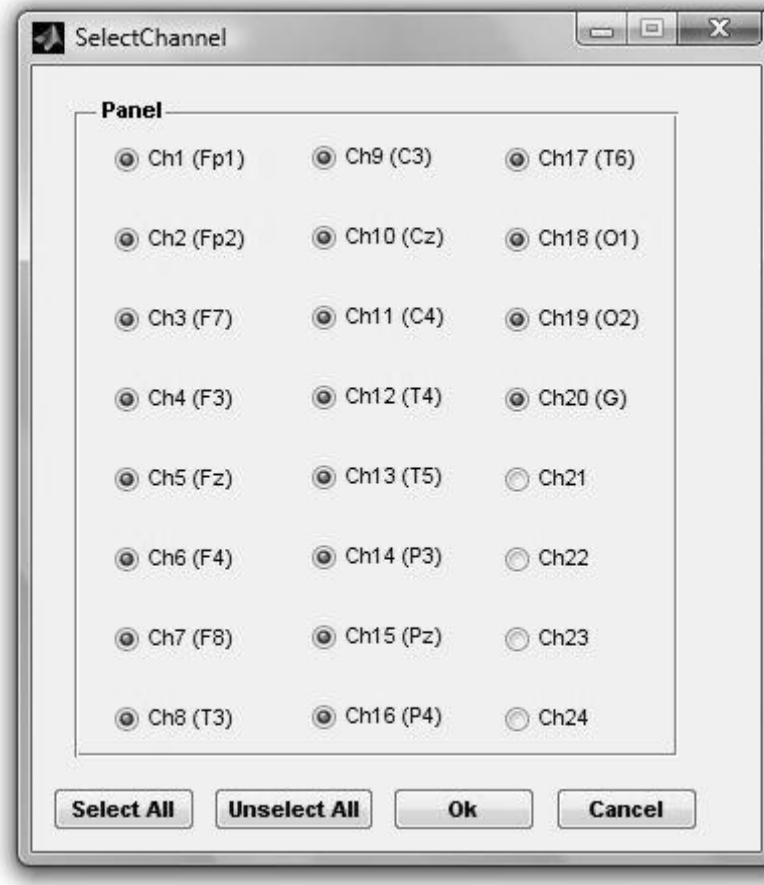

Figure 5.11. IMTE channel selection panel

User has freedom to choose which files are to be used as Train trails or Test trials and store them in either TRAIN or TEST folders accordingly. IMTE reads data directly from above folders and necessary data structures and arrays are constructed.

The signal processing and classification methods implemented in this software have been discussed in detail in Chapter 04. In the preprocessing stage, IMTE carries out bandpass filtering on each channel using Butterworth filters. User can enter low frequency and high frequency values of the bandpass filter at runtime.

Three methods are available for construction of feature vectors; Bandpower, Downsampling and Principal Component Analysis (PCA). IMTE provides facility to enter number of data points and size of the band to be used in Bandpower calculations. Downsampling can be carried out with or without scaling with IMTE. In the case of Downsampling with scaling, user can enter downsampling rate and the range of scaling. User can enter downsampling rate for the case of downsampling without scaling. For Principal Component Analysis, IMTE uses for PCA routings available in MATLAB.

In the classification stage user has following options to be used for optimizing the performance.

1. Under Discriminant Analysis, IMTE consists of Linear Discriminant Analysis, Diaglinear analysis, Quadratic analysis, Diagquadratic analysis, and Mahalanobis analysis.

2. Under Support Vector Machine, available options are Linear kernel or dot product, Quadratic kernel, Polynomial kernel, Gaussian radial basis function kernel (Rbf), and Multilayer perceptron kernel. For the Polynomial kernel case user can enter the order of the polynomial. For Rbf, sigma value can be entered at runtime.

3. K - Nearest Neighbor (KNN) classification: Number of nearest neighbors can be entered at runtime. Metric to be used with KNN can be chosen. The available choices given in IMTE are Euclidean, Cityblock, Cosine, and Correlation. Rules for KNN can also be entered at runtime. The options are Nearest, Random and Consensus.

4. Critical distance classification: This classification method has been developed and implemented by us for investigating possible improvements in classifying mental tasks which we have newly introduced. In this simple method user enters the critical distance $d_c$ and IMTE calculates the Euclidean distances between the test feature vector and each and every training feature vectors in the training set. Then this classification method claims that the mental task corresponding to the test feature vector should belong to the class which has the majority of feature vectors inside the hyper-sphere with radius $d_c$.

Following steps are followed by IMTE software when classifying mental tasks.

I. IMTE reads both testing and training data while it is being loaded to the memory.

II. It shows all the default parameters on GUI. Then user can change the parameters if necessary. Then user can press the CALCULATE button to start the calculation.

III. IMTE first performs preprocessing, and constructs feature vectors according to instructions given in the GUI.

IV. Then classification of mental tasks is carried out as to the chosen methods and parameters.

V. The outcome of the classification is displayed on GUI of IMTE. User has the option to save the outcome alone with methods and parameters used in the calculation to a text file.

There are many methods and parameters available to the user for optimizing the performance of classification. Therefore user may have to tryout many combination of methods and parameters before obtaining the optimal solution. Manually choosing and modifying signal processing techniques and parameters to be used in every calculation are time consuming.

As a result we have developed non-GUI based IMTE version call AutoIMTE (AIMTE) which accepts commands and parameters from a script file. One of the advantage of having commands and parameters in a script file is that multiple (can be hundreds or thousands) calculations corresponding to several parameter options can be carried out with a single input script file in a single run. The list of commands and parameters in the script file must have following format,

1. The first line of the script file is considered as the *Title* of the calculation.

2. If the user would like to enter more information or any other text, he/she could use ***Details*** command. This information can be entered multiple lines and should be terminated with the ***End*** command.

   e.g.
   ***Details***
   *Recorded Dates:*
   *Subject information:*
   *Description about how the data is recorded*
   *Any other information*
   ***End***

3. User can write any comment anywhere in the script file by starting a line with **%** and AIMTE will ignore lines starting with **%**. Further the software will ignore blank lines and multiple spaces while processing the input script file.

4. ***Channels*** is the next command in the script file. With the ***Channels*** command user can enter the channels to be used in the calculation. Multiple channels should be separated by spaces. Line followed by the command ***Channels*** should contain a number indicating how many channels ought to be chosen in a single run from the channels entered in the previous line.

Example -1,

*Channels 1 2 11 12*

*4*

This instructs the software to choose all the channels above (1, 2, 11, and 12) to be used in the calculation. Example -2,

*Channels 1 2 11 12*

*3*

This instructs the software to choose three channels out of *1, 2, 11,* and *12* at a time in the calculation. i.e., the software will automatically perform four separate runs for each of the following combinations:

1 2 11

1 11 12

1 2 12

2 11 12

Further, user does not have to type consecutive channels explicitly. The consecutive channels can be entered with **:** symbol. As an example, the line

*Channels 3 4 5 6 7 10 12*

can be written using **:** symbol as,

*Channels 3:7 10 12*

In particular $n$ **:** $m$ means $n$  $n+1$  $n+2$ …… $m-1$  $m$. Therefore use of **:** will make command lines shorter.

5. The next command AIMTE will accept is ***pprocess***. With this command, the user can instruct the preprocessing method which should be used in the calculation. The required parameters should be entered in the proceeding lines. As an example, if the bandpass filter should be used with 1 Hz as the low frequency and 30 Hz as the high frequency (i.e., frequency range for the bandpass filtering is [1 , 30] Hz, the command line  should be

*pprocess  bandpass*

*1*

*30*

If the user wants to do multiple calculations for different frequency ranges, with single set of commands can be used to instruct the program. As an example, the command

*pprocess  bandpass*

*5 8 9*

*30 31*

will instruct the software to carryout multiple calculations corresponding to the frequency ranges [5 , 30], [5 , 31], [8 , 30], [8 , 31], [9 , 30], and [ 9 , 31].

With the ***pprocess*** command the **:** symbol can also be used. As an example,

*pprocess  bandpass*

*3 4 5 10*

*20 21 22*

and,

*pprocess  bandpass*

*3:5 10*

*20:22*

are equivalent.

6.  In a similar manner commands ***feature*** and ***class*** can be used to provide the feature vector construction method and the classification method to be used in the calculation. With these commands parameters and the **:** symbol can be used as before.

7. The results of the calculations are saved in an output file alone with all the parameters used in the each calculation. A typical input script file alone with the output is given in the Appendix.

8. The AIMTE assumes that the input script file is placed in the INPUT folder and it saves the out file in the OUTPUT folder. If there are more than one input script file in the INPUT folder, it executes each script file separately and produces corresponding output files in the OUTPUT folder.

## 5.5 Preparation of subjects and recording of EEG

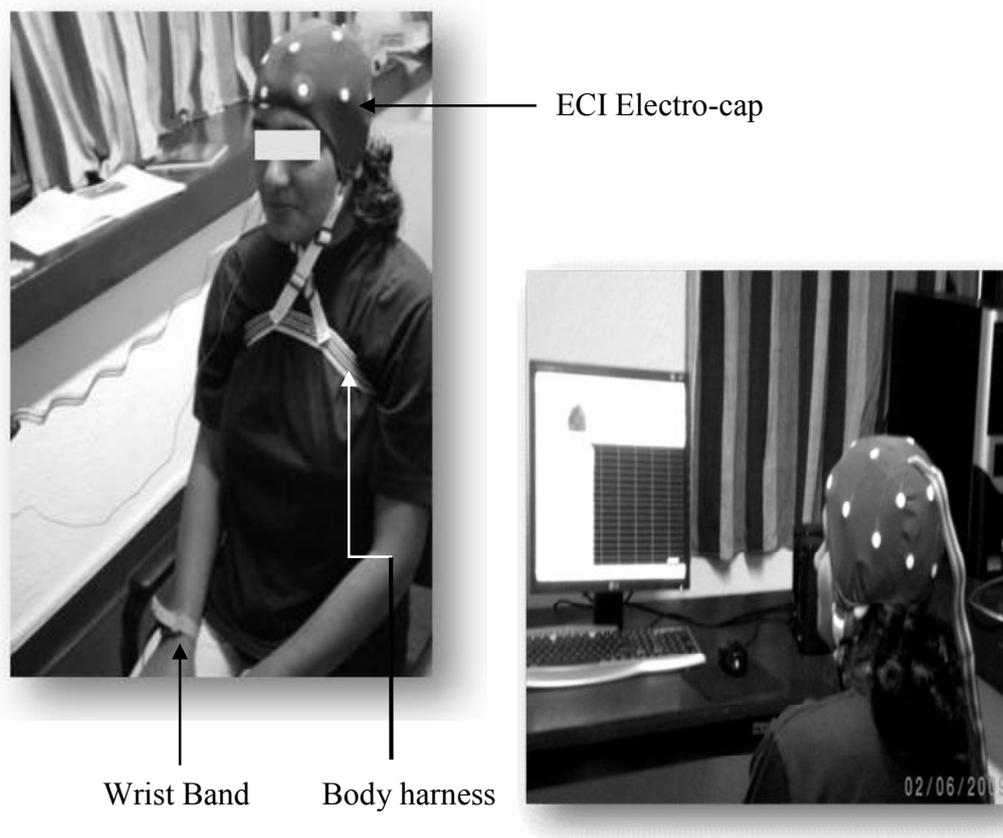

Figure 5.12. Subject and names of the various parts.

In this section first we present steps which are followed for preparing subjects for recording of EEG. The major steps followed are as follows (Figure 5.12),

i.    Put the harness on the subject.

ii.   Choose a proper size electro-cap for the subject.

iii.  Attach the ear electrodes: add conducting gel and abrade the skin.

iv.   Slip the cap on to the subject head and attach it to the body harness.

v.    Using the conducting gel electric connection to the scalp will be made. This is achieved by filling each electrode cavity with conducting gel and rocking the syringe rapidly back and forth.

vi.   Until the impedance between reference electrodes (ear electrodes) and individual electrodes in the cap are below 2.5 K$\Omega$, previous step will be followed.

vii.  Move the subject to a chair which has been placed in front of a blank wall.

viii. Electro-cap connector will be connected to the amplifier alone with ear electrodes.

ix.   Fix the wrist band to the subject firmly and connect the other end to the ground. This step is very important and all the electrical wires should be connected properly. This will reduce noise present in the EEG signal drastically.

Now the subject is ready for the recording session. Before starting the recordings, few test recordings are made to find out whether there is any apparent noise in the EEG signal. The recording sessions will start when test recordings show clean EEG signals.

Alarm Program which was described previously in the software section of this Chapter is used to inform the subject about the mental tasks he/she has to perform. Then after a certain amount of time (called a preparation period) the software will make a beep to inform the subject that he/she should start carrying out the mental task which was given before the preparation period. After a predetermined duration, the alarm program will again make another beep to inform the subject to stop the mental task. Those beeps will also inform the operator to start and stop the recording EEG signals receiving from the amplifier system. This procedure is repeated for each trial. Subjects were given time to relax and rest frequently as necessary.

Experimental methodology and the results will be discussed in the next Chapter.

# CHAPTER 6

# EXPERIMENTAL METHODOLOGY AND RESULTS

## 6.1 Experimental Sessions

As mention in the previous Chapter, in the preliminary studies, we found that most promising mental tasks are in the HS. We have carried out detail investigations of HS and compared it with performance of MI. HS series consists of four different mental tasks which are labeled as Right Hit (RH), Left Hit (LH), Up Hit (UH), and Down Hit (DH). The details of these labels are as follows (Table 6.1), MI consists of two mental tasks namely Left Middle Finger Movement (LFM) and Right Middle Finger Movement (RFM).

Table 6.1. Mental task labels and showing the images of the arrows.

| MENTAL TASKS LABELS | IMAGINATION | IMAGE |
|---|---|---|
| RH | Imagining a right arrow hitting a square from the left | 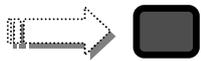 |
| LH | Imagining a left arrow hitting a square from the right | 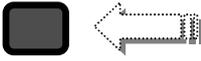 |
| UH | Imagining an up arrow hitting a square from the bottom | 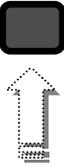 |
| DH | Imagining a down arrow hitting a square from the top | 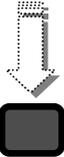 |

We have used similar settings, same recording parameters and same subjects for both sets of mental tasks (HS and MI). Following parameters and settings were used in all the recording sessions (Table 6.2).

Table 6.2. Parameters and settings used in all the recording sessions.

| | |
|---|---|
| Channels used for recording | All 20 channels according to 10 - 20 system |
| Sample rate | 256 samples per second |
| Block size | 768 bytes per block per channel |
| Preparation period used in alarm program | 7 seconds |
| Recording duration | 9 seconds |
| Number of Subjects | 03 |
| Number of Test trials per mental task per subject | 30 trials |
| Number of Train trials per mental task per subject | 90 trials |
| Total number of mental tasks including Baseline(BL) | 08 |
| Total number of test trials per subject | 240 trials |
| Total number of train trials per subject | 720 trials |
| Total number of trials per subject | 960 trials |

Subjects: Two male, one right-handed and one left handed subjects and one female left-handed subject (age between 22-32) took part in this study.

Before starting the experiment we calibrated the amplifier system as described in the previous Chapter. The EEG laboratory at Institute of Fundamental Studies has a separate EEG recording room which is subdivided in to two sections such that the subject and the recordist are well separated. The tube lights in the recording room were turned off during the recording sessions to remove the electrical noise produced by them which may contaminate the EEG signals. However, the room is dimly-lit due to light coming from the adjacent area of the laboratory. Also we made sure that EEG amplifier and the computer used for recording EEG are kept as far apart as possible to reduce the noise contamination due to the computer.

### 6.2 Recording procedure

Now we describe the procedure followed for recording the EEG corresponding to mental tasks in the Hit Series (HS). Subjects were seated in an armchair looking at a black square pasted on a white blank screen which was placed approximately one and half meters away from the subject. They were asked to keep their arms and hands relaxed and to avoid eye movements during the recordings. Each trial started with the alarm program informing the subject verbally what mental task should be imagined in the upcoming trial. Mental tasks are randomly chosen while keeping the total number of trials of each mental task in a recording session to a fixed value of 20. Seven seconds later, a short beep is made to inform the subject to start the mental task while keeping his/her eyes open. If it is observed that eye blinks or eye moments occurred during the recording of the trial, data recorded in the trial is discarded and the trial is repeated. Duration of a trial is nine seconds and at the end of nine seconds another beep is produced by the alarm program informing the subject to stop the mental task. We have carried out six recording sessions each consist of 100 trials for all three subjects.

For Motor imaginary (MI) sessions we followed a similar procedure as for HS with few exceptions. In this setup the black square on the white screen was removed and the subjects were asked not to pay attention to the white board but to concentrate on mental task they are performing instead.

After completing all the recordings of each subject, the data files were converted to the format that could be used for analyzing with IMTE software. Using both IMTE and AIMTE, we have analyzed all the data files recorded from all three subjects.

Performance of mental tasks were calculated as a percentage (6.1),

$$Performance = \frac{Number\ of\ successfully\ identified\ mental\ tasks}{Total\ number\ of\ mental\ tasks\ to\ be\ identified} \times 100 \quad (6.1)$$

Since always less number of EEG channels is favorable in BCI, we optimized the performance while keeping number of channels in the analysis to be less than six. During the analysis of EEG data from all the subjects, it was observed that downsampling methods and PCA performed poorly with all classification schemes. Moreover, classification methods in the discriminant analysis also performed poorly except diaglinear and diagquadratic. With the exception of linear and polynomial kernels all the other kernels in the Support Vector Machine (SVM) classification methods performed very poorly (less than 65%).

Therefore we will be presenting the results of the analysis carried out with feature vectors, constructed by bandpower method and classified by the classification methods given below.

(a) Diaglinear and Diagquadratic Discriminant analysis.

(b) Linear and Polynomial SVM kernels.

(c) Critical Distance classifier.

(d) k - Nearest Neighbor classifier (kNN).

Further, all the methods tested in this study could not identify the difference between RH and LH and similarly DH and UH. The best performance of classification for RH and LH is less than 65% while for DH and UH it is less than 70%. Therefore the results presented below are for BL, DH, and RH (Table 6.3, 6.4, 6.5). Further, the classifications carried out with certain combination of parameters for SVM linear and SVM polynomial did not converge and the best performance presented below are chosen only from the set of converged classifications.

Table 6.3. Best performance of the three subjects for Baseline and DownHit. Number of data points used in the calculation was 2048.

| SUBJECT 1 | | | | |
|---|---|---|---|---|
| Channels | Preprocessing | Feature Vectors | Classification | Performance |
| 1, 2, 10, 11 | Bandpass (1, 36) | BP (39) | Diaglinear | 90% |
| 1, 2, 10, 15 | Bandpass (1, 37) | BP (36) | Diagquadratic | 72% |
| 11, 13, 15, 18 | Bandpass (1, 31) | BP (33) | Linear SVM | 82% |
| 1, 2, 10, 13 | Bandpass (1, 43) | BP (48) | Polynomial SVM (3) | 93% |
| 2, 10, 11, 15 | Bandpass (1, 31) | BP (37) | Critical Distance (0.2) | 95% |
| 1, 2, 15, 19 | Bandpass (1, 39) | BP (34) | kNN ($k = 4$) | 98% |
| SUBJECT 2 | | | | |
| 1, 2, 13 | Bandpass (1, 35) | BP (33) | Diaglinear | 98% |
| 1, 2, 13 | Bandpass (1, 41) | BP (33) | Diagquadratic | 98% |
| 1, 2, 11 | Bandpass (1, 30) | BP (33) | Linear SVM | 93% |
| 11, 12, 13 | Bandpass (1, 34) | BP (35) | Polynomial SVM (3) | 98% |
| 2, 13, 18 | Bandpass (1, 44) | BP (38) | Critical Distance(0.25) | 100% |
| 2, 13, 18 | Bandpass (1, 42) | BP (41) | kNN ($k = 9$) | 100% |
| SUBJECT 3 | | | | |
| 2, 10, 13 | Bandpass (1, 34) | BP (48) | Diaglinear | 72% |
| 10, 13, 16 | Bandpass (1, 35) | BP (49) | Diagquadratic | 68% |
| 2, 6, 10 | Bandpass (1, 35) | BP (34) | Linear SVM | 82% |
| 2, 6, 10 | Bandpass (1, 34) | BP (35) | Polynomial SVM (3) | 85% |
| 2, 6, 10 | Bandpass (1, 38) | BP (48) | Critical Distance(0.25) | 85% |
| 9, 10, 13 | Bandpass (1, 32) | BP (31) | kNN ($k = 1$) | 88% |

Table 6.4. Best performance of the three subjects for Baseline and RightHit. Number of data points used in the calculation was 2048.

| SUBJECT 1 | | | | |
|---|---|---|---|---|
| **Channels** | **Preprocessing** | **Feature Vectors** | **Classification** | **Performance** |
| 2, 10, 11, 15 | Bandpass (1, 55) | BP (36) | Diaglinear | 87% |
| 2, 10, 13, 15 | Bandpass (1, 37) | BP (35) | Diagquadratic | 68% |
| 2, 10, 11, 12 | Bandpass (1, 45) | BP (36) | Linear SVM | 82% |
| 2, 10, 11, 15 | Bandpass (1, 47) | BP (42) | Polynomial SVM (3) | 87% |
| 2, 10, 11, 13 | Bandpass (1, 45) | BP (40) | Critical Distance (0.2) | 80% |
| 2, 11, 13, 19 | Bandpass (1, 37) | BP (43) | kNN ($k = 7$) | 88% |
| SUBJECT 2 | | | | |
| 11, 13, 18 | Bandpass (1, 30) | BP (30) | Diaglinear | 100% |
| 11, 13, 18 | Bandpass (1, 30) | BP (49) | Diagquadratic | 100% |
| 13, 18, 19 | Bandpass (1, 30) | BP (33) | Linear SVM | 98% |
| 1, 11, 13 | Bandpass (1, 30) | BP (38) | Polynomial SVM (3) | 100% |
| 11, 13, 18 | Bandpass (1, 39) | BP (33) | Critical Distance (0.2) | 100% |
| 11, 13, 18 | Bandpass (1, 38) | BP (34) | kNN ($k = 10$) | 100% |
| SUBJECT 3 | | | | |
| 1, 3, 13 | Bandpass (1, 30) | BP (43) | Diaglinear | 72% |
| 1, 5, 19 | Bandpass (1, 45) | BP (33) | Diagquadratic | 67% |
| 1, 5, 6 | Bandpass (1, 32) | BP (43) | Linear SVM | 85% |
| 1, 13, 19 | Bandpass (1, 42) | BP (41) | Polynomial SVM (3) | 87% |
| 5, 6, 13 | Bandpass (1, 36) | BP (33) | Critical Distance (0.2) | 90% |
| 6, 13, 18 | Bandpass (1, 38) | BP (45) | kNN ($k = 8$) | 88% |

Table 6.5. Best performance of the three subjects for RightHit and DownHit. Number of data points used in the calculation was 2048.

| SUBJECT 1 | | | | |
|---|---|---|---|---|
| **Channels** | **Preprocessing** | **Feature Vectors** | **Classification** | **Performance** |
| 1, 10, 11, 15 | Bandpass (1, 31) | BP (36) | Diaglinear | 82% |
| 1, 10, 11, 15 | Bandpass (1, 31) | BP (29) | Diagquadratic | 72% |
| 1, 2, 10, 11 | Bandpass (1, 30) | BP (35) | Linear SVM | 80% |
| 1, 2, 11, 15 | Bandpass (1, 30) | BP (42) | Polynomial SVM (3) | 82% |
| 1, 2, 10, 11 | Bandpass (1, 30) | BP (20) | Critical Distance (0.2) | 93% |
| 1, 2, 18, 19 | Bandpass (1, 34) | BP (42) | kNN ($k = 10$) | 93% |
| SUBJECT 2 | | | | |
| 1, 2, 11 | Bandpass (1, 30) | BP (49) | Diaglinear | 93% |
| 1, 2, 11 | Bandpass (1, 45) | BP (44) | Diagquadratic | 83% |
| 11, 18, 19 | Bandpass (1, 35) | BP (44) | Linear SVM | 87% |
| 1, 2, 11 | Bandpass (1, 34) | BP (43) | Polynomial SVM (3) | 98% |
| 1, 2, 11 | Bandpass (1, 30) | BP (40) | Critical Distance (0.2) | 98% |
| 1, 2, 11 | Bandpass (1, 30) | BP (44) | kNN ($k = 3$) | 100% |
| SUBJECT 3 | | | | |
| 3, 13, 18 | Bandpass (1, 30) | BP (45) | Diaglinear | 67% |
| 3, 10, 13 | Bandpass (1, 48) | BP (40) | Diagquadratic | 70% |
| 5, 13, 18 | Bandpass (1, 38) | BP (32) | Linear SVM | 80% |
| 5, 18, 19 | Bandpass (1, 46) | BP (45) | Polynomial SVM (3) | 75% |
| 3, 5, 19 | Bandpass (1, 38) | BP (44) | Critical Distance (0.2) | 78% |
| 5, 13, 18 | Bandpass (1, 38) | BP (38) | kNN ($k = 4$) | 83% |

The results presented above show the best performance of each subject. It is evident from the above table that the channels and the parameters used in preprocessing, feature vector construction, and classification have to be changed to achieve the optimal performance for each mental task. Further these parameters vary with the subjects as well.

However, in a real BCI system, channels, parameters and methods used for classifying the mental tasks have to be the same for a given subject. In a real BCI system, the channels, classification methods, and optimal parameters can be determined at the training stage automatically for a given subject. Therefore, the best performance of each subject for fixed channels, methods and parameters used in classifying the mental tasks is given in tables (6.6) below.

Table 6.6. Overall best performance of subjects for fixed channels, methods and parameters for Baseline and DownHit, Baseline and RightHit, and RightHit and DownHit. Number of data points used in the calculation was 2048.

| Channels | Mental Tasks | Preprocessing | Feature Vectors | Classification | Individual Performance | Overall |
|---|---|---|---|---|---|---|
| **SUBJECT 1** | | | | | | |
| 1, 2, 18, 19 | BL & RH | Bandpass (1, 40) | BP ( 42) | kNN ($k = 10$) | 75% | 89% |
| | BL & DH | | | | 90% | |
| | RH & DH | | | | 93% | |
| **SUBJECT 2** | | | | | | |
| 2, 11, 12 | BL & RH | Bandpass (1, 44) | BP ( 43) | kNN ($k = 4$) | 100% | 99% |
| | BL & DH | | | | 98% | |
| | RH & DH | | | | 100% | |
| **SUBJECT 3** | | | | | | |
| 6, 13, 18 | BL & RH | Bandpass (1, 38) | BP ( 45) | kNN ($k = 8$) | 88% | 78% |
| | BL & DH | | | | 77% | |
| | RH & DH | | | | 70% | |

For MI, we present the performance results in the same format as above to make comparisons (Table 6.7, 6.8, 6.9, 6.10) between MI and HS clear.

Table 6.7. Best performance of the three subjects for Baseline and Right Middle Finger Movement (RFM). Number of data points used in the calculation was 2048.

| SUBJECT 1 | | | | |
|---|---|---|---|---|
| **Channels** | **Preprocessing** | **Feature Vectors** | **Classification** | **Performance** |
| 8, 9, 12, 17 | Bandpass (1, 30) | BP (35) | Diaglinear | 70% |
| 9, 11, 12, 17 | Bandpass (1, 30) | BP (32) | Diagquadratic | 72% |
| 2, 13, 17, 18 | Bandpass (1, 41) | BP (41) | Linear SVM | 77% |
| 2, 12, 13, 17 | Bandpass (1, 33) | BP (48) | Polynomial SVM (3) | 77% |
| 8, 11, 12, 13 | Bandpass (1, 36) | BP (40) | Critical Distance(0.35) | 78% |
| 2, 3, 12, 17 | Bandpass (1, 34) | BP (35) | kNN ($k = 9$) | 78% |
| **SUBJECT 2** | | | | |
| 13, 16, 18 | Bandpass (1, 35) | BP (38) | Diaglinear | 98% |
| 1, 2, 8 | Bandpass (1, 36) | BP (37) | Diagquadratic | 93% |
| 1, 2, 8 | Bandpass (1, 30) | BP (37) | Linear SVM | 97% |
| 1, 2, 13 | Bandpass (1, 30) | BP (34) | Polynomial SVM (3) | 98% |
| 8, 11, 12 | Bandpass (1, 32) | BP (37) | Critical Distance (0.2) | 100% |
| 8, 11, 12 | Bandpass (1, 31) | BP (37) | kNN ($k = 7$) | 100% |
| **SUBJECT 3** | | | | |
| 5, 7, 16 | Bandpass (1, 30) | BP (37) | Diaglinear | 93% |
| 5, 7, 14 | Bandpass (1, 30) | BP (39) | Diagquadratic | 95% |
| 5, 7, 13 | Bandpass (1, 30) | BP (31) | Linear SVM | 97% |
| 1, 4, 5 | Bandpass (1, 32) | BP (41) | Polynomial SVM (3) | 97% |
| 5, 7, 14 | Bandpass (1, 31) | BP (37) | Critical Distance(0.25) | 97% |
| 1, 5, 7 | Bandpass (1, 37) | BP (46) | kNN ($k = 7$) | 98% |

Table 6.8. Best performance of the three subjects for Baseline and Left Middle Finger Movement (LFM). Number of data points used in the calculation was 2048.

| SUBJECT 1 | | | | |
|---|---|---|---|---|
| **Channels** | **Preprocessing** | **Feature Vectors** | **Classification** | **Performance** |
| 2, 3, 11, 12 | Bandpass (1, 30) | BP (36) | Diaglinear | 58% |
| 3, 9, 12, 13 | Bandpass (1, 54) | BP (37) | Diagquadratic | 60% |
| 8, 9, 11, 17 | Bandpass (1, 50) | BP (45) | Linear SVM | 65% |
| 8, 9, 11, 17 | Bandpass (1, 41) | BP (31) | Polynomial SVM (3) | 62% |
| 3, 9, 11, 12 | Bandpass (1, 31) | BP (32) | Critical Distance (0.2) | 77% |
| 11, 12, 13, 17 | Bandpass (1, 32) | BP (32) | kNN ($k = 1$) | 80% |
| **SUBJECT 2** | | | | |
| 8, 13, 16 | Bandpass (1, 30) | BP (40) | Diaglinear | 95% |
| 1, 2, 8 | Bandpass (1, 49) | BP (40) | Diagquadratic | 83% |
| 1, 2, 8 | Bandpass (1, 30) | BP (30) | Linear SVM | 95% |
| 8, 13, 16 | Bandpass (1, 40) | BP (39) | Polynomial SVM (3) | 97% |
| 8, 10, 16 | Bandpass (1, 30) | BP (40) | Critical Distance (0.2) | 98% |
| 1, 8, 11 | Bandpass (1, 30) | BP (40) | kNN ($k = 3$) | 100% |
| **SUBJECT 3** | | | | |
| 3, 5, 15 | Bandpass (1, 30) | BP (42) | Diaglinear | 92% |
| 7, 13, 19 | Bandpass (1, 30) | BP (38) | Diagquadratic | 90% |
| 3, 5, 7 | Bandpass (1, 30) | BP (42) | Linear SVM | 95% |
| 7, 10, 19 | Bandpass (1, 46) | BP (39) | Polynomial SVM (3) | 97% |
| 3, 7, 10 | Bandpass (1, 38) | BP (39) | Critical Distance (0.2) | 97% |
| 3, 7, 15 | Bandpass (1, 35) | BP (31) | kNN ($k = 4$) | 97% |

Table 6.9. Best performance of the three subjects for Left Middle Finger Movement (LFM) and Right Middle Finger Movement (RFM). Number of data points used in the calculation was 2048.

| | | | | |
|---|---|---|---|---|
| **SUBJECT 1** | | | | |
| **Channels** | **Preprocessing** | **Feature Vectors** | **Classification** | **Performance** |
| 2, 9, 11, 12 | Bandpass (1, 30) | BP (31) | Diaglinear | 62% |
| 2, 3, 11, 12 | Bandpass (1, 55) | BP (40) | Diagquadratic | 62% |
| 2, 8, 12, 13 | Bandpass (1, 41) | BP (31) | Linear SVM | 65% |
| 2, 8, 12, 13 | Bandpass (1, 56) | BP (40) | Polynomial SVM (3) | 75% |
| 3, 11, 13, 17 | Bandpass (1, 30) | BP (31) | Critical Distance(0.25) | 73% |
| 2, 8, 12, 13 | Bandpass (1, 30) | BP (31) | kNN ($k = 4$) | 75% |
| **SUBJECT 2** | | | | |
| 1, 8, 16 | Bandpass (1, 50) | BP (33) | Diaglinear | 60% |
| 8, 13, 16 | Bandpass (1, 39) | BP (33) | Diagquadratic | 62% |
| 8, 13, 16 | Bandpass (1, 49) | BP (49) | Linear SVM | 80% |
| 1, 11, 16 | Bandpass (1, 46) | BP (42) | Polynomial SVM (3) | 77% |
| 2, 11, 16 | Bandpass (1, 40) | BP (30) | Critical Distance(0.15) | 70% |
| 2, 11, 16 | Bandpass (1, 41) | BP (43) | kNN ($k = 5$) | 77% |
| **SUBJECT 3** | | | | |
| 3, 5, 19 | Bandpass (1, 30) | BP (44) | Diaglinear | 72% |
| 4, 5, 15 | Bandpass (1, 31) | BP (31) | Diagquadratic | 78% |
| 3, 4, 5 | Bandpass (1, 30) | BP (32) | Linear SVM | 80% |
| 3, 4, 5 | Bandpass (1, 35) | BP (49) | Polynomial SVM (3) | 82% |
| 3, 5, 15 | Bandpass (1, 38) | BP (33) | Critical Distance(0.15) | 78% |
| 4, 5, 15 | Bandpass (1, 32) | BP (33) | kNN ($k = 6$) | 82% |

Table 6.10. Overall best performance of subjects for fixed channels, methods and parameters for Baseline (BL) and Right Middle Finger Movement (RFM), Baseline and Left Middle Finger Movement (LFM), and RFM and LFM. Number of data points used in the calculation was 2048.

| Channels | Mental Tasks | Preprocessing | Feature Vectors | Classification | Individual Performance | | Overall |
|---|---|---|---|---|---|---|---|
| **SUBJECT 1** | | | | | | | |
| 2, 8, 12, 13 | BL & RFM | Bandpass (1, 30) | BP ( 34) | kNN ($k = 10$) | 78% | | 69% |
| | BL & LFM | | | | 65% | | |
| | RFM & LFM | | | | 65% | | |
| **SUBJECT 2** | | | | | | | |
| 1, 8, 11 | BL & RFM | Bandpass (1, 30) | BP ( 40) | kNN ($k = 3$) | 97% | | 87% |
| | BL & LFM | | | | 100% | | |
| | RFM & LFM | | | | 63% | | |
| **SUBJECT 3** | | | | | | | |
| 4, 5, 15 | BL & RFM | Bandpass (1, 33) | BP ( 45) | kNN ($k = 6$) | 93% | | 88% |
| | BL & LFM | | | | 90% | | |
| | RFM & LFM | | | | 82% | | |

The next Chapter contains a summary of the research work carried out in this study and conclusions made based on the results presented above.

# CHAPTER 7

# DISCUSSION AND CONCLUSIONS

## 7.1 Discussion

BCI is one of the most active areas in Artificial Intelligence and Machine Learning. Although the main use of BCI is considered as a helping tool for severely handicapped people, some BCI research laboratories concentrate on its use in virtual reality, rehabilitation, multimedia communication and entertainment / relaxation.

There are several reasons why BCI has become an active area of research. BCI is a new neuroscience paradigm that might help to understand how the human brain works in terms of plasticity and reorganization, learning memory, attention, thinking, and motivation, etc. BCI research can also assist developing a new class of bioengineering control devices and robots to provide daily life assistance to handicapped and elderly people. BCI also has a potential for rehabilitation, treating emotional disorders, easing chronic pain, and overcoming movement disabilities due to strokes or accidents.

The BCI community is multidisciplinary. It is composed of physicians, clinicians, engineers, technicians and so on. The field is however strongly dominated by the medical staff. Research groups are distributed all over the world, but mainly from United States and Europe. Asia is emerging and considered to be seemingly very promising as shown by the results of the last world wide BCI classification competition where they won most of the contests. Historically, United States is biased towards medical field, while Asia is biased towards machine learning methods and Europe is more or less in between (geographically and technically speaking).

In this thesis we presented results of BCI research carried out for the MPhil degree project. In this research, our aim was twofold. The first aim was to find new (non-motor) mental tasks which can alter EEG signals and hence can be used in BCI systems effectively. The second aim was to investigate several computational techniques for analyzing and classifying EEG signals and find suitable preprocessing methods, effective and efficient feature vector construction techniques and most accurate classification methods for recognizing mental tasks.

In Chapter 5 we presented the performance of various mental tasks, pre-processing methods, feature vector construction techniques and classification methods. It was evident from the tables that same signal processing techniques do not provide optimal performance for all mental tasks. They vary with the mental tasks and signal processing methods have to be optimized according to the mental task. The best performance of mental tasks which are reported in the literature [1] varies from 52.6% to 98.6%. The 52.6% is corresponding to $M_{10}$ & $M_{11}$ & $M_{12}$ and $M_{13}$ (as labeled in Chapter 5) and they have used Bandpower as the feature vector construction method and Hidden Markov Method for classification (Table 5.3). On the other hand mental tasks $M_2$ & $M_3$ & $M_4$ & $M_5$ gave the 98.6% performance when they were analyzed with PSD with Welch's periodogram for feature vector construction and Multilayer Perceptron for classification (Table 5.1).

In this study we have examined several new mental tasks for potential use in BCI systems. Among all, we found that HS is the most promising set of mental tasks to be used in BCI systems. Since MI is one of commonly used mental tasks in the BCI research and best performance of two classes of MI is 90% (See Table 5.3), we have analyzed the mental tasks in two classes of MI using EEG data recorded from the subjects who were participated in the recordings of HS.

## 7.2 Conclusions

It is evident from the results presented in the Chapter 6 that compared to MI, HS performed better. Optimal performance was achieved with three EEG channels for two subjects (Subject 2 and Subject 3) while four EEG channels for the other (Subject 1). The Subject 2 achieved ideal best performance of 100% for HS while 92% for MI. Further the Subject 2 performed 100% for many combinations of parameters in preprocessing and feature vector construction methods. On the other hand Subject 1 performed very well (88% - 98%) for the mental tasks in HS, while performed poorly (75% - 80%) for MI. Best performance of Subject 3 for HS was good (80% - 90%) while the same for MI is much better (82% - 98%) than performance for HS.

Bandpass filtering for preprocessing and Bandpower for feature vector construction have been used in all the calculations as their overall performance were found to be very good. For classification, Linear and Polynomial SVM kernels, Diaglinear and Diagquadratic Discriminant analysis, Critical Distance classifier, and K - Nearest Neighbor classifier (KNN) have been utilized. It was found that KNN out performed all the other classification schemes in almost all mental tasks in both HS and MI. The second best classification performance was observed with Critical Distance classifier which is related to KNN. Linear and Polynomial SVM kernel classifies showed good performance although several calculations failed to converge.

When we combine performances of all the subjects, new mental tasks (HS) introduced in this thesis performed better than the most widely used MI. One of the major advantages of using mental tasks RH and DH (in HS) for moving a cursor on a computer screen is that compared to other mental tasks such as multiplication, letter composing, LFM or RFM in MI, it is very easy and natural for a user to imagine RH to move the cursor to the right while think of DH to move the cursor down. On the other hand, LFM and RFM can also be used similar manner to simulate clicking of left mouse button and right mouse button respectively. Therefore Hybrid system of MI and HS is ideal for controlling computers through icons in real time BCI systems.

# Appendix

Given below is an example of an Input script file for AutoIMTE.

```
Classifications of BL and DH (HS)
Details
Recorded Date: 16/10/2009
Subject:GW
Sample Rate 256 and Recording Duration 8 Seconds
end

channels 1 2 15 18 19
4

pprocess    bandpass
1
38 39

feature bandpower
2048
34

class knn
4:6
1
1
```

Output file generated by AutoIMTE for the input script file given above.

```
********************************************************************
  AUTO IDENTIFICATION OF MENTAL TASKS THROUGH ELECTROENCEPHOLOGRAPHY
                    (EEG). AutoIMTE BCI Software
********************************************************************
Analysis of EEGData recorded in EEGLAB at Institute of Fundamental
Studies

Developed by - A.Nanayakkara and S.Zahmeeth
Date : 19-Nov-2009
Time : 17:39:55
```

```
                      Details of the Dataset
                      ----------------------
Data : DW_BaseL DW_Downt 256
Subject Name : DW
Recorded Duration: 8 Sec.
Recorded Number of Channels: 24
Recorded Sample rate: 256 Hz
Number of Data Points (RecordTime*SampleRate): 2048
Number of Train Trials: 180
Number of Test Trials: 60
Recorded Date: 16/10/2009
Subject:GW
Sample Rate 256 and Recording Duration 8 Seconds
```

```
======================================================================
Classifications of BL and DH (HS)
======================================================================
Start Calculation.............

Chosen Channels:  1 2 15 18 19
Preprocessing method: bandpass
Low Frequencies:  1
High Frequencies:  38 39
Feature Vector method: bandpower
Data points:  2048
Interval:  34
Classification method: knn
Order:  4 5 6
KNN Metric:  1
Rules for KNN:  1

======================================================================
                              RESULTS
                              -------

Channels: 1 2 15 18      bandpass(1,38)    knn(4,1,1)   91.666667 %

Channels: 1 2 15 18      bandpass(1,38)    knn(5,1,1)   88.333333 %

Channels: 1 2 15 18      bandpass(1,38)    knn(6,1,1)   90.000000 %

Channels: 1 2 15 18      bandpass(1,39)    knn(4,1,1)   91.666667 %

Channels: 1 2 15 18      bandpass(1,39)    knn(5,1,1)   88.333333 %

Channels: 1 2 15 18      bandpass(1,39)    knn(6,1,1)   90.000000 %

Channels: 1 2 15 19      bandpass(1,38)    knn(4,1,1)   95.000000 %

Channels: 1 2 15 19      bandpass(1,38)    knn(5,1,1)   91.666667 %

Channels: 1 2 15 19      bandpass(1,38)    knn(6,1,1)   93.333333 %

Channels: 1 2 15 19      bandpass(1,39)    knn(4,1,1)   98.333333 %

Channels: 1 2 15 19      bandpass(1,39)    knn(5,1,1)   91.666667 %

Channels: 1 2 15 19      bandpass(1,39)    knn(6,1,1)   95.000000 %

Channels: 1 2 18 19      bandpass(1,38)    knn(4,1,1)   83.333333 %

Channels: 1 2 18 19      bandpass(1,38)    knn(5,1,1)   86.666667 %

Channels: 1 2 18 19      bandpass(1,38)    knn(6,1,1)   86.666667 %

Channels: 1 2 18 19      bandpass(1,39)    knn(4,1,1)   81.666667 %

Channels: 1 2 18 19      bandpass(1,39)    knn(5,1,1)   83.333333 %

Channels: 1 2 18 19      bandpass(1,39)    knn(6,1,1)   86.666667 %

Channels: 1 15 18 19     bandpass(1,38)    knn(4,1,1)   88.333333 %
```

```
Channels: 1 15 18 19    bandpass(1,38)    knn(5,1,1)  86.666667 %

Channels: 1 15 18 19    bandpass(1,38)    knn(6,1,1)  83.333333 %

Channels: 1 15 18 19    bandpass(1,39)    knn(4,1,1)  88.333333 %

Channels: 1 15 18 19    bandpass(1,39)    knn(5,1,1)  88.333333 %

Channels: 1 15 18 19    bandpass(1,39)    knn(6,1,1)  85.000000 %

Channels: 2 15 18 19    bandpass(1,38)    knn(4,1,1)  81.666667 %

Channels: 2 15 18 19    bandpass(1,38)    knn(5,1,1)  73.333333 %

Channels: 2 15 18 19    bandpass(1,38)    knn(6,1,1)  81.666667 %

Channels: 2 15 18 19    bandpass(1,39)    knn(4,1,1)  81.666667 %

Channels: 2 15 18 19    bandpass(1,39)    knn(5,1,1)  75.000000 %

Channels: 2 15 18 19    bandpass(1,39)    knn(6,1,1)  81.666667 %

~~~~~~~~~~~~~~~~~~~~~~~~~~~~~~~~~~~~~~~~~~~~~~~~~~~~~~~~~~~~~~~~~~~~~~
                        General Information
                        -------------------
10-20 system followed in this analysis as follows :

Ch(1) --- FP1                   Ch(2) --- FP2
Ch(3) --- F7                    Ch(4) --- F3
Ch(5) --- FZ                    Ch(6) --- F4
Ch(7) --- F8                    Ch(8) --- T3
Ch(9) --- C3                    Ch(10)--- CZ
Ch(11)--- C4                    Ch(12)--- T4
Ch(13)--- T5                    Ch(14)--- P3
Ch(15)--- PZ                    Ch(16)--- P4
Ch(17)--- T6                    Ch(18)--- O1
Ch(19)--- O2                    Ch(20)--- Ground(G)
Ch(21)-Ch(24)  --- Auxiliary 1 - 4 respectively

Electrode Placement Information:

Left side of the Brain (LHSB)   Right side of the Brain (RHSB)
Ch(1) --- FP1                   Ch(2) --- FP2
Ch(3) --- F7                    Ch(6) --- F4
Ch(4) --- F3                    Ch(7) --- F8
Ch(8) --- T3                    Ch(11)--- C4
Ch(9) --- C3                    Ch(12)--- T4
Ch(13)--- T5                    Ch(16)--- P4
Ch(14)--- P3                    Ch(17)--- T6
Ch(18)--- O1                    Ch(19)--- O2

Middle of the Brain (MB)        Neglected Channels
Ch(5) ---FZ                     Ch(21) ---Auxiliary 1
Ch(10)---CZ                     Ch(22) ---Auxiliary 2
Ch(15)---PZ                     Ch(23) ---Auxiliary 3
Ch(20)---Ground(G)              Ch(24) ---Auxiliary 4

                   ~~~ END OF THE PROCESS ~~~
```